\documentclass[
aps, 
reprint, 
superscriptaddress, 
floatfix,
longbibliography, 
]{revtex4-2}

\usepackage{graphicx}
\usepackage{dcolumn}
\usepackage{bm}

\usepackage{amsthm}
\usepackage{amsmath,amssymb} 
\usepackage[utf8]{inputenc}
\usepackage[english]{babel}
\usepackage{dsfont}
\usepackage{xcolor} 
\usepackage{mathtools}
\usepackage{comment}
\usepackage{tabularx}
\usepackage{algpseudocode}
\usepackage{physics}
\usepackage{float}
\usepackage{booktabs}
\usepackage{array}
\usepackage{multirow}
\usepackage{xspace}
\usepackage{subfigure}
\usepackage{empheq}
\usepackage[group-separator={,}]{siunitx}
\usepackage[columnwise]{lineno}
\usepackage{hyperref}
\hypersetup{
colorlinks=true,
urlcolor= blue,
citecolor=blue,
linkcolor= blue,
}

\newcommand{\bl}[1]{\textcolor{blue}{{#1}}}

\newcommand{\Met}{\textcolor{blue}{\emph{Methods}}\xspace} 
\newcommand{\SM}{\textcolor{blue}{\emph{Supplementary Information}}\xspace}

\newcommand{\p}{\rm{p}}
\newcommand{\q}{\rm{q}}
\newcommand{\at}{\rm{a}}
\newcommand{\df}{\rm{d}}
\newcommand{\ra}{\mathbf{r}_{\at}}
\newcommand{\rd}{\mathbf{r}_{\df}}
\newcommand{\rp}{\mathbf{r}_{\p}}
\newcommand{\rqq}{\mathbf{r}_{\q}}
\newcommand{\rtp}{\tilde{\mathbf{r}}_{\p}}
\newcommand{\rtq}{\tilde{\mathbf{r}}_{\q}}
\newcommand{\vat}{\mathbf{v}_{\at}}
\newcommand{\vdf}{\mathbf{v}_{\df}}
\newcommand{\vp}{\mathbf{v}_{\p}}
\newcommand{\eo}{\mathbf{e}_1}
\newcommand{\et}{\mathbf{e}_2}
\newcommand{\ta}{\theta_{\at}}
\newcommand{\td}{\theta_{\df}}
\newcommand{\tp}{\theta_{\p}}
\newcommand{\taup}{\tau_{\p}}
\newcommand{\taua}{\tau_{\at}}
\newcommand{\taud}{\tau_{\df}}
\newcommand{\fp}{f_{\p}}
\newcommand{\fa}{f_{\at}}
\newcommand{\fd}{f_{\df}}
\newcommand{\Fp}{\mathbf{F}_{\p}}
\newcommand{\Fa}{\mathbf{F}_{\at}}
\newcommand{\Fd}{\mathbf{F}_{\df}}

\newcommand{\epsp}{\epsilon_{\p}}

\newcommand{\hatra}{\hat{\mathbf{r}}_{\at}}
\newcommand{\hatrd}{\hat{\mathbf{r}}_{\df}}
\newcommand{\hatvat}{\hat{\mathbf{v}}_{\at}}
\newcommand{\hatvdf}{\hat{\mathbf{v}}_{\df}}

\newcommand{\dvatdt}{\dot{\mathbf{v}}_{\at}}
\newcommand{\dvdfdt}{\dot{\mathbf{v}}_{\df}}

\newcommand{\rr}{\mathbf{r}}
\newcommand{\vv}{\mathbf{v}}
\newcommand{\RR}{\mathbf{R}}
\newcommand{\VV}{\mathbf{V}}
\newcommand{\er}{\mathbf{e}_r}
\newcommand{\es}{\mathbf{e}_s}

\newcommand{\rhoa}{\rho_{\at}}
\newcommand{\rhod}{\rho_{\df}}
\newcommand{\rhop}{\rho_{\p}}
\newcommand{\psia}{\psi_{\at}}
\newcommand{\psid}{\psi_{\df}}

\newcommand{\phia}{\phi_{\at}}
\newcommand{\phid}{\phi_{\df}}
\newcommand{\phip}{\phi_{\p}}
\newcommand{\Psia}{\Psi_{\at}}
\newcommand{\Psid}{\Psi_{\df}}
\newcommand{\Psip}{\Psi_{\p}}

\newcommand{\opts}{\mathcal{S}}

\newcommand{\pnl}[1]{\textbf{#1}}

\newcommand{\MetSec}{\SM \bl{Sec.}~\bl{S1}\xspace}

\newcommand{\SIFigSec}{\SM \bl{Sec.}~\bl{S3}\xspace}

\newcommand{\AppFig}[1]{Appendix Fig.~\ref{#1}}
\newcommand{\AppTwoFigs}[2]{Appendix Figs.~\ref{#1} and \ref{#2}}

\newcounter{extdatafig}
\renewcommand{\theextdatafig}{\arabic{extdatafig}}

\newcommand{\exfigcaption}[2][\linewidth]{
  \refstepcounter{extdatafig}
  \par\medskip
  \noindent\parbox{#1}{
    \raggedright
    \textbf{Appendix Figure~\theextdatafig.} #2
  }
  \par\medskip
}

\newcommand{\totfiltered}{\num[group-separator={,}]{18088}\xspace}
\newcommand{\tauemp}{2.97\xspace}
\newcommand{\eventdurationemp}{2.52\xspace}
\newcommand{\totunfiltered}{\num[group-separator={,}]{19231}\xspace}
\newcommand{\fracmatchingsigns}{85\%\xspace}

\addto\captionsenglish{}
\addto\captionsenglish{}

\begin{document}

\title{A behavioral principle underlying attacker--defender interactions in soccer}

\author{Issei Yamazaki}
\thanks{Contributed equally with H.G.}
\affiliation{Graduate School of Advanced Mathematical Sciences, Meiji University, Nakano, Tokyo 164-8525, Japan}

\author{Hirotaka Goto}
\thanks{Contributed equally with I.Y.; corresponding author}
\affiliation{Graduate School of Advanced Mathematical Sciences, Meiji University, Nakano, Tokyo 164-8525, Japan}
\affiliation{Department of Biology, University of Pennsylvania, Philadelphia, PA 19104, United States}

\author{Kojiro Otoguro}
\affiliation{Meiji Institute for Advanced Study of Mathematical Sciences, Meiji University, Nakano, Tokyo 164-8525, Japan}

\author{Hiraku Nishimori}
\affiliation{Meiji Institute for Advanced Study of Mathematical Sciences, Meiji University, Nakano, Tokyo 164-8525, Japan}

\author{Masashi Shiraishi}
\affiliation{Meiji Institute for Advanced Study of Mathematical Sciences, Meiji University, Nakano, Tokyo 164-8525, Japan}
\affiliation{Graduate School of Information Sciences, Hiroshima City University, 3-4-1 Ozukahigashi, Asaminami-ku, Hiroshima 731-3194, Japan}

\author{Takuma Narizuka}
\email[Corresponding author]{}
\affiliation{Faculty of Data Science, Rissho University, Kumagaya, Saitama 360-0194, Japan}

\date{July 6, 2026}

\begin{abstract}

Soccer is widely popular for its simple rules and complex yet coordinated play that unfolds on the pitch. 
Nevertheless, the fundamental mechanisms governing such play are not well understood: 
what shapes player interactions on the pitch? 
What short-term goals guide players' decisions about their movements over the next few seconds? 
We address these questions by focusing on one-on-one settings in open play, in which the attacker, in possession of the ball and typically dribbling, faces a defender aiming to stop or delay the attacker's actions over a short period. 
Here we develop a mathematical model of attacker--defender interactions and analyze 306 professional soccer games. 
Synthesizing the large-scale dataset with an analysis of the model reveals a simple behavioral principle that may underlie these interactions: the defender seeks to minimize their future relative speed to the attacker, whereas the attacker initiates their movements to preempt the defender's objective. 
This principle, relative-speed minimization, provides a consistent and unified account of the empirical data. 
Since our framework depends little on soccer-specific details, this principle may govern diverse pursuit--evasion scenarios as well as other invasion team sports. 

\end{abstract}

\maketitle

\begin{quote}
    The ball is round and the game lasts 90 minutes. Everything else is pure theory.
\end{quote} 
As German soccer coach Sepp Herberger famously remarked \cite{americanphilatelicsociety,stephens2022no}, soccer (association football) is notoriously unpredictable. 
Nonetheless, among the world's most recognized sports, it has been studied extensively by physicists, revealing various statistical laws and patterns within and across games \cite{mendes2007statistics, clauset2015safe, marcelino2020collective, yamamoto2025polya}.

In recent decades, modern team sports have provided powerful testbeds for the study of collective behaviors arising from mobile agents \cite{gudmundsson2017spatio, fujii2018prediction, koudela2025investigating, merritt2014scoring, araujo2016team, sarlis2020sports, bai2021sports, yamamoto2021preferential, chacoma2022simple, chacoma2023probabilistic, nguyen2025fractional, li2025machine, parker2024framework, cheng2025information, nakahara2023action, fujii2025machine, ichinose2020winning, yeung2024strategic, fewell2012basketball, ribeiro2017team, heuer2010soccer, chacoma2021stochastic, chacoma2022complexity}. 
Coordination among players is often well captured at a macroscopic level. 
However, a mechanistic understanding of individual interactions between players remains limited in most invasion sports, let alone in soccer, which stands out for its continuous flow of play over vast spatial scales. 
Even more elusive is whether these interactions are governed by a simple behavioral principle.

Previous studies have revealed long-range spatiotemporal organization among soccer players, including analysis of ball passing 
\cite{narizuka2014statistical, ichinose2021robustness, morishita2025tactical} 
and player positioning 
\cite{spearman2018beyond, narizuka2019clustering, narizuka2021space, fernandez2018wide}.  
Studies have also analyzed highly structured scenarios, ranging from isolated player motion 
\cite{keller1973theory, fajen2003behavioral, fujimura2005geometric, lopez2020touch, narizuka2023validation} to set pieces 
\cite{chiappori2002testing, wang2024tacticai}. 
By contrast, direct \emph{attacker--defender interactions} have received comparatively little attention \cite{narizuka2016statistical}. 
These interactions are neither completely structured nor spread across full temporal and spatial scales. 
Studying one-on-one interactions spanning such short but meaningful timescales---typically a few seconds---also demands large, high-resolution datasets, which were not readily available until recently \cite{pappalardo2019public, bassek2025integrated}. 
Despite the limited focus to date, attacker--defender interactions could be the right unit for understanding player motion in soccer.

In this paper, using an extensive dataset from a professional soccer league, we study one-on-one encounters and ask the following question: 
is there a simple universal principle underlying player motion and decision-making that has been obscured by the complexity of the game? 
A one-on-one interaction arises when two opposing players---the attacker (in possession of the ball, typically dribbling) and the defender (the closest opposing player)---execute tactical maneuvers to achieve their own goals: attackers to invade the opponent's territory aiming to score, whereas defenders to delay and preempt their actions. 
Any subsequent event, such as passing, interception, or shooting, ends the interaction, as the circumstances of the game rapidly change; 
the cycle repeats. 
Attacker--defender interactions are therefore fundamental building blocks of player dynamics that give rise to high-level team tactics and strategies \cite{goes2021unlocking}.

A previous study developed a mathematical model of attacker--defender interactions in the presence of a goal \cite{brink2023measuring}. 
We simplify that description further to identify a behavioral principle underlying them. 
Using the simplified model, we analyze a large dataset and obtain the distribution of driving-force angles in attacker-defender interactions. 
Combining empirical results with a detailed analysis of the model, we show that a defender responds so as to minimize their future relative speed to the attacker, whereas the attacker anticipates the defender's objective and counteracts their actions. 
This simple principle provides a unified account of the consistent empirical patterns, while capturing the tactical intuition underlying territorial invasion in the sport. 
We also examine an alternative explanation based on attacker--defender distance, highlighting the commonalities and differences between these two principled accounts. 
Because our framework depends little on soccer-specific details, it may also extend to other sports and human behaviors, offering a theoretical basis for the physics of invasion team sports.

\section{\label{sec:data_model} Model}

Let \( \rp(t), \vp(t) \in \mathbb{R}^2 \) be the position and velocity of player \( \p \) at time \( t \), respectively, where the subscript \( \p \in \{\at, \df \} \) denotes the attacker (\( \at \)) or the defender (\( \df \)) of a one-on-one dribbling event. 
Our dynamical model is given by the two coupled equations of motion 
\begin{align} \label{eq:oad}
    \begin{dcases}
        \dvatdt = - \frac{\vat}{\taua}  + \fa \left[ \cos \ta \, \eo(\ra, \rd) + \sin \ta \, \et(\ra, \rd) \right],   \\ 
        \dvdfdt = - \frac{\vdf}{\taud}  + \fd \left[ - \cos \td \, \eo(\ra, \rd) + \sin \td \, \et(\ra, \rd) \right], 
    \end{dcases}
\end{align}
where the overdot represents a time derivative, \( \taup \) is the relaxation time for player \( \p \)'s velocity, and \( \fp \) sets the magnitude of the driving acceleration. 
The unit vector \( \eo(\ra,\rd) \coloneqq (\rd - \ra)/\|\rd - \ra \| \) points from the attacker to the defender. 
To ensure a well-defined basis, we require \( \ra \neq \rd \) for all \( t \). 
We define \( \et \) as the unit vector obtained by rotating \( \eo \) by \( 90^\circ \) clockwise, so that \( \{ \eo, \et \} \) forms a co-moving orthonormal basis. 
Let \( \ta \) and \( \td \) denote the attacker's and defender's driving angles, measured from the reference vectors \( \eo \) and \( - \eo \), respectively. 
These angles remain constant throughout a one-on-one event. 
This assumption reflects the expectation that a player's orientation toward the opponent remains stable over the few-second timescale of a typical one-on-one dribbling event. 
Consequently, the driving angle \( \tp \) remains constant in the player's reference frame, whereas \( \eo \) and \( \et \) rotate in Cartesian coordinates as the players move (see Fig.~\ref{fig:empirical_distribution}a and Table~\ref{tab:symbols}).

\begin{table}[t]
    \centering
    \begin{tabular}{clcc}
        \toprule
        Symbol & Description & Unit  \\ 
        \midrule
        \( \rp \) & player \( \p \)'s position & \si{\meter} \\  
        \( \vp \) & player \( \p \)'s velocity & \( \si{\meter\per\second}  \) \\  
        \( \eo \) & directed from the attacker to the defender & -- \\ 
        \( \et \) & perpendicular to \( \eo \), rotated clockwise & -- \\
        \( \taup \) & exponential decay time for \( \vp \) & \( \si{\second} \) \\
        \( \tp \) & player \( \p \)'s driving direction & \( \mathrm{rad} \) \\
        \( \fp \) & magnitude of \( \p \)'s driving force per unit mass & \( \si{\meter\per\second\squared} \) \\
        \bottomrule
    \end{tabular}
    \caption{
    Notation used in the model. 
    The subscript \( \p \in \{\at, \df\} \) denotes either the attacker (\( \at \)) or the defender (\( \df \)). 
    Units are provided where applicable. 
    The parameters \( \taup \) and \( \fp \) are, by definition, positive for all \( \p \). 
    }
    \label{tab:symbols}
\end{table}

\section{\label{sec:results} Results}

\subsection{\label{subsec:empirical_data} Empirical data}

\begin{figure*}[t]
    \centering
    \includegraphics[width=\linewidth]{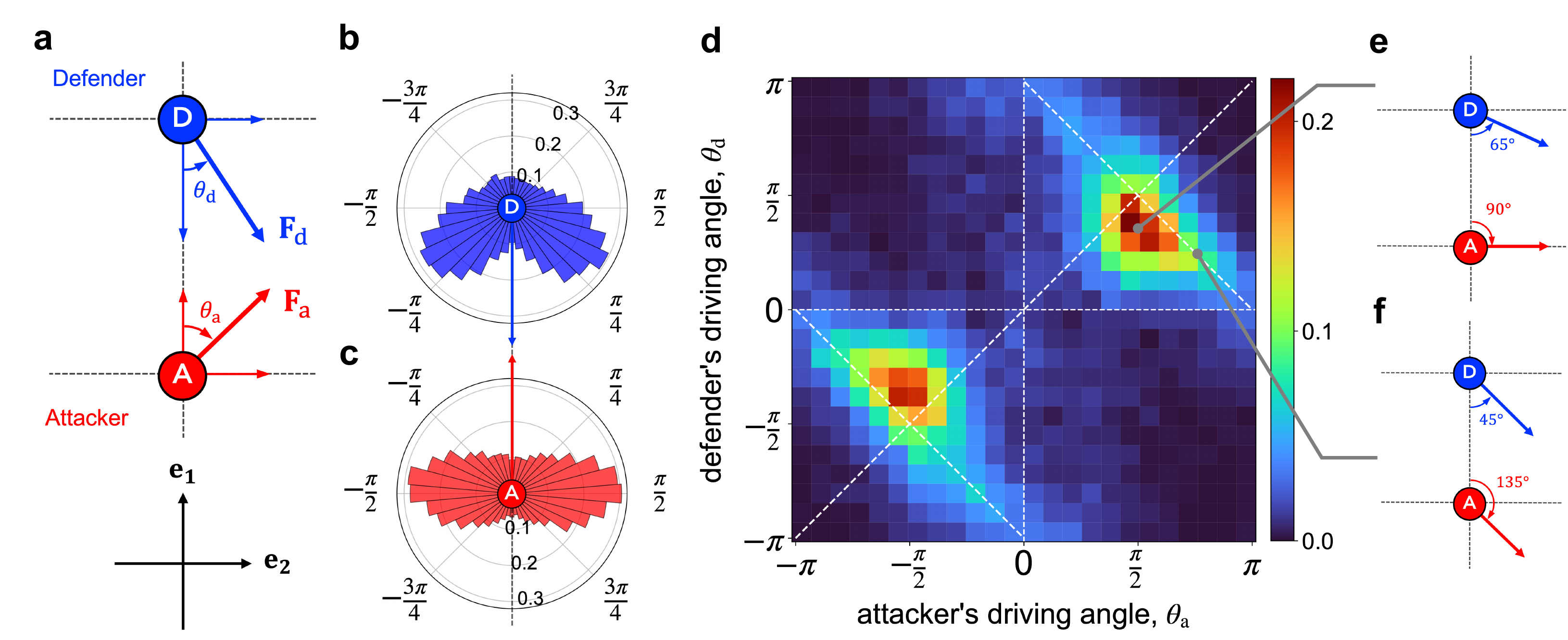}
    \caption{
    \pnl{a,} 
    A schematic illustration of the attacker--defender interaction in the model. 
    The attacker (defender) is driven by a force of constant magnitude, \( \Fa \) (\( \Fd \)), at angle \( \ta \) (\( \td \)) relative to the defender (attacker), where \( \Fa \coloneqq \fa (\cos \ta \eo + \sin \ta \et) \) and \( \Fd \coloneqq \fd (-\cos\td \eo + \sin \td \et) \). 
    \pnl{b,c,} 
    Marginal probability densities of the defender's (\( \td \)) and attacker's (\( \ta \)) driving angles, where the blue and red arrows indicate \( -\eo \) and \( \eo \) respectively. 
    \pnl{d,} The corresponding estimated joint probability density, 
    which exhibits two peaks at \( (\ta, \td) \approx (1.61, 1.14) \) and \( (-1.62, -1.12) \) nearly symmetric around the origin, 
    implying that, on average, defenders tend to be driven slightly toward the attacker (\( \approx\pm 65^\circ \)), whereas attackers accelerate almost sideways relative to the defender (\( \approx \pm 90^\circ \)); 
    see the caption of Fig.~\ref{fig:peak_motion} and \Met for details on how these values were estimated. 
    The joint distribution also shows that the data tend to concentrate moderately along the anti-diagonals \( \ta + \td = \pm\pi \). 
    This implies that players tend to be driven instantaneously in directions parallel to those of their opponents during one-on-one dribbles. 
    \pnl{e,f,} 
    The two schematic figures beside the joint distribution illustrate typical scenarios corresponding to the symmetric peaks of modal angles [panel (\pnl{e})] and the anti-diagonal lines along which the data moderately concentrate [panel (\pnl{f})]. 
    All angles are represented modulo \( 2 \pi \), with the fundamental domain \( (-\pi, \pi] \); note, however, that the driving angles are estimated without enforcing such constraints and converted afterwards. 
    Data are from \totfiltered one-on-one events. 
    }
    \label{fig:empirical_distribution}
\end{figure*}

We obtained data for 306 games in Japan's top professional soccer league from DataStadium~Inc. 
The dataset includes both tracking (two-dimensional positional data for all players on the field) and event data (e.g., on-ball actions with timestamps and player IDs). 
Combining these sources allows us to identify the player in possession of the ball (attacker) and the nearest opposing player (defender). 
We define a ball-possession event as the time interval during which a player remains in possession of the ball. 
To isolate one-on-one dribbling situations from these ball-possession events, we impose several additional criteria, including minimum duration and travel distance and the initial configuration of the attacker and defender relative to the goal. 
This filtering yields \totunfiltered events, which we refer to as one-on-one (dribbling) events (see \Met and \MetSec for more details on the data and preprocessing).

Using smoothed trajectories, we fit our model to the data. 
We separately solve the attacker's and defender's equations of motion (Eq.~\ref{eq:oad}) using the other player's empirical trajectory data as input. 
This greatly reduces computational cost and avoids numerical instability associated with high-dimensional parameter estimation. 
We then estimate the model parameters \( ( \tp, \taup, \fp ) \) by minimizing the discrepancy between empirical and model trajectories. 
Retaining only events with fitting errors below a threshold leaves \totfiltered events for analysis 
(see \Met and \MetSec for details on numerical implementation).

Figures~\ref{fig:empirical_distribution}b--\ref{fig:empirical_distribution}d show the marginal and joint probability distributions of attackers' and defenders' driving angles, obtained from the events filtered above. 
The marginal distributions exhibit two symmetric peaks (Figs.~\ref{fig:empirical_distribution}b and \ref{fig:empirical_distribution}c). 
However, these distributions peak at distinct values: attackers' directions concentrate approximately at \( \pm 90^\circ \), whereas defenders' directions concentrate approximately at \( \pm 65^\circ \). 
This implies that the attacker tends to accelerate almost laterally relative to the defender, whereas the defender tends to accelerate slightly toward the attacker (Fig.~\ref{fig:empirical_distribution}e). 
Because the model is symmetric under exchange of the attacker and defender labels, these distinct modal angles suggest that the two players exhibit systematically different preferences for relative orientation during one-on-one interactions.

Furthermore, the joint probability distribution shows a moderate concentration along the anti-diagonals \( \ta + \td = \pm\pi \) (Fig.~\ref{fig:empirical_distribution}d). 
This observation suggests that the attacker and defender are driven instantaneously in parallel directions (see Fig.~\ref{fig:empirical_distribution}f for an illustration). 
The instantaneous alignment of the players' acceleration directions is reminiscent of a well-known defensive technique called ``jockeying,'' which we briefly discuss in Sec.~\ref{sec:discussion}.

Taken together, these empirical findings raise deeper questions: 
why are player angles localized in the parameter space at all? 
Why do the data show moderate concentration along the anti-diagonals \( \ta + \td = \pm \pi \)? 
Can these observations be decomposed into finer-grained structures? 
And, critically, is there a behavioral principle that explains all of these salient patterns in the data? 
In what follows, we address these questions by developing a unifying theoretical framework.

\subsection{\label{subsec:analysis} Mathematical Analysis}

\begin{figure*}[t]
    \centering
    \includegraphics[width=\linewidth]{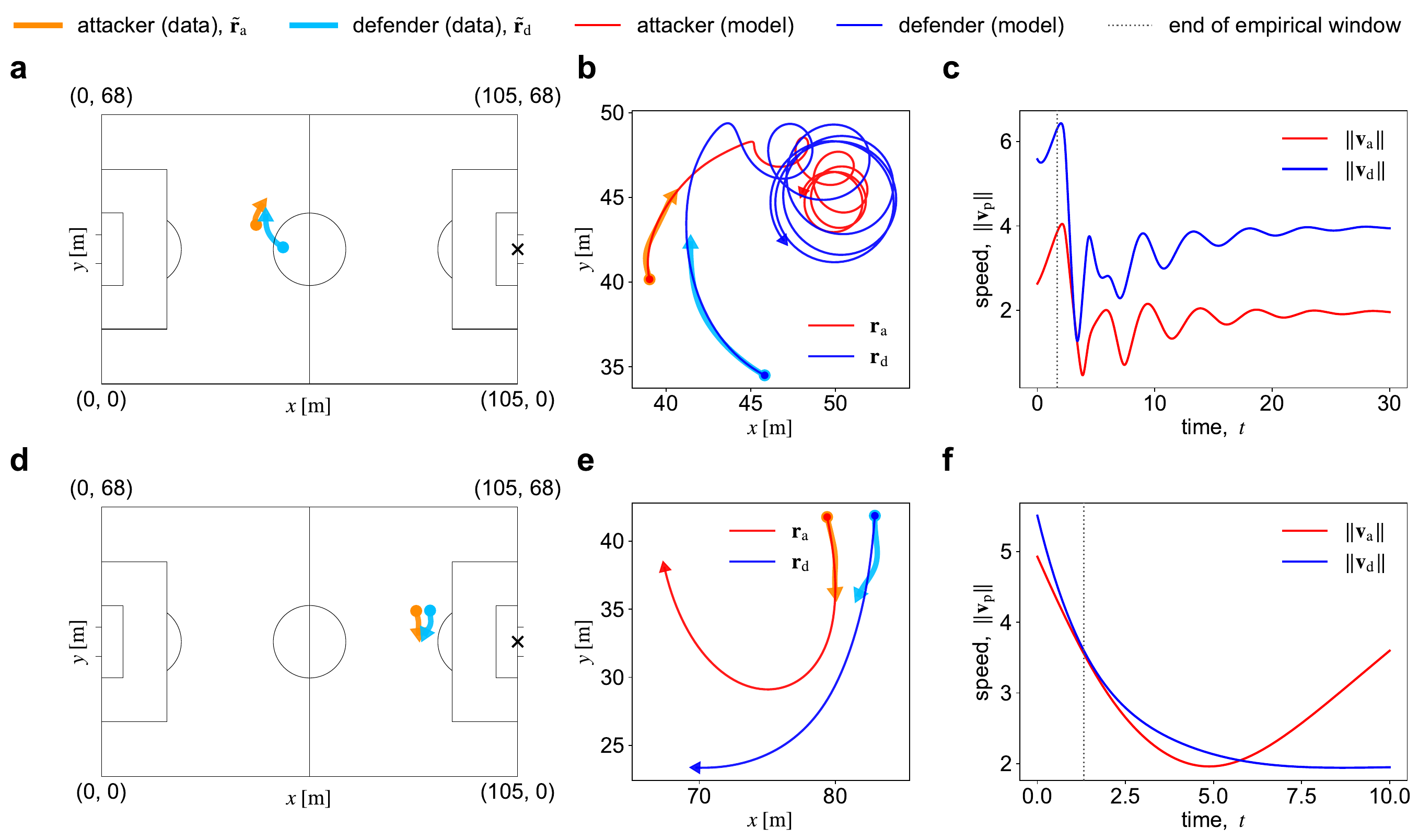}
    \caption{
     \pnl{a,d,} Empirical (semi-transparent) and best-fit model (opaque) trajectories of the attackers (orange, red) and defenders (cyan, blue) in one-on-one events. 
     Their initial positions are marked by filled circles; directions of motion are indicated by arrows; and the attacking goal (the goal that belongs to the defender's team) is indicated by the black cross marker. 
     Both equations of motion are solved simultaneously to obtain  model trajectories. 
    \pnl{b,e,} We extended the model trajectories [(\pnl{a}) and (\pnl{d})] beyond the empirical intervals of the one-on-ones. 
    These trajectories exhibit (\pnl{b}) limit cycles and (\pnl{e}) outward logarithmic spirals as \( t \to \infty \) (See \AppFig{fig:ex:trajectories_extended} for extended model trajectories that exhibit asymptotic behaviors). 
    \pnl{c,f,} Timeseries data of player speeds in the model. The vertical dotted lines indicate empirical time windows. 
    Parameters and initial conditions: 
    \pnl{a--c,} 
    \( \fa = 2.19 \), \( \fd = 4.71 \), \( \ta = -1.42 \), \( \td = -0.92 \), \( \taua = 4.66 \), \( \taud = 2.17 \), \( \ra(0) = (39.02, 40.17) \), \( \rd(0) = (45.81, 34.50) \), \( \vat(0) = (-0.60, 2.56) \), \( \vdf(0) = (-4.94, 2.59) \);
    \pnl{d--f,} 
    \( \fa = 0.95 \), \( \fd = 1.16 \), \( \ta = -2.55 \), \( \td = 1.13 \), \( \taua = 10.66 \), \( \taud = 1.75 \), \( \ra(0) = (79.39, 41.78) \), \( \rd(0) = (82.90, 41.87) \), \( \vat(0) = (1.00, -4.82) \), \( \vdf(0) = (-0.41, -5.49) \). 
    }
    \label{fig:empirical_model_trajectories}
\end{figure*}

\begin{figure*}[t]
    \centering
    \includegraphics[width=\linewidth]{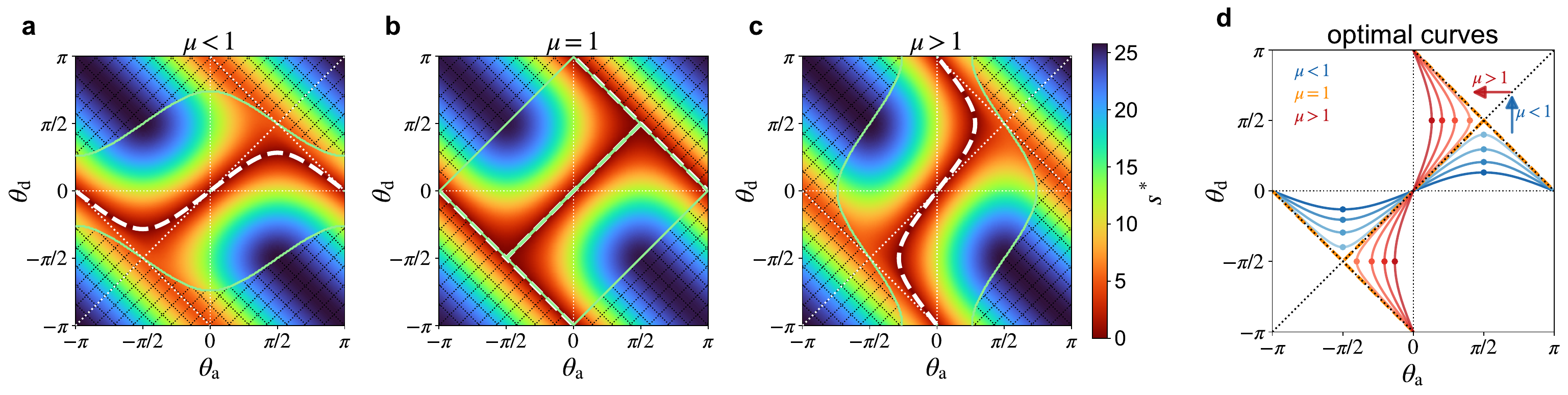}
    \caption{
    Phase diagrams in the angle space \( (\ta, \td) \in [-\pi,\pi)^2 \) for three distinct regimes based on the value of \( \mu = \fa/\fd \): 
    \pnl{a,} \( \mu < 1 \), \pnl{b,} \( \mu = 1 \), and \pnl{c,} \( \mu > 1 \). 
    The color scale represents asymptotic relative speed \( s^* \). 
    The parameter regions enclosed by the solid light green lines indicate the limit-cycle (LC) regime, whereas the crosshatched regions (dotted black lines) indicate the logarithmic-spiral (LS) regime. 
    The thick dashed white line represents the optimal curve \( \opts \). 
    The thin dotted white lines serve as visual guides. 
    For \( \mu = 1 \), \( \opts \) lies on the boundaries of the LC regime but is nonetheless included for reference. 
    \pnl{d,} Optimal curves \( \opts \) for varying \( \mu \), illustrating its geometric dependence on \( \mu \) (\( \mu \in \{ 0.4, 0.6, 0.8, 0.95, 1, 1.05, 1.25, 1.67, 2.5 \} \)). 
    The critical points of \( \opts \) shift vertically along \( \ta=\pm\pi/2 \) away from \( \td = 0\) as \( \mu \) increases to \( \mu = 1 \); 
    thereafter, the optimal curve \( \opts \) flips along the diagonal line \( \ta = \td \) and the critical points move horizontally along \( \td=\pm\pi/2 \) toward \( \ta = 0 \) as \( \mu \) increases above \( \mu = 1 \). 
    The blue and red arrows illustrate the geometric shift of these points in Quadrant I as \( \mu \) increases. 
    Only the branch of \( \opts \) passing through the origin is plotted. 
    Parameters: 
    \pnl{a,} \( \fa=7 \), \( \fd=9 \), \( \taua=1.5 \), \( \taud=1.7 \);
    \pnl{b,} \( \fa=8 \), \( \fd=8 \), \( \taua=1.6 \), \( \taud=1.6 \); and   
    \pnl{c,} \( \fa=9 \), \( \fd=7 \), \( \taua=1.7 \), \( \taud=1.5 \). 
    Blank regions denote parameter combinations for which the asymptotic classification does not assign a unique state: either no admissible asymptotic solution is identified, or more than one admissible solution is possible. In the three parameter regimes shown here, however, no multiple-solution cases occurred and the fraction of unclassified cases is negligible; therefore, blank regions are not visible in the plots (see \Met for theoretical details). 
    }
    \label{fig:relative_speed}
\end{figure*}

We analyze our model focusing on its asymptotic behaviors and show that a simple principle underlying defensive behavior provides a unified account of the observed distribution of driving angles. 
Here we outline the analysis and describe our main results (see \Met for a full analytical treatment). 
First, we decompose the model into relative and center-of-mass motions and express them in polar coordinates. 
We then show that both relative and center-of-mass motions admit two asymptotic forms: \emph{limit cycles} corresponding to uniform circular motion or \emph{outward logarithmic spirals}. 
The asymptotic trajectories of both the attacker and defender then take the form of uniform circular motion or outward logarithmic spirals. 
In Figs.~\ref{fig:empirical_model_trajectories}b and \ref{fig:empirical_model_trajectories}e, 
the best-fit model trajectories produce either limit cycles or asymptotic logarithmic spirals when extended beyond the empirical time windows of one-on-ones.

We now set the stage for showing that a simple principle governing relative motion provides a unifying explanation of the empirical patterns in the distribution of driving angles. 
In \Met we derive the asymptotic relative speed for both the limit-cycle (LC) and logarithmic-spiral (LS) regimes. 
Let \( s(t) \coloneqq \| \vv(t) \| \) denote the relative speed at time \( t \), where \( \vv = \vdf - \vat \). 
For notational convenience, we define 
\( \alpha = (\taud^{-1} + \taua^{-1})/2 \), \( \beta = (\taud^{-1} - \taua^{-1})/2 \), 
and 
\begin{align}
    F_1 &= - \fd \cos \td - \fa \cos \ta, \\
    F_2 &= - \fd \sin \td + \fa \sin \ta, \\
    G_1 &= - \fd \cos \td + \fa \cos \ta, \\
    G_2 &= - \fd \sin \td - \fa \sin \ta. 
\end{align}
The asymptotic relative speed \( s^* \coloneqq \lim_{t \to \infty} \| \vv(t) \|\) is then given by 
\begin{equation}
    s^* = \label{eq:s}
    \begin{dcases}
        \frac{\left|-\alpha F_1 + \beta G_1 + \omega F_2\right|}{2\alpha |\omega|},  \;& \text{(LC)} \\
        \frac{\sqrt{(\alpha F_1 - \beta G_1)^2 + (\alpha F_2 - \beta G_2)^2}}{\left|\alpha^2-\beta^2\right|}, \;& \text{(LS)} 
    \end{dcases}
\end{equation}
where the angular speed of a limit cycle, \( \omega \), satisfies the cubic equation
\begin{multline} 
    F_2 \omega^3 + (\alpha F_1 + \beta G_1) \omega^2 \\ 
    + \left[F_2(\alpha^2 +\beta^2) - 2\alpha\beta G_2 \right] \omega \\
    + (\alpha^2 - \beta^2)(\alpha F_1 - \beta G_1) = 0. \label{eq:omega_cubic}
\end{multline}
We refer the reader to \Met for a full derivation. 
Let us introduce a central principle of our analysis: 
defenders choose \( \td \) to minimize the asymptotic relative speed \( s^* \) in response to attackers' driving angles \( \ta \). 
We refer to this behavioral principle as the \emph{relative-speed minimization principle}.

Figure~\ref{fig:relative_speed} illustrates the asymptotic relative speed \( s^* \) in the angle space. 
The LS and LC regimes are indicated by the crosshatched regions and those enclosed by light green lines, respectively. 
Notably, the parameter regions with near-zero \( s^* \) are concentrated along a characteristic S-shaped curve (Figs.~\ref{fig:relative_speed}a and \ref{fig:relative_speed}c). 
This curve extends along either the x- or y-axis depending on the value of \( \mu \coloneqq \fa/\fd \), the ratio of the attacker's driving force magnitude to the defender's. 
When \( \mu = 1 \), the small-\( s^* \) region has the shape of the letter H rotated by \( 45^\circ \) (Fig.~\ref{fig:relative_speed}b). 
We refer to this locus as the \emph{optimal curve}, denoted \( \opts \) hereafter. 
In \Met we derive an analytical expression for \( \opts \) within the LC regime: 
\begin{align}
    \fa \sin \ta = \fd \sin \td. \label{eq:opts}
\end{align} 
When this relation holds exactly, however, the asymptotic relative speed \( s^* \) becomes zero and the radius of relative motion also vanishes as \( (\ta, \td) \to \opts \), implying that a limit cycle does not occur in this case. 
Accordingly, \( \opts \) should be more precisely interpreted as the locus along which \( s^* \) can be made arbitrarily small. 
In the LS regime, by contrast, the minimum asymptotic speed is achieved along \( \ta + \td = \pm\pi \), given by 
\begin{align}
    s^* = | \taua \fa - \taud \fd |, \label{eq:s_min_LS}
\end{align}
which is strictly positive provided \( \taua \fa \neq \taud \fd \). 
Therefore, in general, the global infimum of the asymptotic relative speed can be approached only in the LC regime along \( \opts \) (see \Met for details).

\begin{figure*}[t]
    \centering
    \includegraphics[width=\linewidth]{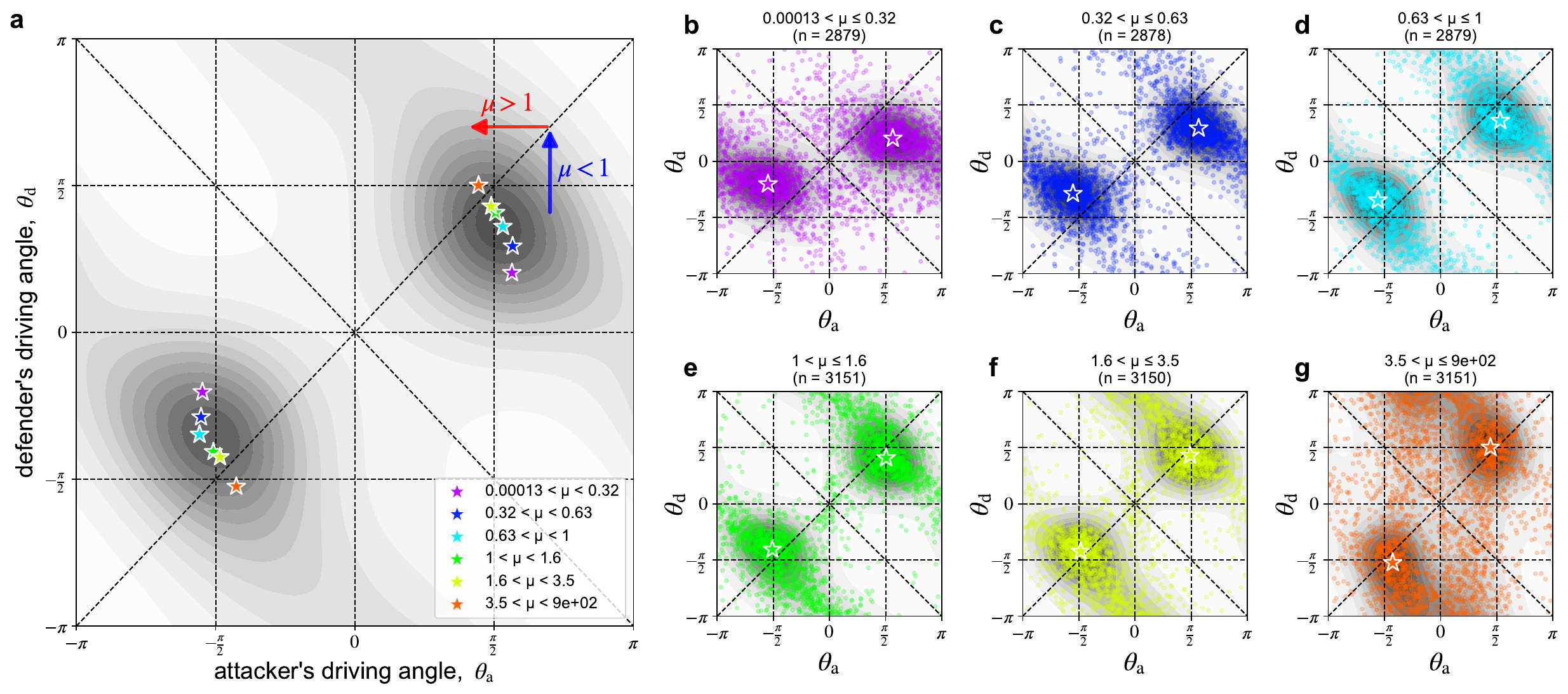}
    \caption{
    \pnl{a,} The estimated probability density function for all data (\totfiltered one-on-one events), shown as a grayscale contour plot.  
    The colored stars indicate the modal angle pairs of each of six data subsets categorized according to the value of \( \mu \). 
    We first split all events into two groups based on whether \( \mu \) is greater or smaller than \( 1 \), each of which is further divided into three subgroups---based on the value of \( \mu \)---of roughly equal size. 
    The \( \mu \)-dependent motion of the modal angles is qualitatively consistent with the theoretical prediction (Sec.~\ref{subsec:analysis} and Fig.~\ref{fig:relative_speed}\pnl{d}).
    \pnl{b--g,} Scatter plot of each data subset (semitransparent unfilled colored circles) and the corresponding estimated probability density function conditioned on \( \mu \) (grayscale contour plot). 
    The modal angles were identified using a von Mises kernel assuming uncorrelated axes and an isotropic angle space with a smoothing factor fixed across all panels. 
    The smoothing factor was varied to verify that the peak structure remained stable (see \Met for details). 
    Using the same method, we identified the peaks in the full dataset (Fig.~\ref{fig:empirical_distribution}). 
    In all contour plots, darker gray shading represents higher estimated probability density; see \AppFig{fig:ex:angle_joint_by_mu_bins} for the histograms of these partitioned data subsets and \AppFig{fig:ex:log_mu_histogram} for the distribution of rescaled \( \mu \). 
    Finally, although the \( \mu \)-dependent motion of the modal angles across angle space is qualitatively consistent with theory, the observed modal values deviate from \( \opts \) and are less sensitive to \( \mu \), suggesting that finite-time effects (dependence on initial conditions; see \Met and \AppTwoFigs{fig:ex:timeseries_phase_diagram_1}{fig:ex:timeseries_phase_diagram_2} for more details) or additional mechanisms limit the \( \mu \)-dependence of angle selection. 
    Moreover, our theory alone does not predict the defender's response when \( \mu > 1 \), which remains an open problem. 
    } 
    \label{fig:peak_motion}
\end{figure*}

Next we examine how the structure of the optimal curve \( \opts \) is modulated by \( \mu \) (Figs.~\ref{fig:relative_speed}a--\ref{fig:relative_speed}c, thick white dashed lines; see Fig.~\ref{fig:relative_speed}d for an illustration). 
For \( \mu < 1 \), \( \opts \) is defined by \( \td = \arcsin(\mu \sin \ta) \), meaning that as \( \mu \) increases toward \( 1 \), \( \opts \) stretches along the \( \td \) direction and its critical points shift vertically. 
For \( \mu > 1 \), \( \opts \) is given by \( \ta = \arcsin(\mu^{-1} \sin \td) \), meaning that as \( \mu \) increases above \( 1 \), \( \opts \) contracts along the \(\ta\) direction and its critical points shift horizontally. 
For \( \mu = 1 \), \( \opts \) lies on the boundaries of the LC regime (where the radius of relative motion vanishes) and is therefore not physically realizable. 
Nonetheless, this regime provides useful intuition for one-on-ones under relative-speed minimization.
The optimal curve for \( \mu = 1 \) can be decomposed as \( \opts \equiv \mathcal{S}_1 \cup \mathcal{S}_2 \), where \( \mathcal{S}_1 \) denotes the line \(  \ta = \td, \, |\ta| < \pi/2 \), and \( \mathcal{S}_2 \) denotes \( \ta + \td = \pm\pi, \, |\ta| < \pi \). 
Along \( \mathcal{S}_1 \), a more aggressive attacking angle \( \ta \) is met by a more direct defensive response \( \td \) toward the attacker, suggesting an attempt to pressure and possibly win possession of the ball; 
\( \mathcal{S}_1 \) therefore indicates a \emph{risk-seeking} defensive strategy. 
By contrast, along \( \mathcal{S}_2 \), the defender chooses their angle such that their direction of motion parallels the attacker's, suggesting an effort to mirror the attacker's motion to avoid being outmaneuvered; 
\( \mathcal{S}_2 \) therefore reflects a more conservative, \emph{risk-averse} defensive strategy.

The relative-speed minimization principle states that the defender chooses their angle to minimize relative speed in response to the attacker's angle. 
We now turn to the attacker's point of view. 
Within this framework, an attacker seeks to choose an angle that preemptively limits the defender's response---but can they?

For \( \mu < 1 \), \( \opts \) extends along the \( \ta \) axis, implying that defenders can make asymptotic relative speed arbitrarily small for any given \( \ta \)---that is, defenders can neutralize any \( \ta \) chosen by attackers. 
Nevertheless, \( \ta \approx \pm\pi/2 \) may be strategically favorable for attackers because these critical points of \( \opts \) force defenders to switch between \emph{risk-seeking} (\( \mathcal{S}_1 \)-like branch) and \emph{risk-averse} (\( \mathcal{S}_2 \)-like branch) strategies under small perturbations in \( \ta \). 
Attackers therefore may favor these angles to exploit the branching points for tactical advantage, even though defenders can still minimize the asymptotic relative speed (Fig.~\ref{fig:relative_speed}).

For \( \mu > 1 \), by contrast, the attacker's choice of driving angle directly influences the attainable asymptotic relative speed. 
By choosing \( |\ta| > \ta^* \), where \( \ta^* \coloneqq \arcsin(\mu^{-1}) \), the attacker can prevent asymptotic relative speed from approaching zero, regardless of how the defender subsequently chooses \( \td \). 
This preemptive strategy gives the attacker a measurable advantage. 
The critical angle \( \ta^* \) decreases as \( \mu \) increases above \( 1 \), expanding the range of \( \ta \) that attackers can choose from without losing the advantage (Fig.~\ref{fig:relative_speed}d). 
Within \( |\ta| > \ta^* \), however, attackers may still favor sharper angles because defenders typically lie between the attacker and their own goal---or more generally, the open space the attacker seeks to invade. 
Therefore, we predict that attackers are incentivized to choose \( |\ta| \) larger than \( \ta^* \) to preempt relative-speed minimization, but their preferred \( |\ta| \) should nonetheless decrease as \( \mu \) increases above \( 1 \) to capitalize on any potential offensive advantage associated with smaller driving angles.

Overall, the difference in driving force magnitudes governs which player has the advantage in one-on-one interactions. 
For \( \mu < 1 \), the defender's stronger acceleration enables them to adapt effectively to the attacker's motion, and therefore to make the asymptotic relative speed arbitrarily small for any given \( \ta \). 
For \( \mu > 1 \), by contrast, the attacker's stronger acceleration allows them to maintain non-zero asymptotic relative speed by strategically choosing \( \ta \) in advance. 
The resulting critical angle pairs for different values of \( \mu \) are summarized in Fig.~\ref{fig:relative_speed}d.

\subsection{\label{subsec:empirical_data_explained} A theoretical account of empirical data}

Here we show that the empirical data reported earlier are in good agreement with theory. 
First, the relative-speed minimization principle postulates that the defender's decision follows the attacker's choice of driving angle. 
In the analyzed events, approximately \fracmatchingsigns of the angle pairs \( (\ta, \td) \) have matching signs. 
Given the attacker's and defender's roles, this provides reasonable empirical support for the core assumption, although the evidence remains circumstantial and context-dependent.  
Next, we examine how the empirical distribution of driving angles changes with \( \mu \). 
To this end, we partition our data into six non-overlapping groups based on the value of \( \mu \). 
Figure~\ref{fig:peak_motion} shows the estimated joint probability density function of all data (\ref{fig:peak_motion}a), as well as of each data group (\ref{fig:peak_motion}b--\ref{fig:peak_motion}g), which all exhibit two roughly symmetric peaks at different locations. 
The modal angles identified from the density estimates for each \( \mu \)-group are overlaid in Fig.~\ref{fig:peak_motion}a (colored stars). 
The symmetric pairs of angle modes show a characteristic \( \mu \)-dependent shift across the parameter space: vertically for \( \mu < 1 \) and horizontally toward the \( \td \)-axis for \( \mu > 1 \). 
This qualitative behavior is consistent with the prediction in Fig.~\ref{fig:relative_speed}d.

The moderate concentration of data along \( \ta + \td = \pm\pi \) (or \( \mathcal{S}_2 \)) reported in Fig.~\ref{fig:empirical_distribution} also coincides with the near-optimal locus predicted for the LS regime (Sec.~\ref{subsec:analysis}). 
This suggests that the LS-optimal locus \( \mathcal{S}_2 \) represents a defender's fallback strategy, which may help to balance the risk of being outmaneuvered against marginal reductions in relative speed, depending on the circumstances of the game. 
This tendency is in fact pronounced near the goal (\SM Fig.~\bl{S6}).

Our theoretical account is based purely on the model's limiting behaviors. 
One potential caveat is that these asymptotic states are never realized during one-on-ones. 
In \Met we nonetheless show that they often provide a reasonable approximation to the finite-time evolution of phase diagrams over the course of a typical one-on-one event. 
Therefore, our account does not necessarily hinge on unrealizable future outcomes of the model; 
instead, it relies on proxies for short-term outcomes that players can anticipate and use to choose their best angles at the outset of one-on-one interactions (see \Met and \AppTwoFigs{fig:ex:timeseries_phase_diagram_1}{fig:ex:timeseries_phase_diagram_2}).

\subsection{\label{subsec:alternative_accounts} Alternative behavioral principles}

\begin{figure*}
    \centering
    \includegraphics[width=\linewidth]{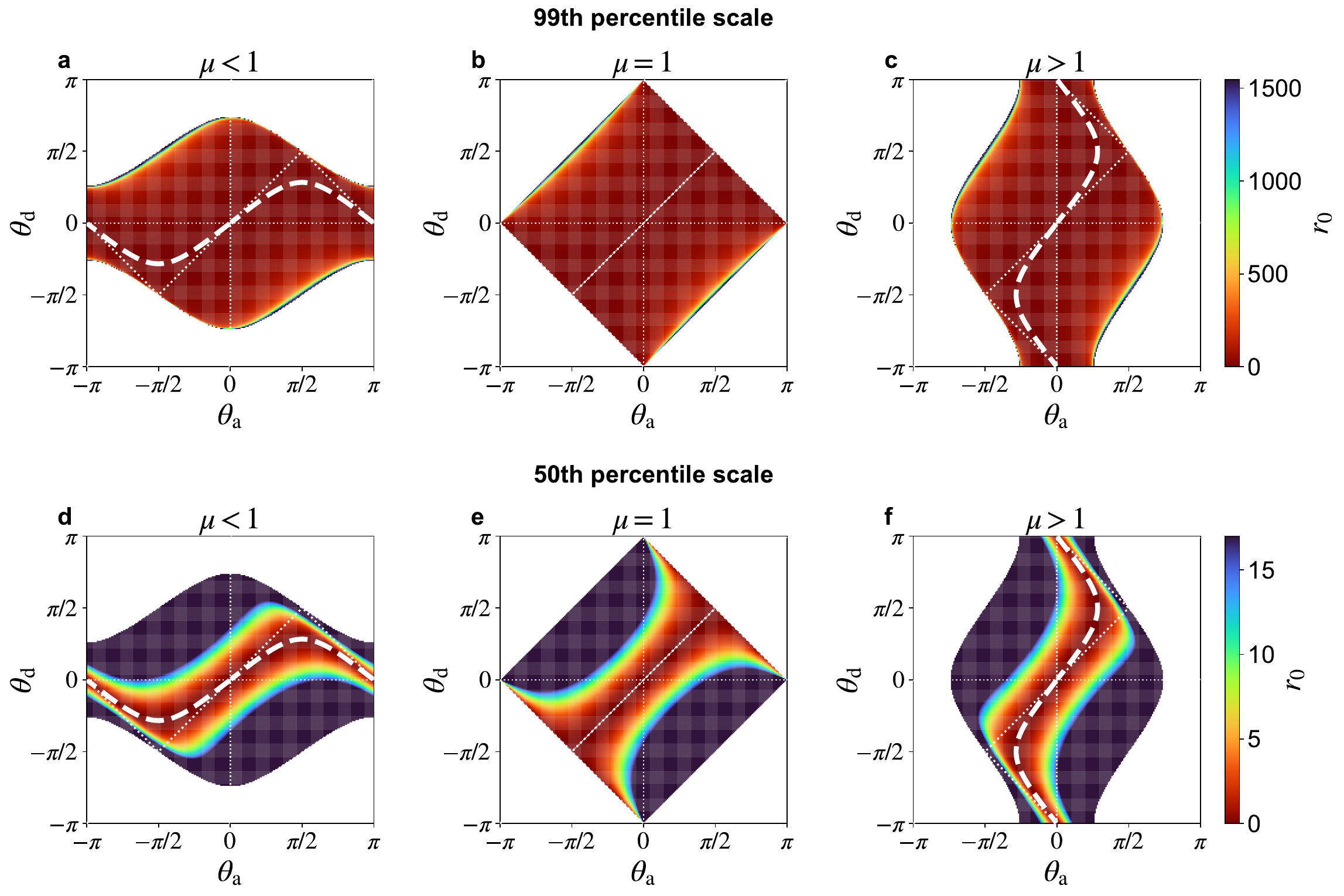}
    \exfigcaption[\linewidth]{
    \pnl{a--c,} Asymptotic attacker--defender distance \( r_0 \) obtained from theoretical analysis. 
    Values are shown only where theory predicts a unique limit cycle (uniform circular motion) and no outward logarithmic spiral; regions outside this regime are left blank. 
    Color scale is capped at the 99th percentile of the data presented, as values near the boundaries of the LC regime can be disproportionately large. 
    The optimal curve \( \opts \) is indicated by the white dashed line within the LC regime. 
    White dotted lines are guides to the eye. 
    \pnl{d--f,} Same as (\pnl{a--c}) but with the color scale capped at the 50th percentile. 
    These panels highlight the optimal curve \( \opts \) more clearly. 
    Parameters: 
    \pnl{a,} \( \fa=7 \), \( \fd=9 \), \( \taua=1.5 \), \( \taud=1.7 \);
    \pnl{b,} \( \fa=8 \), \( \fd=8 \), \( \taua=1.6 \), \( \taud=1.6 \);  
    \pnl{c,} \( \fa=9 \), \( \fd=7 \), \( \taua=1.7 \), \( \taud=1.5 \). 
    }
    \label{fig:ex:r0_panels}
\end{figure*}

\begin{figure*}
    \centering
    \includegraphics[width=\linewidth]{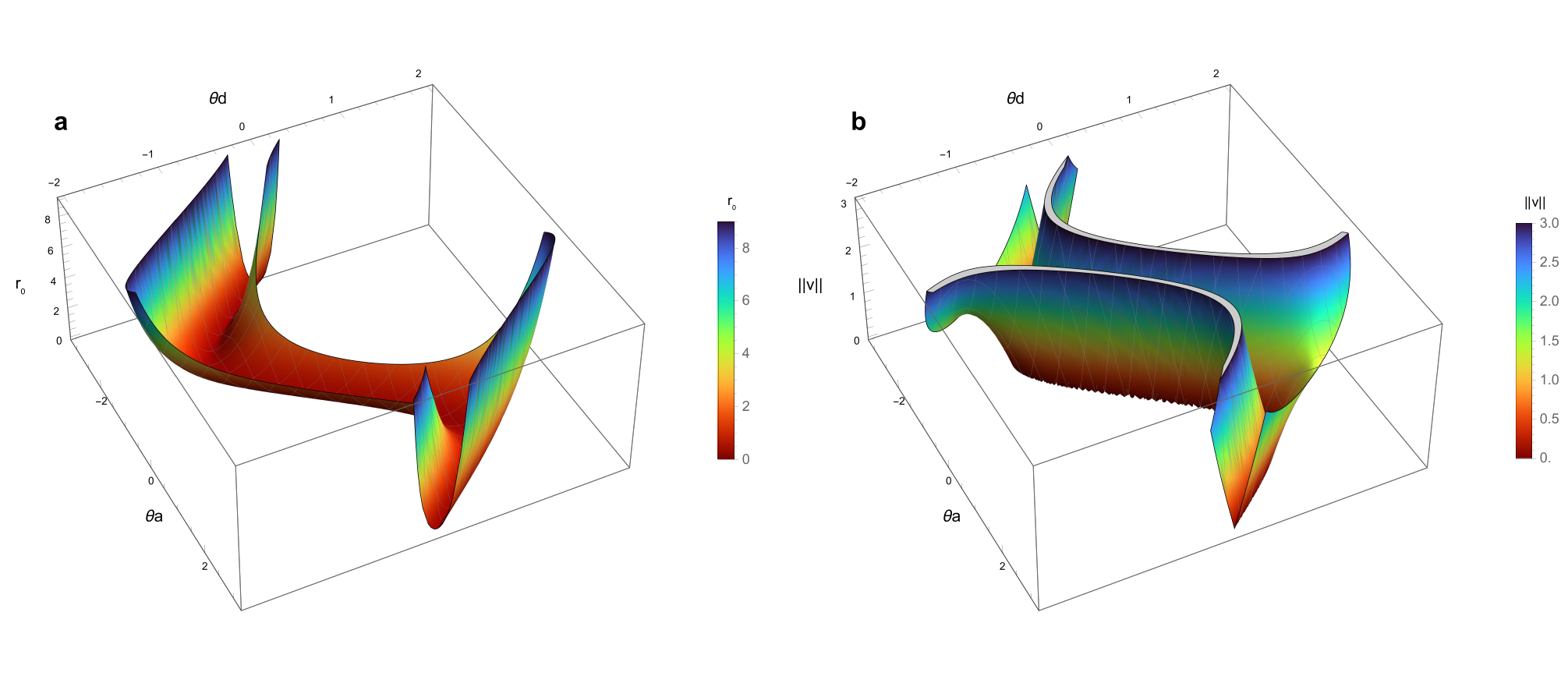}
    \exfigcaption[\linewidth]{
    Landscapes of objective functions near the optimal curve \( \opts \). 
    \pnl{a,} Asymptotic attacker--defender distance. 
    \pnl{b,} Asymptotic relative speed. 
    Asymptotic distance exhibits smooth curvature near \( \opts \), whereas asymptotic speed has a sharp local structure, yielding stronger gradient signals near \( \opts \) for local optimization. 
    Note that the zero level set \( \gamma = 0 \) yields \( \opts \), where \( \gamma(\ta, \td) = - \fd \sin \td + \fa \sin \ta \). 
    The normal distance from \( \opts \) to a nearby level set \( \gamma = c \), denoted by \( \Delta d \), is approximately given by \( \Delta d \approx |c|/\|\nabla \gamma \| \); therefore, near \( \opts \), it is proportional to the level value \( c \) provided \( \|\nabla\gamma\| \neq 0 \), although the proportionality coefficient may vary along \( \opts \). 
    Parameters: 
    \( \fa = 4 \), \( \fd = 5 \), \( \taua = 1.9 \), \( \taud = 1.8 \). 
    This regime (\( \mu < 1 \)) corresponds to that of Fig.~\ref{fig:relative_speed}a. 
    }
    \label{fig:ex:landscapes}
\end{figure*}

So far, we have argued that the relative-speed minimization principle provides a basis for understanding attacker--defender interactions. 
In \Met we show that asymptotic distance is also minimized along \( \opts \) (see also \AppFig{fig:ex:r0_panels}). 
This means that the same angle pairs emerge from optimizing either objective, yielding alternative interpretations of the optimal curve and underlying structure of the model. 
However, their landscapes are substantially different and exhibit qualitatively distinct behavior around \( \opts \) (\AppFig{fig:ex:landscapes}): in the limit-cycle regime, the asymptotic distance \( r_0 \) converges quadratically \( O(\epsilon^2) \) to zero along \( \opts \), whereas the asymptotic relative speed \( s^* \) exhibits linear convergence \( O(\epsilon) \) (see \Met). 
This geometric distinction implies that these two objectives play complementary roles in local optimization search near \( \opts \). 
Distance minimization is better suited to identifying angle pairs that are ``good enough'' because marginal gains in optimization are largest far from \( \opts \) but vanish locally as the landscape flattens. 
Conversely, relative-speed minimization is better suited to locating ``best among good'' action pairs, as its landscape provides a consistent proportional advantage all the way toward \( \opts \), although it may be less robust in noisy local search. 
Therefore, players may benefit from switching between or even combining these two objectives depending on the precision and robustness required in the situation at hand, as \( \opts \) remains intact irrespective of the choice of objective function. 
We nevertheless argue that relative-speed minimization has greater explanatory power and is more useful operationally because asymptotic relative speed is globally well-behaved across the action space, whereas asymptotic distance diverges in the LS regime, in which case the near-optimal lines \( \mathcal{S}_2 \) for \( \mu \neq 1 \) cannot be identified.

\section{\label{sec:discussion} Discussion}

Our key result is that the simple principle of relative speed minimization captures the striking patterns in the empirical distribution of driving angles during one-on-one interactions. 
This explanatory power suggests that the defender's preference for relative orientation is goal-oriented and strategically adaptive.

Attacker--defender interactions resemble classical pursuit--evasion behaviors, including predator--prey chase--escape dynamics widely observed in nature \cite{kamimura2010group, nahin2012chases, tsutsui2019underlying, fajen2007behavioral}. 
These systems, including fish and insects, have often been studied using geometric models  \cite{domenici2011animal, kawabata2023multiple, jiao2023evaluating, szopa2022responsive}. 
However, our framework is distinct from classical pursuit--evasion in the objective that underlies behavioral decisions: 
whereas most studies postulate that a pursuer seeks to reach the evader, here the defender aims to minimize future relative speed to the attacker.

The principle of relative-speed minimization is consistent with intuition in soccer. 
\emph{Jockeying} is a tactical maneuver in which a defender backs off and moves in parallel with the attacker to remain between the attacker and the goal while slowing down the attacker. 
This resembles \( \mathcal{S}_2 \) in spirit---the near-optimal fallback strategy predicted by our theory. 
Moreover, the \( \mu \)-dependence of modal angles for \( \mu < 1 \) suggests adaptive defensive strategies in one-on-one encounters in response to relative advantage in acceleration strength: 
when highly advantaged (\( \mu \ll 1 \)), defenders tend to accelerate toward the attacker (\( |\td| \ll \pi/2 \)) to leverage their advantage despite the associated risk, perhaps to win possession of the ball. 
Conversely, when only marginally advantaged (\( \mu \approx 1 \)), defenders avoid such high-stakes maneuvers and instead adopt lateral acceleration (\( |\td| \approx \pi/2 \)) because their acceleration advantage is insufficient to substantially reduce future relative speed if they attempt to ``chase'' the attacker. 
The moderate concentration of data along \( \mathcal{S}_2 \) is also reminiscent of the previously reported tendency for directional alignment between nearest opposing players \cite{narizuka2016statistical}. 
Overall, relative-speed minimization provides a simple unifying account of these empirical observations. 
Since the model is largely independent of soccer-specific details, similar principles may also arise in other invasion sports, such as basketball and hockey 
\cite{sarlis2020sports}.

Combined with the attacker's attempt to counteract the defender's aim, this principle also reveals a simple game-theoretic structure underlying player interactions in dynamic environments. 
In contrast to prior game-theoretic studies of structured settings, we analyze the competition between two moving opponents, providing a basis for understanding player interactions in dynamic, open play settings.

Our study is certainly not without limitations. 
First, our principled account of the empirical data rests on some nontrivial assumptions. 
In our theoretical framework, all but angle parameters are fixed at the start of each event, likely determined by preceding events, followed by the choice of \( \ta \) and \( \td \). 
Although this assumption is plausible over short timescales and broadly supported by the data, it would benefit from additional empirical validation. 
Second, multiple defenders may press the ball carrier \cite{andrienko2017visual, yokoyama2018social}, suggesting multi-player interactions as a natural extension of the model. 
Another promising---and perhaps most important---direction is to relate this one-on-one framework to higher-level phenomena \cite{marcelino2020collective}, ranging from overall game status to more nuanced, fine-grained observations. 
For example, does the ``winner'' of a one-shot interaction under the relative-speed minimization principle outperform the opponent in terms of on-ball success? 
How do the outcomes of such interactions carry over to subsequent play? 
Addressing these questions will help bridge player motion and collective outcomes, laying the foundation for understanding coordination and tactics in team sports.

Finally, the complementary roles of distance and relative speed in optimization add important subtlety to our findings. 
One possibility is that skill development may involve a transition from distance-based to relative-speed-based evaluation, allowing players to sharpen their internal representation of the payoff landscape. 
Identifying the mechanisms underlying rapid orientational decisions, however, requires detailed analysis of motor perception and cognition, suggesting new directions for future research.

First-principles approaches are not common in studies of human behavior. 
Yet sports data provide powerful opportunities to quantify emergent regularities and discover unifying principles underlying human behavior \cite{radicchi2012universality}. 
Herberger's maxim still resonates: unpredictability makes soccer worth playing and watching---but perhaps all the more because of the striking patterns beneath the surface.

\bibliography{ref}

\clearpage

\section{\label{sec:matmet} Methods}

\subsection{\label{subsec:data} Data}

Our dataset was provided by DataStadium~Inc., including both event and tracking data from 306 games in the 2023 season of the J1 League (the premier division of the professional soccer league in Japan). 
Event data allows us to identify the player in possession of the ball and the corresponding time interval. 
Tracking data provides two-dimensional positions of all players on the field at centimeter resolution, sampled every \SI{40}{\milli\second}. 
Combining these sources, we isolate ball-possession events across games (a ball possession event is defined in the main text; see \MetSec for more details, including how to identify ball-possession events from event data).

\subsection{\label{subsec:preprocessing} Preprocessing}

We identify one-on-one events among ball-possession events by imposing the following conditions: 
(i) the event duration exceeds \SI{0.5}{\second}; 
(ii) both the attacker and the defender have net displacements greater than \SI{5}{\meter} over the event; 
(iii) the attacker is not a goalkeeper; 
(iv) the angle between \( \rr \) and the vector from the attacker's position to the opponent's goal is less than \( 60^\circ \) at the start of the event; and 
(v) the nearest opposing player remains unchanged throughout the event. 
The angular condition ensures that the defender lies approximately between the attacker and their own goal at the start of the event (see \MetSec for additional details on data preprocessing).

\subsection{\label{subsec:parameter_estimation} Estimating parameters}

We fit our model to the identified one-on-one events and estimate parameters. 
Let \( \{ \rtp (t_n) \}_{n=0, \ldots, T-1} \) denote the smoothed empirical trajectory of player \( \p \in \{\at, \df\} \) during a given event, where \( T \) is the number of recorded frames for that event. 
We fit the model trajectory, \( \rp \), to the empirical trajectory, \( \rtp \), by minimizing the following error function over the parameters \( (\taup, \fp, \tp) \): 
\begin{align}
    \epsilon_{\p} = \frac{ \frac{1}{T-1}\sum_{t=1}^{T-1} \|\tilde{\mathbf{r}}_{\mathrm{p}}(t) - \mathbf{r}_{\mathrm{p}}(t)\|}{\sum_{t=1}^{T-1} \|\tilde{\mathbf{r}}_{\mathrm{p}}(t) - \tilde{\mathbf{r}}_{\mathrm{p}}(t-1)\|},\quad \p \in\{\at, \df\}. 
    \label{eq:err_func}
\end{align}
Here \( \|\cdot \| \) denotes the Euclidean norm. 
The numerator represents the mean pointwise distance between the model and empirical trajectories. 
The denominator provides a discrete approximation to the path length of the empirical trajectory, obtained by summing the Euclidean distances between successive points on the trajectory. 
This quantity represents the mean pointwise distance between the trajectories normalized by the total path length.

We obtain the model trajectory \( \rp \) by numerically evolving the corresponding equation of motion in Eq.~\ref{eq:oad} over a broad range of parameter values, while fixing the trajectory of the opposing player \( \q \neq p \) to its empirical counterpart, namely, \( \rqq \equiv \rtq \).  
Repeating this process for different parameter values, we identify the parameters that minimize \( \epsp \) over bounded parameter ranges for each one-on-one event. 
Our analysis is restricted to events for which errors are sufficiently small. 
More details are provided in \SM~\bl{Secs.~S1 and S2}.

\subsection{\label{subsec:preliminaries} Preliminaries}

Let \( \rr \coloneqq \rd - \ra \) and \( \vv \coloneqq \vdf - \vat \) denote the relative position and velocity, respectively. 
We also define the center-of-mass variables, up to a constant factor, as \( \RR \coloneqq \ra + \rd \) and \( \VV \coloneqq \vat + \vdf \). 
Let \( \er \coloneqq \eo \) and \( \es \coloneqq - \et \). 
The equations governing the relative and center-of-mass motion are given by 
\begin{subequations} \label{eq:eom_relative_com}
\begin{align}
    \dot \vv &= - \alpha \vv - \beta \VV + F_1 \er + F_2 \es, \\
    \dot \VV &= - \beta \vv - \alpha \VV + G_1 \er + G_2 \es,
\end{align}
\end{subequations}
where \( \alpha \), \( \beta \), \( F_1 \), \( F_2 \), \( G_1 \), and \( G_2 \) are defined in the main text. 
We represent the relative position \( \rr \) in polar coordinates \( (\rho_r, \phi_r) \) as 
\begin{align}
    \rr(t) = \rho_r(t)
    \begin{bmatrix}
        \cos \phi_r(t) \\
        \sin \phi_r(t)
    \end{bmatrix}. 
    \label{eq:r_polar}
\end{align}
Similarly, the center-of-mass position \( \RR \) is expressed in the basis \( \{ \er, \es \} \), in polar coordinates \( (\rho_R, \phi_R) \), as 
\begin{multline}
    \RR(t) = \rho_R(t) [\cos (\phi_R(t) - \phi_r(t))\, \er \\
    + \sin (\phi_R(t) - \phi_r(t)) \,\es ]. \label{eq:R_polar}
\end{multline}
For convenience, we omit explicit time dependence where clear from context.

\subsection{\label{subsec:limit_cycle} Limit cycles}

\begin{figure*}
    \centering
    \includegraphics[width=\linewidth]{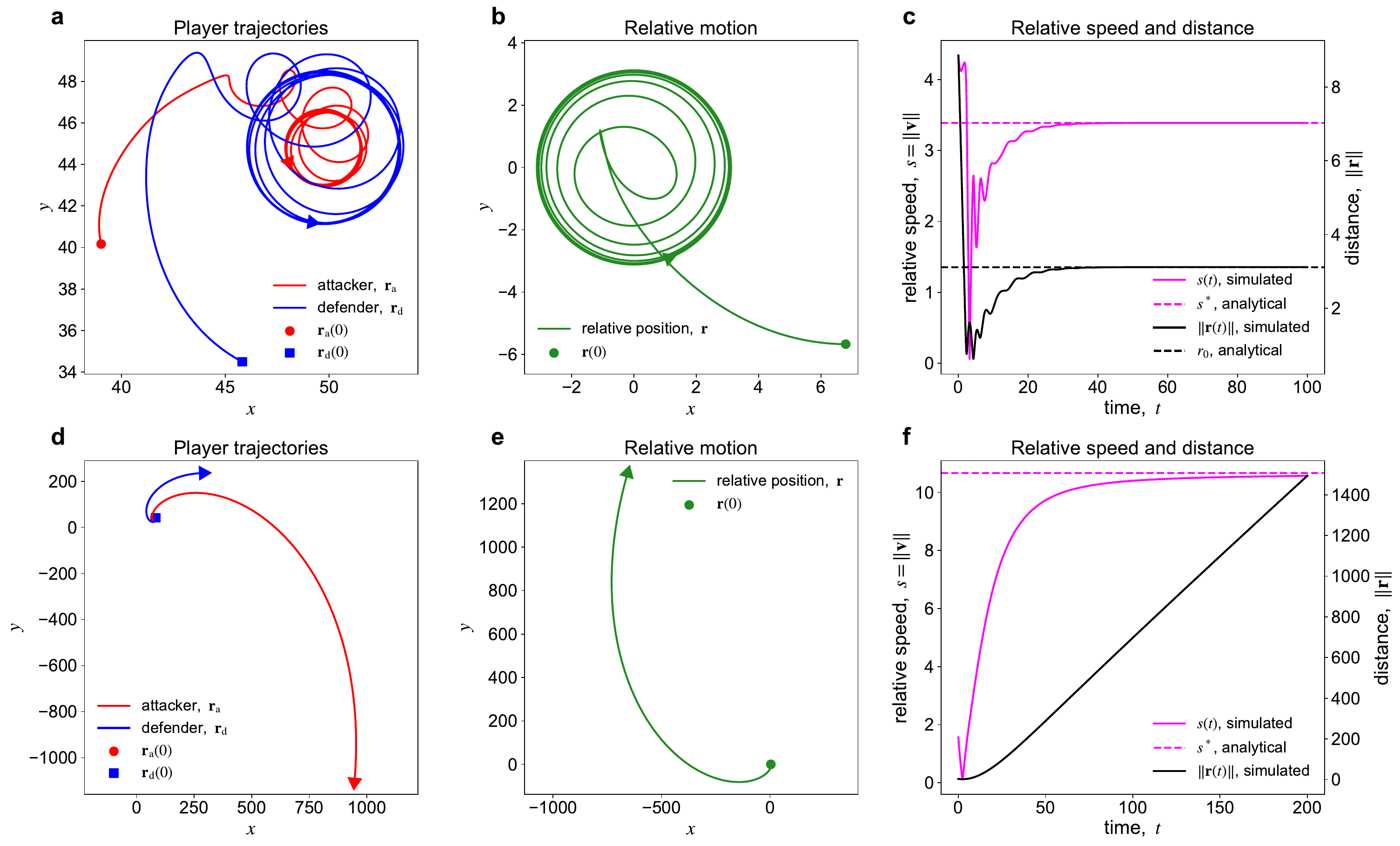}
    \exfigcaption[\linewidth]{
    Model trajectories extended well beyond empirical time windows and the corresponding time evolution of state variables (the model trajectories in Fig.~\ref{fig:empirical_model_trajectories} are extended): 
    \pnl{a--c,} limit cycles converging to uniform circular motion (corresponding to Fig.~\ref{fig:empirical_model_trajectories}b); 
    \pnl{d--f,} logarithmic spirals growing outward (corresponding to Fig.~\ref{fig:empirical_model_trajectories}d). 
    Solutions are extended up to (\pnl{a--c}) \( t = 100 \) and (\pnl{d--f}) \( t = 200 \), so that the asymptotic behaviors are visibly recognizable. 
    \pnl{c,f,} The horizontal dashed lines represent theoretical predictions. 
    Parameters and initial conditions: 
    \pnl{a--c,} 
    \( \fa = 2.19 \), \( \fd = 4.71 \), \( \ta = -1.42 \), \( \td = -0.92 \), \( \taua = 4.66 \), \( \taud = 2.17 \), \( \ra(0) = (39.02, 40.17) \), \( \rd(0) = (45.81, 34.50) \), \( \vat(0) = (-0.60, 2.56) \), \( \vdf(0) = (-4.94, 2.59) \); 
    \pnl{d--f,} 
    \( \fa = 0.95 \), \( \fd = 1.16 \), \( \ta = -2.55 \), \( \td = 1.13 \), \( \taua = 10.66 \), \( \taud = 1.75 \), \( \ra(0) = (79.39, 41.78) \), \( \rd(0) = (82.90, 41.87) \), \( \vat(0) = (1.00, -4.82) \), \( \vdf(0) = (-0.41, -5.49) \). 
    }
    \label{fig:ex:trajectories_extended}
\end{figure*}

Here we derive the asymptotic behaviors of our model from Eq.~\ref{eq:eom_relative_com}, mentioned in Sec.~\ref{subsec:analysis}. 
First, we derive limit cycles and identify the asymptotic relative speed. 
Using the polar representations of \( \rr \) and \( \RR \) (see Eqs.~\ref{eq:r_polar} and \ref{eq:R_polar}), we rewrite the equations of relative and center-of-mass motion. 
Because \( \dot{\mathbf{e}}_r = \dot\phi_r \es \) and \( \dot{\mathbf{e}}_s = - \dot\phi_r \er \), we find that 
\begin{align}
    \vv &= \dot\rho_r \er + \rho_r \dot\phi_r \es, \\
    \dot\vv &= \left[\ddot \rho_r - \rho_r (\dot\phi_r)^2 \right] \er + (2\dot\rho_r \dot\phi_r + \rho_r \ddot\phi_r ) \es. 
\end{align}
Likewise, letting \( \Phi:= \phi_R - \phi_r \), we find 
\begin{multline}
    \VV = \left[ \dot\rho_R \cos \Phi - \rho_R \dot \phi_R \sin \Phi \right] \er \\
    + \left[\dot\rho_R \sin \Phi + \rho_R \dot \phi_R \cos \Phi \right] \es, 
\end{multline}
and 
\begin{multline}
    \dot \VV = \Bigl[ \ddot\rho_R \cos \Phi - 2\dot \rho_R \dot\phi_R \sin \Phi - \rho_R \ddot \phi_R \sin \Phi - \rho_R (\dot\phi_R)^2  \cos \Phi \Bigr] \er \\
    + \Bigl[ \ddot \rho_R \sin \Phi + 2\dot\rho_R \dot\phi_R \cos \Phi + \rho_R \ddot\phi_R \cos \Phi - \rho_R (\dot \phi_R)^2 \sin \Phi \Bigr] \es.
\end{multline}
Then, the relative and center-of-mass motions (Eq.~\ref{eq:eom_relative_com}) are described by 
\begin{align}
&\ddot \rho_r - \rho_r (\dot\phi_r)^2 \nonumber \\
&\quad
= - \alpha \dot \rho_r - \beta \left(\dot \rho_R \cos\Phi - \rho_R \dot\phi_R \sin \Phi\right) + F_1, \label{eq:eom_r_radial}\\
&2\dot\rho_r \dot\phi_r + \rho_r \ddot \phi_r \nonumber \\
&\quad
= - \alpha \rho_r \dot \phi_r - \beta \left(\dot\rho_R \sin \Phi + \rho_R \dot\phi_R \cos \Phi \right)+ F_2, \label{eq:eom_r_tangential}\\
&\ddot\rho_R \cos \Phi - 2\dot \rho_R \dot\phi_R \sin \Phi - \rho_R \ddot \phi_R \sin \Phi 
- \rho_R (\dot\phi_R)^2  \cos \Phi \nonumber \\
&\quad
=- \beta \dot\rho_r - \alpha \left(\dot\rho_R \cos \Phi - \rho_R \dot\phi_R \sin \Phi \right) + G_1, \label{eq:eom_R_radial}\\
&\ddot \rho_R \sin \Phi + 2\dot\rho_R \dot\phi_R \cos \Phi + \rho_R \ddot\phi_R \cos \Phi - \rho_R (\dot \phi_R)^2 \sin \Phi \nonumber \\
&\quad
= - \beta \rho_r \dot\phi_r
- \alpha \left( \dot\rho_R \sin \Phi + \rho_R \dot\phi_R \cos \Phi \right) + G_2. \label{eq:eom_R_tangential}
\end{align}
Suppose both relative and center-of-mass trajectories exhibit uniform circular motions, potentially with different radii and angular frequencies, 
that is, 
\begin{align*}
    &\rho_r(t) = r_0, \quad 
    \phi_r(t) = \omega t, \\ 
    &\rho_R(t) = R_0, \quad 
    \phi_R(t) = \Omega t + \lambda. 
\end{align*}
Here, \( r_0 > 0 \) and \( R_0 > 0 \) are the radii of the relative and center-of-mass circular motions, with angular velocities \( \omega \) and \( \Omega \), respectively (\( \omega, \Omega \neq 0 \)). 
The constant \( \lambda \) accounts for the phase difference between the relative and center-of-mass circular motions. 
We find 
\begin{align*}
    &\ddot\rho_r = \dot\rho_r = 0, \quad 
    \ddot \rho_R = \dot \rho_R = 0, \\ 
    &\dot \phi_r = \omega, \quad 
    \dot \phi_R = \Omega, \quad 
    \ddot \phi_r = \ddot \phi_R = 0. 
\end{align*}
Using these relations, Eqs.~\ref{eq:eom_r_radial} and \ref{eq:eom_r_tangential} become   
\begin{align*}
    -r_0 \omega^2 &= \beta R_0 \Omega \sin (\phi_R-\phi_r) + F_1, \\
    0 &= - \alpha r_0 \omega - \beta R_0 \Omega \cos (\phi_R - \phi_r) + F_2. 
\end{align*}
Because the left-hand sides of the equations above are time-independent, so are the right-hand sides. 
Thus, \( \omega = \Omega \). 
That is, 
\begin{align}
    -r_0 \omega^2 &= \beta R_0 \omega \sin \lambda + F_1, \label{eq:eom_r_radial_UCM} \\
    0 &= - \alpha r_0 \omega - \beta R_0 \omega \cos \lambda + F_2. \label{eq:eom_r_tangential_UCM}
\end{align}
Similarly, given \( \omega = \Omega \), we find 
\begin{align}
    - R_0 \omega^2 \cos \lambda &=  \alpha R_0 \omega \sin \lambda + G_1, \label{eq:eom_R_radial_UCM} \\
    - R_0 \omega^2 \sin \lambda &= - \beta r_0 \omega - \alpha R_0 \omega \cos \lambda + G_2. \label{eq:eom_R_tangential_UCM}
\end{align}
Let \( X = r_0 \omega \), \( Y = R_0 \omega \cos \lambda \), and \( Z = R_0 \omega \sin \lambda \). 
Then, Eqs.~\ref{eq:eom_r_radial_UCM}--\ref{eq:eom_R_tangential_UCM} are given by the coupled algebraic equations 
\begin{align}
\begin{aligned}
    - \omega X &= \beta Z + F_1, \\
    0 &= - \alpha X - \beta Y + F_2, \\
    - \omega Y &= \alpha Z + G_1, \\
    - \omega Z &= - \beta X - \alpha Y + G_2 .
\end{aligned}
\end{align}
After some algebraic manipulations, we obtain  
\begin{align}
    X = \frac{-\alpha F_1 + \beta G_1 + \omega F_2}{2\alpha \omega}. 
\end{align}
Therefore, the speed of this uniform circular motion is given by 
\begin{align}
    s = r_0 |\omega| 
    = |X| = \frac{\left|-\alpha F_1 + \beta G_1 + \omega F_2\right|}{2\alpha |\omega|}.  
\end{align} 
The angular velocity \( \omega \) satisfies the following cubic equation: 
\begin{multline}
    F_2 \omega^3 + (\alpha F_1 + \beta G_1) \omega^2 
    + \left[F_2(\alpha^2 +\beta^2) - 2\alpha\beta G_2 \right] \omega \\
    + (\alpha^2 - \beta^2)(\alpha F_1 - \beta G_1) = 0. 
\end{multline}
We note that if \( \tilde\tau \coloneqq \taua = \taud \) and \( \fa = \fd \), 
\begin{align}
    \omega = \frac{1}{\tilde\tau}\frac{1}{\tan B}, \quad  B = \frac{\ta - \td}{2}. 
\end{align} 
When the relative and center-of-mass trajectories both exhibit uniform circular motions, the attacker and defender are also in uniform circular motion (see Fig.~\ref{fig:empirical_model_trajectories}b or \AppFig{fig:ex:trajectories_extended}).

While not essential for understanding our main results, it is instructive to derive individual player motion in addition to the relative and center-of-mass motion. 
To this end, we identify \( \mathbb{R}^2 \) with \( \mathbb{C} \) and express our solution in the complex plane: 
\begin{align}
    \rr(t) = r_0 e^{i\omega t}, \quad \RR(t) = R_0 e^{i(\omega t + \lambda)},
\end{align}
where \( r_0, R_0 \in \mathbb{R}_{>0} \) and \( i \) denotes the imaginary unit. 
We find 
\begin{align}
    \ra = \frac{e^{i\omega t}}{2} \left( - r_0 + R_0 e^{i\lambda} \right), \quad
    \rd = \frac{e^{i\omega t}}{2} \left( r_0 + R_0 e^{i\lambda} \right). 
\end{align}
Let us define
\begin{align}
    \rhoa &= \frac{1}{2}\sqrt{r_0^2 + R_0^2 - 2r_0 R_0 \cos \lambda}, \\ 
    \psia &= \arg \left(-r_0 + R_0 e^{i \lambda}\right).\\
    \rhod &= \frac{1}{2}\sqrt{r_0^2 + R_0^2 + 2r_0 R_0 \cos \lambda}, \\ 
    \psid &= \arg \left(r_0 + R_0 e^{i \lambda}\right).
\end{align}
Then we obtain 
\begin{align}
    \ra = \rhoa e^{i(\omega t + \psia)}, 
    \quad 
    \rd = \rhod e^{i(\omega t + \psid)},
\end{align}
implying that both the attacker and defender undergo circular motion at constant speed. 
Finally, we note that our model is translationally invariant, i.e., 
\begin{align}
    (\ra', \rd') = (\ra + \bm{c}, \rd + \bm{c})
\end{align}
is also a solution for any constant vector \( \bm{c} \in \mathbb{R}^2 \) if \( (\ra, \rd) \) satisfies Eq.~\ref{eq:oad}. 
Therefore, the attacker and defender do not necessarily rotate around the origin; rather, they may exhibit uniform circular motion centered around any point in the plane, depending on the initial conditions.

\subsection{\label{subsec:logarithmic_spiral} Asymptotic logarithmic spirals}

Next, we derive asymptotic logarithmic spirals that grow outward. 
Let us consider cases where 
\begin{align}
    \label{eq:asymptotic_logarithmic_spiral_condition}
    \dot \rho_r \to a, \quad \rho_r \dot\phi_r \to c, \quad 
    \dot \rho_R \to A, \quad \rho_R \dot\phi_R \to C,
\end{align}
where \( a, A > 0 \) and \( c, C \neq 0 \) are constants as \( t \to \infty \). 
This means that both relative and center-of-mass motions attain constant radial and tangential speeds in the long-time limit. 
Assuming that the radial acceleration of relative motion vanishes asymptotically (\( \ddot \rho_r \to 0 \)), 
we find 
\begin{align}
    \rho_r (\dot\phi_r)^2 \sim \frac{c^2}{\rho_r} \to 0, \quad \dot\rho_r\dot\phi_r \sim a \frac{c}{\rho_r} \to 0. 
\end{align}
Similarly, provided \( \ddot \rho_R \to 0 \), we find 
\begin{align}
    \rho_R (\dot\phi_R)^2 \to 0, \quad \dot\rho_R \dot\phi_R \to 0.
\end{align}
Then, these parameters satisfy the following asymptotic relations: 
\begin{align}
    \label{eq:stationary_condition}
    \begin{cases}
        0 \sim - \alpha a - \beta (A \cos \Phi - C \sin \Phi) + F_1, \\
        0 \sim - \alpha c - \beta (A \sin \Phi + C \cos \Phi) + F_2, \\
        0 \sim - \beta a - \alpha (A\cos \Phi - C \sin \Phi) + G_1, \\
        0 \sim - \beta c - \alpha (A \sin \Phi + C \cos \Phi) + G_2,
    \end{cases}
\end{align}
giving 
\begin{align}
    a = \frac{\alpha F_1 - \beta G_1}{\alpha^2 - \beta^2}, \quad 
    c = \frac{\alpha F_2 - \beta G_2}{\alpha^2 - \beta^2}. 
\end{align}
Therefore, 
\begin{align*}
    \| \vv_{\infty}\| 
    &= \sqrt{a^2 + c^2} \\
    &= \frac{1}{\alpha^2-\beta^2}\sqrt{(\alpha F_1 - \beta G_1)^2 + (\alpha F_2 - \beta G_2)^2} 
\end{align*}
Equivalently, in the dimensionless form (see Sec.~\ref{subsec:dimensionless_form}), it is given by 
\begin{align}
    \|\hat \vv_\infty\| = \frac{\|\vv_\infty\|}{\taud \fd} = \sqrt{1 + (\eta \mu)^2 + 2\eta \mu \cos (\td + \ta)}, \label{eq:LS_speed_asymptotic}
\end{align}
where \( \eta \coloneqq \taua/\taud \) and \( \mu \coloneqq \fa/\fd \). 
Notice that \( \ta + \td = \pm\pi \) minimizes the asymptotic relative speed obtained above, in which case the asymptotic relative speed is given by \( \|\vv_\infty\| = |\taua \fa - \taud \fd| \), i.e., the absolute difference between the attacker's and defender's steady-state speeds. 
Because \( \eta \mu > 0 \), however, this may not be globally minimum in the \( (\ta, \td) \) parameter space; in fact, we will later show that it is suboptimal globally when the limit cycle regime is taken into account and the large-\( \omega \) branch is dynamically relevant.

We derive individual motion from the asymptotic relative and center-of-mass motion. 
We note that 
\begin{align}
    \rr = \rho_r \er, \quad \RR = \rho_R (\cos\Phi \,\er + \sin \Phi \,\es).
\end{align}
Therefore, we find 
\begin{align}
    \ra &= \frac{1}{2} \left[(-\rho_r + \rho_R \cos \Phi) \er + \rho_R \sin \Phi \,\es \right], \\
    \rd &= \frac{1}{2} \left[(\rho_r + \rho_R \cos \Phi) \er + \rho_R \sin \Phi \,\es \right]. 
\end{align}
The radius of each individual motion is thus given by 
\begin{align}
    \rhoa &= \frac{1}{2} \sqrt{\rho_r^2 + \rho_R^2 - 2 \rho_r \rho_R \cos \Phi}, \\
    \rhod &= \frac{1}{2} \sqrt{\rho_r^2 + \rho_R^2 + 2 \rho_r \rho_R \cos \Phi}. 
\end{align}
Let \( \Psia \) and \( \Psid \) be the angles of the attacker and the defender relative to \( \er \), respectively, so that 
\begin{align}
    \ra = \rhoa e^{i \phia}, \quad 
    \rd = \rhod e^{i \phid},
\end{align}
where 
\begin{align}
    \phia = \phi_r + \Psia, \quad 
    \phid = \phi_r + \Psid. 
\end{align}
Since \( \rho_r = at + o(t) \) and \( \rho_R = At + o(t) \), we find
\begin{align}
    \dot\Phi 
    = \dot \phi_R - \dot \phi_r 
    = \frac{1}{t} \left(\frac{C}{A} - \frac{c}{a} \right) + o(1/t),
\end{align}
giving rise to
\begin{align}
    \Phi(t) = \Phi^\ast + \left(\frac{C}{A} - \frac{c}{a} \right) \ln t + o(\ln t).
\end{align}
Consistency of Eq.~\ref{eq:stationary_condition} requires that \( \Phi \) converges to some time-independent value \( \Phi^\ast \) modulo \( 2\pi \) as \( t \to \infty \), implying 
\begin{align}
    k \coloneqq \frac{C}{A} = \frac{c}{a}, 
\end{align}
and 
\begin{align}
    \Phi \to \Phi^*. 
\end{align} 
Provided that \( \rho_r \) and \( \rho_R \) scale with \( t \) for large \( t \), \( \rhoa \) and \( \rhod \) also scale with \( t \). 
Likewise, we find 
\begin{align}
    \Psip \to \Psip^*, \quad \p \in \{ \at, \df \}.
\end{align}
Therefore, in the limit of \( t \to \infty \), we obtain 
\begin{align}
    \phip(t) 
    \sim k \ln \rhop(t),
\end{align}
implying that player \( \p \)'s motion follows a logarithmic spiral asymptotically.

Finally, we explain why Eq.~\ref{eq:asymptotic_logarithmic_spiral_condition} produces asymptotic logarithmic spirals. 
Here we focus on relative motion; the same argument applies to the center-of-mass motion. 
Given \( \dot \rho = a \), we have \( \rho = at + \rho_0 \), where \( \rho_0 \) is the initial distance between the attacker and the defender. 
Because \( \rho \dot \phi = c \), we find 
\begin{align}
    \phi(t) = \int_0^t \frac{c}{as + \rho_0}ds = \frac{c}{a} \left[\ln (as + \rho_0)\right]^{s=t}_{s=0},
\end{align}
yielding 
\begin{align}
    \phi =  \frac{c}{a} \ln \frac{\rho}{\rho_0}, 
\end{align}
which reproduces the definition of a logarithmic spiral.

\subsection{\label{subsec:optimal_curve} Derivation of the optimal curve}

Let us turn to the limit cycle regime, where the asymptotic speed is given by Eq.~\ref{eq:s}. 
Here we show that minimum asymptotic speed is achieved (in the \( \ta \)--\( \td \) plane) only in the limit-cycle regime for any \( \eta > 0 \) and \( \mu > 0 \). 
We also derive the analytical expression for the optimal curve \( \opts \) by a simple perturbation argument.

Our goal is to find the angular frequency \( \omega \) that minimizes the asymptotic relative speed \( s^* \). 
From Eq.~\ref{eq:s}, the asymptotic speed is bounded by 
\begin{align}
    s^* \leq \frac{|-\alpha F_1 + \beta G_1|}{2\alpha |\omega|} + \frac{1}{2\alpha} |F_2|. \label{eq:asymptotic_rel_speed_inequality}
\end{align}
We notice that the right-hand side of Eq.~\ref{eq:asymptotic_rel_speed_inequality} tends to zero as \( |\omega| \to \infty \) and \( |F_2| \to 0 \). 
Since \( \omega \) depends on \( F_2 \), however, we must investigate the limiting behavior of \( |\omega| \) in the limit \( |F_2| \to 0 \), i.e., whether \( |\omega| \) indeed tends to infinity in this limit.

Suppose \( F_2 = O(\epsilon) \), where \( \epsilon \ll 1 \). 
In the limit \( \epsilon \to 0 \), Eq.~\ref{eq:omega_cubic} is singularly perturbed, and this equation has a solution of the order \( \epsilon^{-1} \) (provided the non-degeneracy condition \( \alpha F_1 + \beta G_1 \neq 0 \)). 
The other two solutions are \( O(1) \). 
The large-\( \omega \) branch, \( \omega = O(\epsilon^{-1}) \), blows up as \( \epsilon \) tends to zero. 
Therefore, the asymptotic relative speed in this limit is given by \( s^* = O(\epsilon) \); 
as a corollary, we obtain \( r_0 = O(\epsilon^2) \). 
In other words, while the angular frequency of circular motion tends to infinity, its speed converges to zero because the radius converges to zero at a faster rate than the angular frequency blows up. 
In short, as \( |F_2| \to 0 \), the asymptotic relative speed becomes arbitrarily small, that is, 
\begin{align}
    \forall \zeta > 0, \, \exists (\theta_a, \theta_d): \; 0 < s^*(\theta_a, \theta_d) < \zeta, 
\end{align}
provided that the singular solution is dynamically relevant. 
The analytical expression for the optimal curve (Eq.~\ref{eq:opts}) arises from the limiting condition \( F_2 = 0 \). 
Note that \( |F_2| > 0 \) is required to guarantee \( r_0 > 0 \).

\subsection{Overlap (or lack thereof) between LS and LC regimes}

We discuss whether and when the limit-cycle and logarithmic spiral regimes are mutually exclusive and collectively exhaustive in the \( (\ta, \td) \) space. 
Here we focus on relative motion. 
A necessary condition for an outward logarithmic spiral to emerge is given by \( a > 0 \). More explicitly, 
\begin{align}
    P \coloneqq \alpha F_1 - \beta G_1 > 0.
\end{align}
Similarly, a necessary condition for uniform circular motion to hold is \( r_0 > 0 \), or equivalently,  
\begin{align}
    Q \coloneqq - \alpha F_1 + \beta G_1 + \omega F_2 > 0. 
\end{align}
Here we are interested in whether \( P \) and \( Q \) have opposite signs. 
First, we find 
\begin{align}
    Q = - P + \omega F_2. 
\end{align}
Using these notations, Eq.~\ref{eq:omega_cubic} is given by 
\begin{align}
    F_2 \omega^3 + (\alpha F_1 + \beta G_1) \omega^2 + K \omega + (\alpha^2 - \beta^2)P = 0, \label{eq:omega_cubic_new_notation}
\end{align}
where 
\begin{align}
    K = F_2(\alpha^2 + \beta^2) - 2 \alpha \beta G_2. 
\end{align}
Suppose that \( P \) is small. 
Then, Eq.~\ref{eq:omega_cubic_new_notation} yields the small angular frequency 
\begin{align}
    \omega \approx -\frac{P}{K}(\alpha^2-\beta^2), \quad P \approx 0, \label{eq:small_omega_approx}
\end{align}
for \( K \neq 0 \). 
The induced quadratic equation is 
\begin{align}
    F_2 \omega^2 + (\alpha F_1 + \beta G_1) \omega + K \approx 0,
\end{align}
which has no real solution if and only if 
\begin{align}
    \Delta = (\alpha F_1 + \beta G_1)^2 - 4 F_2 K < 0,
\end{align}
which requires 
\begin{align}
    F_2 K > 0. 
\end{align}
Therefore, when \( P \) is small, if the cubic has indeed only one real solution approximated by Eq.~\ref{eq:small_omega_approx} (hence two other complex conjugates), \( P \) and \( Q \) have opposite signs, because 
\begin{align}
    Q \approx - \left(1 + \frac{\alpha^2-\beta^2}{K} F_2 \right) P. 
\end{align}
If the cubic has only one real root, this result extends beyond the limiting regime: 
that is, the expressions \( P \) and \( Q \) have opposite signs throughout the parameter space generically. 
To see why, we investigate whether \( Q = 0 \) and \( P \neq 0 \) can occur at the same time. 
Suppose this happens assuming that the cubic has only one real root. 
We denote such an \( \omega \) by \( \bar{\omega} \coloneqq P/F_2 \). 
Using \( \bar{\omega} \), we factor the following cubic polynomial: 
\begin{align}
    \varphi(\omega) 
    = F_2 \omega^3 + (\alpha F_1 + \beta G_1) \omega^2 + K \omega + (\alpha^2 - \beta^2)P. 
\end{align}
Since \( \varphi(\bar\omega) = 0 \), we find 
\begin{align}
    \varphi(\omega) = (\omega - \bar\omega)(\mathcal{A}\omega^2 + \mathcal{B}\omega + \mathcal{C}),
\end{align}
where the polynomial coefficients are 
\( \mathcal{A} = F_2 \), 
\( \mathcal{B} = \alpha F_1 + \beta G_1 + \bar\omega F_2 \), and 
\( \mathcal{C} = - (\alpha^2-\beta^2)F_2 \). 
The discriminant of the quadratic polynomial \( \mathcal{A}\omega^2 + \mathcal{B}\omega + \mathcal{C} \) is given by \( \bar\Delta = (\alpha F_1 + \beta G_1 + \bar\omega F_2)^2 + 4 F_2^2 (\alpha^2 - \beta^2) \), which is strictly positive. 
Thus, the cubic equation \( \varphi(\omega) = 0 \) has three real roots, contradicting the initial assumption.

We just showed that if the cubic has only one real root, \( P \) and \( Q \) always have opposite signs. 
When does the cubic have only one real root? 
Below, we present a simple and informative case in which the cubic always has a unique real root. 
Suppose that \( \beta \) is small (i.e., \( \taua \approx \taud \)). 
Up to leading order, the cubic reduces to 
\begin{align}
    (\omega^2 + \alpha^2) (F_2 \omega + F_1 \alpha) = 0, 
\end{align}
which has exactly one real solution given by 
\begin{align}
    \omega = - \frac{F_1}{F_2}\alpha,
\end{align}
when \( F_2 \neq 0 \). 
Therefore, when \( \taua \approx \taud \), 
it is guaranteed that \( P \) and \( Q \) have opposite signs, implying that the two asymptotic regimes for relative motion are mutually exclusive and collectively exhaustive in the angle space 
(except in some measure-zero regions), if the necessary conditions discussed thus far suffice to characterize each asymptotic regime.

\subsection{\label{subsec:characteristic_timescale} An estimate of the exponential decay time}

Here we derive the intrinsic damping time of the system. 
Let \( \Fa(\ra, \rd) \coloneqq \fa [\cos \ta \eo + \sin\ta \et] \) and \( \Fd(\ra, \rd) \coloneqq \fd [-\cos\td \eo + \sin\td \et] \). 
Since 
\begin{multline}
    \vp(t) = \vp(0) e^{-\frac{t}{\taup}} \\
    + \int_0^t \Fp(\ra(s), \rd(s)) e^{-\frac{1}{\taup}(t-s)} ds,
\end{multline}
we obtain 
\begin{align*}
    \| \vp(t) \| 
    &\leq \|\vp(0)\| e^{-\frac{t}{\taup}} \\
    &\quad\quad + \int_0^t \|\Fp(\ra(s), \rd(s))\| e^{-\frac{1}{\taup}(t-s)} ds \\
    &= \|\vp(0)\| e^{-\frac{t}{\taup}} + \fp\taup (1 - e^{-\frac{t}{\taup}}). 
\end{align*}
Therefore, the intrinsic damping time for \( \vp \) is \( \taup \). 
Similarly, since 
\begin{align*}
    \| \vv(t) \| &= \|\vdf(t) - \vat(t) \| \leq \| \vdf(t) \| + \| \vat(t) \| \\
    &\leq \sum_{\p \in \{\at, \df \}} \fp \taup + (\|\vp(0)\| - \fp \taup) e^{-t/\taup} ,
\end{align*}
we find that the intrinsic damping time for \( \vv \) is given by \( \tau \coloneqq \max(\taua, \taud) \). 
In other words, the effect of the initial condition disappears over the timescale of \( \tau \). 
Note, however, that this only provides the exponential convergence time for \( \vv \) to reach the ultimate bound \( \fa\taua + \fd\taud \), which is different from orbital convergence.

\subsection{\label{subsec:dimensionless_form} Dimensionless formulation} 

We introduce the nondimensional parameters 
\begin{align}
    \eta = \frac{\taua}{\taud}, \quad \mu = \frac{\fa}{\fd}. 
\end{align}
Under the following change of variables, 
\begin{align*}
    &\hat t = \frac{t}{\taud}, \quad
    \hatra = \frac{\ra}{\taud^2 \fd}, \quad
    \hatrd = \frac{\rd}{\taud^2 \fd}, \\
    &\hatvat = \frac{\vat}{\taud \fd}, \quad
    \hatvdf = \frac{\vdf}{\taud \fd}, 
\end{align*}
Eq.~\ref{eq:oad} is written as  
\begin{align*}
    \begin{dcases}    
        \frac{d \hatvat}{d\hat t} = - \frac{\hatvat}{\eta}  + \mu \left[ \cos \ta \, \eo(\hatra, \hatrd) + \sin \ta \, \et(\hatra, \hatrd) \right], \\     
        \frac{d \hatvdf}{d\hat t} = - \hatvdf  + \left[ - \cos \td \, \eo(\hatra, \hatrd) + \sin \td \, \et(\hatra, \hatrd) \right].
    \end{dcases}
\end{align*}
This implies that only four parameters are relevant---namely, \( \ta \), \( \td \), \( \eta > 0 \), and \( \mu > 0 \). 
All of our analytical results apply by resetting 
\begin{align}
    \taua = \eta, \quad \taud = 1, \quad \fa = \mu, \quad \fd = 1. 
\end{align}

\subsection{\label{subsec:peak_identification} Identifying modal angles}

We identify modal angles in a given dataset using the toroidal kernel density estimation \cite{di2011kernel} and local gradient-based optimization. 
First, we estimate the probability density function of angles in a given dataset, using a product of von Mises kernels and assuming independent angular components and an isotropic smoothing parameter: 
\begin{align}
    \rho(\ta,\td) \propto \sum_j \exp \left\{ \kappa \left[\cos(\ta - \ta^j) + \cos(\td - \td^j) \right] \right\},
\end{align}
where \( \tp^j \) is the angle for player \( \p \) in event \( j \), and \( \kappa > 0 \) is the smoothing factor. 
We then minimize the negative log of the estimated density using the L-BFGS-B algorithm (SciPy implementation) over an unconstrained parameter space, with angles mapped to \( [-\pi, \pi) \) via periodic wrapping. 
Optimization is initialized with points on a uniform grid over \( [-\pi, \pi)^2 \), as well as all individual data in the dataset. 
We exclude peaks that are in the vicinity of previously identified peaks within a threshold Euclidean distance on the torus (set to \( 0.2 \)), and retain the top two peaks with the highest densities. 
Implementing this process over all subsets of data, we identify six pairs of peaks corresponding to the various ranges of \( \mu \). 
We also examine whether the observed geometric structure remains robust when \( \kappa \) is varied (\( \kappa \in [2, 10] \)) (see \SM Figs.~\bl{S7--S14}); 
\( \kappa = 4 \) was applied in Figs.~\ref{fig:empirical_distribution} and \ref{fig:peak_motion}.

\subsection{\label{subsec:finite_time_landscapes} Finite-time evolution of objective landscapes} 

Here we show that asymptotic behaviors provide a reasonable approximation to the finite-time evolution over the course of a typical one-on-one event. 
First, the empirical median of the intrinsic decay time is \( \tau_{\text{med}} \approx \tauemp \), across all filtered one-on-one dribbling events. 
By comparison, the median duration of a one-on-one is approximately \eventdurationemp seconds. 
Therefore, the players' velocities decay exponentially, typically on a timescale comparable to the ball-possession duration. 

\begin{figure*}
    \centering
    \includegraphics[width=\linewidth]{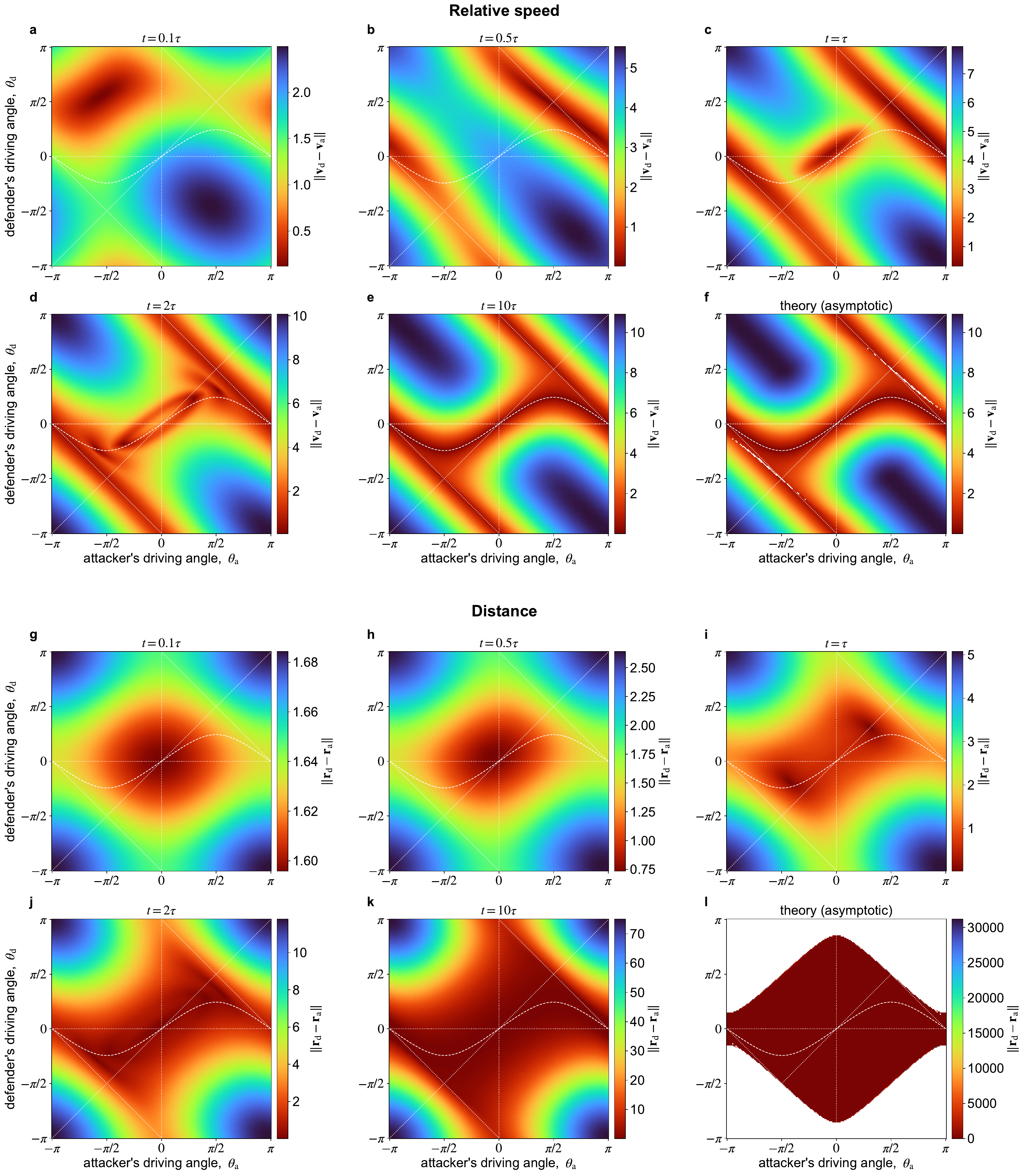}
    \exfigcaption[\linewidth]{
    Timeseries data of 
    (\pnl{a--e}) relative speed 
    and 
    (\pnl{g--k}) attacker--defender distance across angle space, computed from ODE simulations. 
    \pnl{f,l,} Corresponding asymptotic values. 
    Unclassified cases are indicated by blank regions; see the caption of Fig.~\ref{fig:relative_speed} for details. 
    The optimal curve \( \opts \) is indicated by the white dashed line within the LC regime. 
    White dotted lines serve as visual guides. 
    The finite-time phase diagram converges to the asymptotic prediction as time progresses; however, its overall structure is already well captured by the asymptotic prediction as early as \( t \approx \tau \), when the effects of initial conditions mostly disappear. 
    This suggests that asymptotic behavior provides a reasonable approximation to a finite-time objective landscape. 
    In panel \pnl{l}, values are shown only where theory predicts a unique limit cycle and no logarithmic outward spiral. 
    Parameters and initial conditions: \( \fa = 7.06 \), \( \fd = 10.17 \), \( \ta = 2.70 \), \( \td = 0.43 \), \( \taua = 0.73 \), \( \taud = 0.57 \), \( \ra(0) = (42.71, 17.92) \), \( \rd(0) = (44.24, 18.56) \), \( \vat(0) = (-4.51, -2.72) \), \( \vdf(0) = (-5.46, -1.56) \). 
    These values were taken or estimated from a real one-on-one event in the dataset. 
    See \SIFigSec for more examples.  
    }
    \label{fig:ex:timeseries_phase_diagram_1}
\end{figure*}

We next compute the relative speed at \( t = h \tau \) across the angle space (\( h > 0 \)). 
\AppTwoFigs{fig:ex:timeseries_phase_diagram_1}{fig:ex:timeseries_phase_diagram_2} visualize the finite-time relative speed and distance for a wide range of \( h \), with the asymptotic results in the lower-right panels for comparison. 
While the predicted asymptotic behavior (either a limit cycle or an outward logarithmic spiral) may not be evident yet, finite-time relative speed already shows meaningful variation across the angle space sufficient to reveal predicted optimal regions, typically as early as \( t \approx \tau \). 
This suggests that asymptotic behavior is a good predictor of the short-term evolution of relative speed on the timescale of the initial exponential relaxation of players' velocities---hence, of the typical ball-possession duration.

It is also worth noting that \( \mathcal{S}_2 \)-like lines can appear optimal at early times in the relative-speed landscape, despite not being asymptotically so (\AppTwoFigs{fig:ex:timeseries_phase_diagram_1}{fig:ex:timeseries_phase_diagram_2}).

\section*{Data Availability} 
The dataset used in this study was provided by DataStadium~Inc.~under institutional agreements and cannot be made publicly available due to contractual nondisclosure terms. 
The full dataset may be obtained directly from DataStadium~Inc.~for a fee.

\begin{acknowledgments}

H.G.~acknowledges support from the Japan Society for the Promotion of Science (JSPS) through the JSPS Research Fellowship for Young Scientists. 
T.N.~acknowledges support from a JSPS Grant-in-Aid for Early-Career Scientists (Grant No.~JP23K16729). 
H.N.~acknowledges support from JSPS KAKENHI (Grant No.~26K00677) and the MEXT Promotion of Distinctive Joint Usage/Research Center Support Program (Grant No.~JPMXP0724020292). 
The authors acknowledge DataStadium~Inc.~for providing access to proprietary data under nondisclosure agreements (access granted through paid licenses). 
This study was originally conceived for the 2025 Sports Data Science Competition, organized in part by the School of Statistical Thinking at the Institute of Statistical Mathematics, Research Organization of Information and Systems. 

\end{acknowledgments}

\section*{Author Contributions}

All authors contributed to conceptualization and study design, interpretation of the results, and editing of the paper; 
I.Y., H.G., K.O., and T.N. contributed to data curation and analysis; 
H.G. performed mathematical analysis with input from K.O.; 
I.Y. and H.G. designed and generated figures; 
H.G. led the writing of the paper with literature input from T.N., and prepared the Analysis and Supplementary Figure sections in Supplementary Information; 
H.G. and K.O. wrote the Methods section in Supplementary Information. 
I.Y. and H.G. contributed equally to this work.

\begin{figure*}
    \centering
    \includegraphics[width=\linewidth]{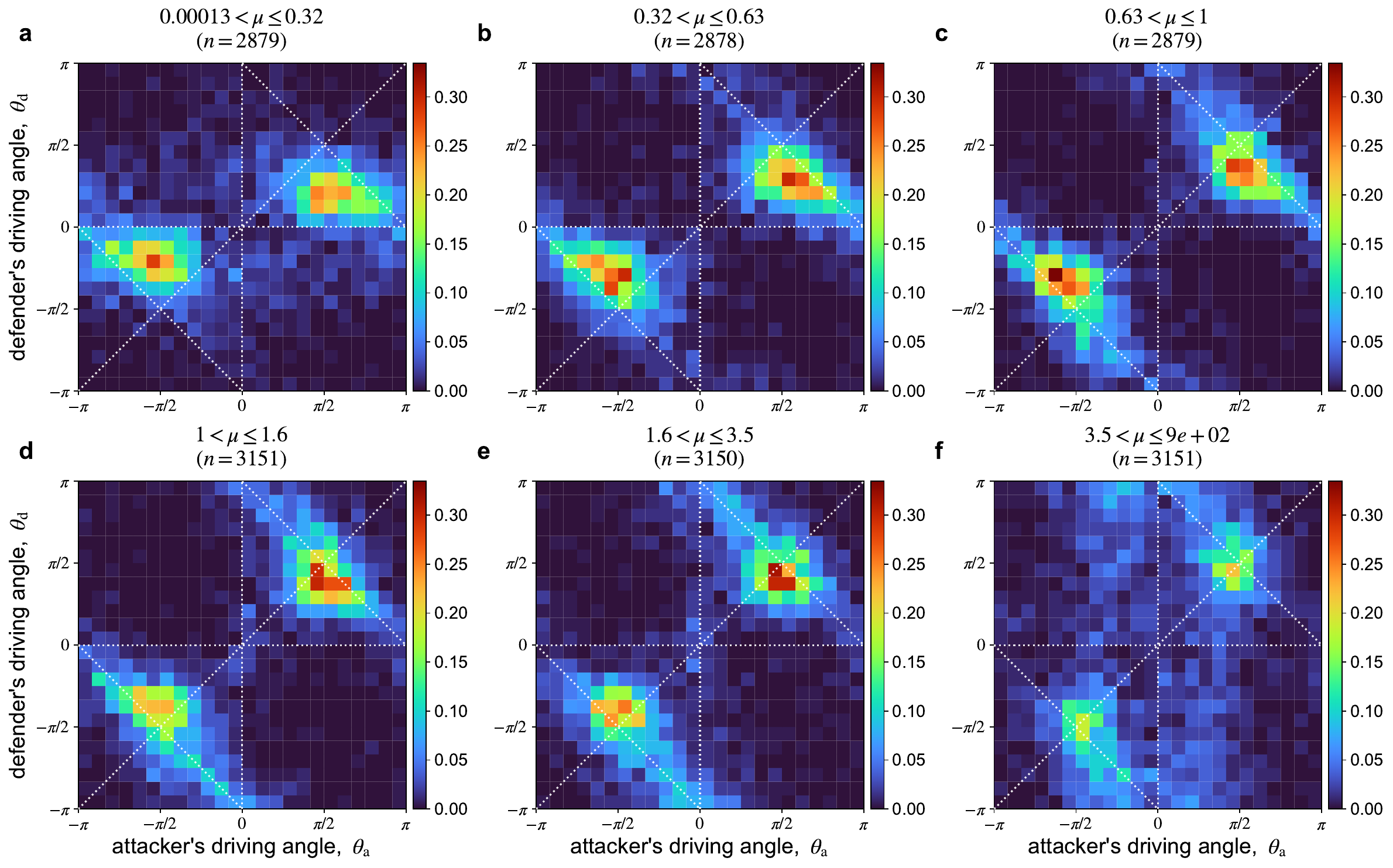}
    \exfigcaption[\linewidth]{
    \pnl{a--f,} 
    Joint distributions of angle pairs for the six data groups partitioned by \( \mu \). 
    These panels correspond to those on the right side of Fig.~\ref{fig:peak_motion}. 
    Color scale denotes joint probability density and is fixed across all panels. 
    }
    \label{fig:ex:angle_joint_by_mu_bins}
\end{figure*}

\begin{figure*}
    \centering
    \includegraphics[width=0.5\linewidth]{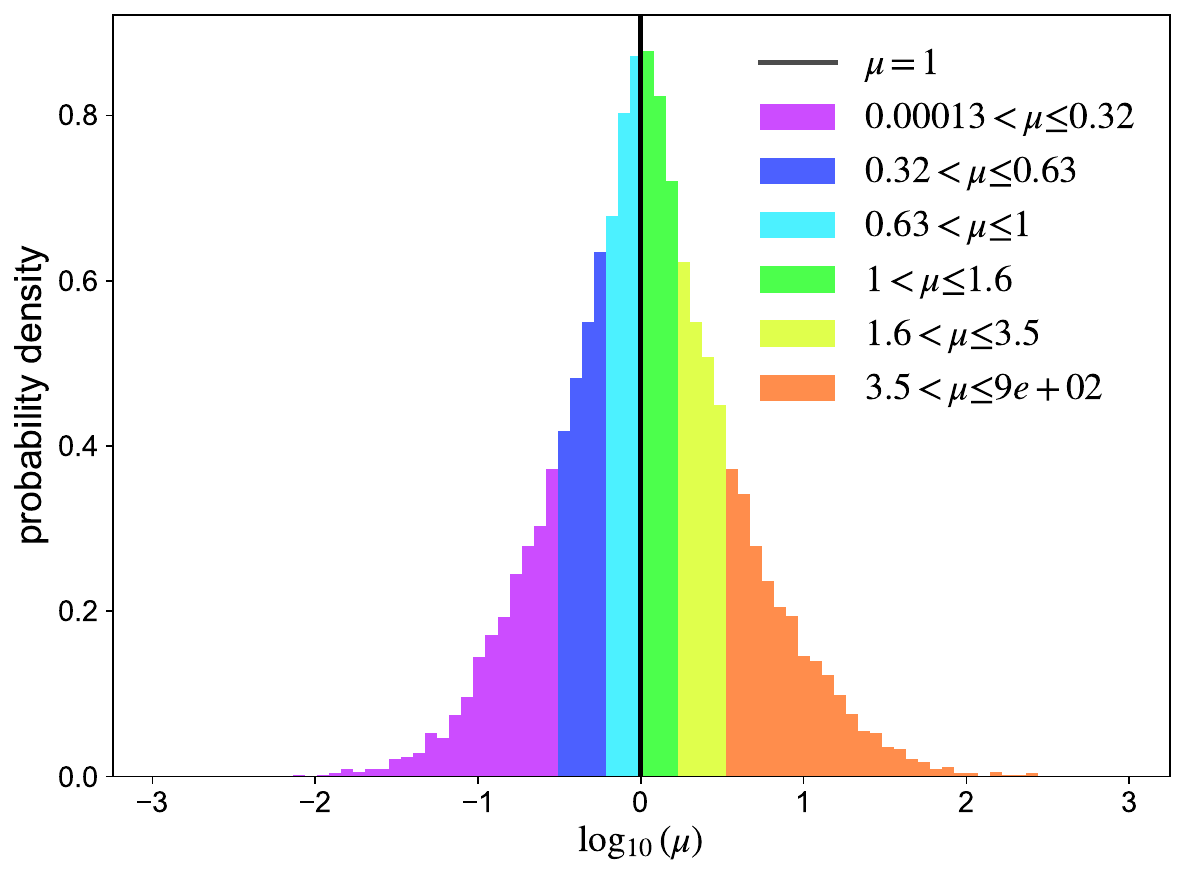}
    \exfigcaption[\linewidth]{
    Probability distribution of log-rescaled \( \mu \). 
    Data groups correspond to those used for kernel density estimation; colors match those in the right panels of Fig.~\ref{fig:peak_motion}. 
    The symmetric case, \( \mu = 1 \), is indicated by the black solid line. 
    }
    \label{fig:ex:log_mu_histogram}
\end{figure*}

\begin{figure*}
    \centering
    \includegraphics[width=\linewidth]{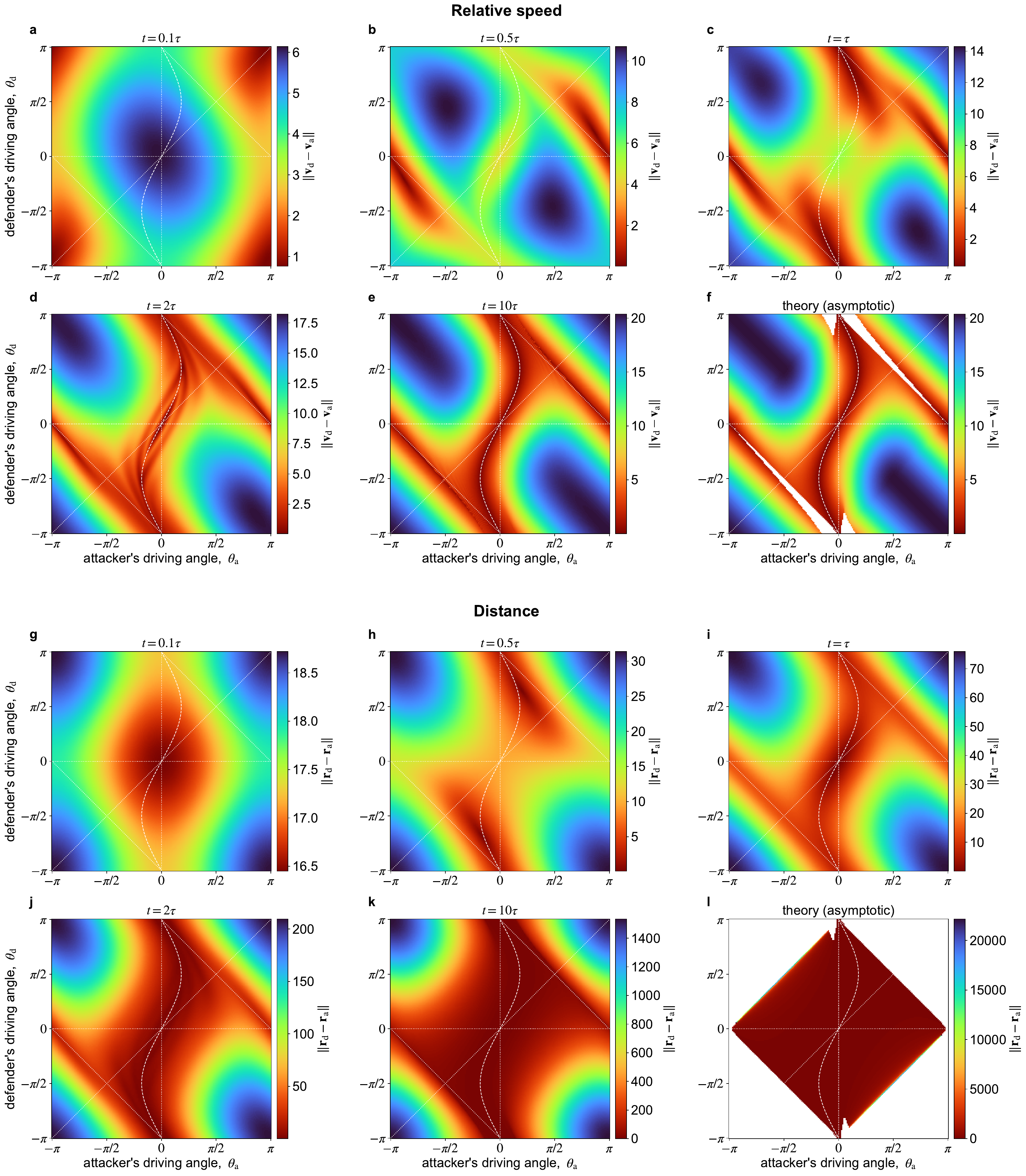}
    \exfigcaption[\linewidth]{
    Timeseries data of 
    (\pnl{a--e}) relative speed 
    and 
    (\pnl{g--k}) attacker--defender distance across angle space, computed from ODE simulations. 
    \pnl{f,l,} Corresponding asymptotic values. 
    Plotting details are the same as in \AppFig{fig:ex:timeseries_phase_diagram_1}, with parameters and initial conditions taken from a different event. 
    Parameters and initial conditions: \( \fa = 2.29 \), \( \fd = 1.23 \), \( \ta = -0.71 \), \( \td = -1.41 \), \( \taua = 4.46 \), \( \taud = 8.24 \), \( \ra(0) = (62.98, 36.18) \), \( \rd(0) = (82.59, 30.16) \), \( \vat(0) = (-0.75, 1.85) \), \( \vdf(0) = (-4.43, 2.47) \). 
    These values were taken or estimated from a real one-on-one event in the dataset. 
    See \SIFigSec for more examples. 
    }
    \label{fig:ex:timeseries_phase_diagram_2}
\end{figure*}

\end{document}


\title{Supplementary Information for  \\
A behavioral principle underlying attacker--defender interactions in soccer}

\author{Issei Yamazaki}
\thanks{Contributed equally with H.G.}
\affiliation{Graduate School of Advanced Mathematical Sciences, Meiji University, Nakano, Tokyo 164-8525, Japan}

\author{Hirotaka Goto}
\thanks{Contributed equally with I.Y.; corresponding author}
\affiliation{Graduate School of Advanced Mathematical Sciences, Meiji University, Nakano, Tokyo 164-8525, Japan}
\affiliation{Department of Biology, University of Pennsylvania, Philadelphia, PA 19104, United States}

\author{Kojiro Otoguro}
\affiliation{Meiji Institute for Advanced Study of Mathematical Sciences, Meiji University, Nakano, Tokyo 164-8525, Japan}

\author{Hiraku Nishimori}
\affiliation{Meiji Institute for Advanced Study of Mathematical Sciences, Meiji University, Nakano, Tokyo 164-8525, Japan}

\author{Masashi Shiraishi}
\affiliation{Meiji Institute for Advanced Study of Mathematical Sciences, Meiji University, Nakano, Tokyo 164-8525, Japan}
\affiliation{Graduate School of Information Sciences, Hiroshima City University, 3-4-1 Ozukahigashi, Asaminami-ku, Hiroshima 731-3194, Japan}

\author{Takuma Narizuka}
\email[Corresponding author ]{}
\affiliation{Faculty of Data Science, Rissho University, Kumagaya, Saitama 360-0194, Japan}

\maketitle

\tableofcontents

\newpage

\section{\label{sec:methods} Data Processing Methods and Numerical Implementation}

\subsection{\label{subsec:preprocessing} Preprocessing: identifying ball-possession and one-on-one events}

We use player-tracking and event data from the Japan Professional Football League, provided by DataStadium~Inc. 
Tracking data provide two-dimensional positions of all players on the field, obtained from video recordings at 25 frames per second (fps) and using optical tracking. 
Player positions were estimated automatically from the footage in a two-dimensional coordinate system at centimeter resolution. 
The data were corrected manually by the company when tracking failed, for example when players overlapped in the recordings. 
Event data includes action types (Table~\ref{tab:event_list}), associated player IDs, and timestamps sampled at 30 fps. 
We first round the annotated timestamps in the event data to discrete times sampled at 25 fps, so that both tracking and event data align in time. 
Subsequently, by limiting our analysis to the sequences of two consecutive actions listed in Table~\ref{tab:event_list}, we identify \emph{ball-possession events}.  
A ball-possession event refers to the continuous time interval during which a player remains in possession of the ball (also defined in the main text). 
Table~\ref{tab:event_list} provides additional methodological details on the identification of ball-possession events. 
In addition, events with missing data are excluded; positional data must be complete throughout each event, and sufficient positional data beyond the event duration must also be available for estimating player velocities at the event onset using the method described in Sec.~\ref{subsec:velocity_estimation}.

\begin{table}[b]
    \centering
    \begin{tabular*}{\columnwidth}{@{\extracolsep{\fill}}ll@{\qquad}ll}
        \toprule
        Start event & Description & End event & Description \\
        \midrule
        \texttt{trap}      & Bringing the ball under control upon receiving it  
                  & \texttt{home\_pass} & Pass by a player on the home team \\

        \texttt{dribble}   & Moving the ball while maintaining possession 
                  & \texttt{away\_pass} & Pass by a player on the away team \\

        \texttt{tackle}    & Attempting to dispossess an opposing player 
                  & \texttt{cross}      & Cross \\

        \texttt{block}     & Blocking an opponent's pass or shot  
                  & \texttt{through}    & Through ball \\

        \texttt{intercept} & Intercepting an opponent's pass 
                  & \texttt{shoot}      & Shot (an attempt to score a goal) \\

                   & 
                  & \texttt{clear}      & Removing the ball from an area of danger \\
        \bottomrule
    \end{tabular*}
    \caption{
    Taxonomy of action types used to identify ball-possession events. 
    The short descriptions are provided by the authors, as they are not included in the dataset. 
    Any sequential combinations of the action types listed (start and end events) define the continuous time intervals corresponding to ball-possession events. 
    The dataset also includes other action types, including additional on-ball actions (e.g., shots), ball-out-of-play events (e.g., throw-ins), and offenses or stoppages during ball-in-play periods (e.g., offsides). 
    However, because these events are not relevant to the present study, an exhaustive list of action types is not presented here. 
    When several actions occur simultaneously, they are ordered according to the sequence recorded in the dataset. 
    }
    \label{tab:event_list}
\end{table}

To isolate one-on-one dribbling events among ball-possession events, we impose five additional conditions (also described in \Met), independently of order. 
Specifically: 
\begin{enumerate}
    \item The event duration exceeds \SI{0.5}{\second}. \label{cond:duration}
    \item The attacker and defender both have net displacements exceeding \SI{5}{\meter} between the start and end of the event. \label{cond:net_displacement}
    \item The attacker is not a goalkeeper. \label{cond:no_goalkeeper}
    \item The angle between \( \rr = \rd - \ra \) and the vector from the attacker's position to the opponent's goal is less than \( 60^\circ \) at the start of the event.  \label{cond:angle}
    \item The nearest opposing player remains unchanged throughout the event. \label{cond:same_defender}
\end{enumerate}
These criteria are motivated as follows: 
interactions between the attacker and defender should occur over sufficiently large temporal and spatial scales so that trajectory fitting is feasible and meaningful (Conditions~\ref{cond:duration} and \ref{cond:net_displacement}). 
Goalkeepers may behave differently and therefore are not included in our analysis (Condition~\ref{cond:no_goalkeeper}). 
The defender lies approximately between the attacker and the defender's own goal at the start of the event, so that the event reflects a direct attacker--defender interaction (Condition~\ref{cond:angle}). 
We restrict our analysis to one-on-one interactions between nearest opposing players with fixed identities throughout the event (Condition~\ref{cond:same_defender}). 
Note that a net displacement is different from the distance traveled by a player (see Condition~\ref{cond:net_displacement}). 
We use smoothed positional data to identify one-on-one events (see \ref{subsec:velocity_estimation}). 
For readers interested in additional details, Tables~\ref{tab:ball_possession_events} and \ref{tab:1v1_breakdown} provide a detailed breakdown of ball-possession and one-on-one events based on the sequence of two consecutive recorded actions.

\begin{figure}[b]
\centering
\begin{tikzpicture}[
    node distance=0.25in,
    >=Latex,
    every node/.style={
        draw,
        align=left,
        rounded corners,
        minimum height=0.5in,
        text width=1.55in,
        font=\small, 
        inner sep=8pt,
    }
]

\node (filter) [text width=1.8in] {
\textbf{Preprocessing tracking data} \\
\textbullet\ smooth positional data \\
\textbullet\ estimate velocities \\
};

\node (pre) [right=of filter, text width=1.2in] {
\textbf{Preprocessing event data} \\
\textbullet\ adjust frame rate \\
\textbullet\ no missing data \\
\textbullet\ isolate one-on-ones \\
};

\node (fit) [right=of pre, text width=1.3in] {
\textbf{Model fitting} \\
\textbullet\ optimize parameters 
};

\node (thresh) [right=of fit, text width=1.1in] {
\textbf{Thresholding} \\
\textbullet\ \(\max( \epsa^*, \epsd^*) < 0.1 \)
};

\draw[->, thick] (filter) -- (pre);
\draw[->, thick] (pre) -- (fit);
\draw[->, thick] (fit) -- (thresh);

\end{tikzpicture}

\caption{
Overview of preprocessing, parameter estimation, and event selection. 
Only selected events are included in the empirical analysis. 
}
\label{fig:pipeline}
\end{figure}
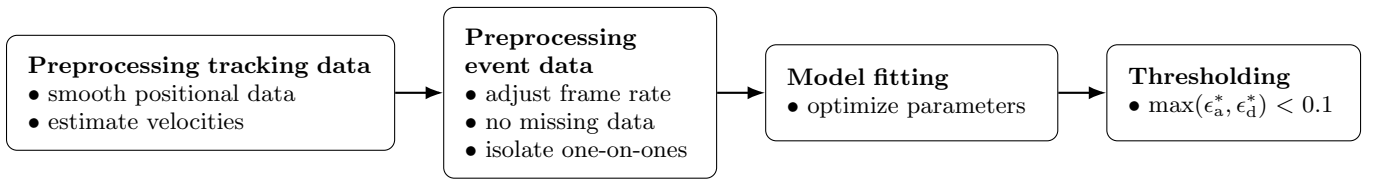

\subsection{\label{subsec:velocity_estimation} Estimating player velocities}

We now estimate player velocities from the tracking data in one-on-one events, following a previous study \cite{brink2023measuring}. 
First, we smooth the positional data by applying the Savitzky--Golay (SG) filter to each coordinate independently. 
We use a polynomial of order 2 and a 21-point window, yielding the smoothed positional data \( (\hat x, \hat y) \). 
We fit a cubic spline to each coordinate of \( (\hat x, \hat y) \), obtaining continuous position functions. 
We differentiate them with respect to time and evaluate them at discrete times \( t_n \), obtaining \( (v_x(t_n), v_y(t_n)) \). 
We apply the SG filter again to \( (v_x(t_n), v_y(t_n)) \), using the same polynomial order and window size. 
We finally obtain the smoothed velocities \( (\hat v_x(t_n), \hat v_y(t_n)) \). 
In our study, only the player velocity at the event onset, \( \vp(0) \), is required to fit the model to empirical trajectories. 
This nonetheless requires positional data extending beyond event duration, tying back to the earlier discussion of events with missing data in Sec.~\ref{subsec:preprocessing}.

\subsection{\label{subsec:parameter_estimation} Estimating parameters}

Next, we fit our model to each empirical event (see the main text for the full description of the model). 
Let 
\begin{align} \label{eq:transformation}
    \Aa \coloneqq \fa \cos \ta, \quad 
    \Ba \coloneqq \fa \sin \ta, \quad 
    \Ad \coloneqq \fd \cos \td, \quad 
    \Bd \coloneqq \fd \sin \td. 
\end{align}
The model then becomes 
\begin{subequations} \label{eq:oad}
    \begin{align} 
        \dvatdt &= - \frac{\vat}{\taua} + \Aa \eo(\ra, \rd) + \Ba \et(\ra, \rd), \label{eq:oad:at} \\ 
        \dvdfdt &= - \frac{\vdf}{\taud} - \Ad \eo(\ra, \rd) + \Bd \et(\ra, \rd). \label{eq:oad:df}
    \end{align}
\end{subequations} 
In addition, we restrict model parameters to the ranges  
\begin{align}
    0.2 \le \taup, \quad
    0 \le \fp\taup \le 10.2.
\end{align}
The first condition ensures that the relaxation rates \( 1/\taup \) remain bounded. 
The second condition sets an upper bound on the steady-state speed \cite{brink2023measuring}. 
We reparameterized \( \chip \coloneqq (\taup, \fp, \tp) \) as \( \xip \coloneqq (\taup, \Ap, \Bp) \) to avoid singular behavior near the origin in a polar representation.
The inverse transformation is given by 
\begin{align} \label{eq:inverse_transformation} 
    \fp = \sqrt{\Ap^2 + \Bp^2}, \qquad
    \tp = 
    \begin{cases}
        0 & \sqrt{\Ap^2 + \Bp^2} < 10^{-12},\\
        \operatorname{Arg}(\Ap + i\Bp) & \text{otherwise}. \\
    \end{cases}
\end{align}

We fit the model trajectory, \( \rp(t_n) \), to the empirical trajectory, \( \rtp(t_n) \), while fixing the trajectory of the opposing player \( \q \neq p \) to its empirical counterpart, 
namely, \( \rqq(t_n) \equiv \rtq(t_n) \) for \( n \in \{0, 1, \ldots, T-1 \} \), where \( T \) denotes the number of frames in the event. 
For each player \( \p \in \{ \at, \df \} \) in a given one-on-one event, we estimate parameter values by solving the following optimization problem: 
\begin{align}
    \label{eq:optimization}
    \minimize_{\xip \in \Xi} \epsp \left ( \{\rp(t_n; \xip), \rtq(t_n) \}_{n=0, \ldots, T-1} \right), 
\end{align}
where 
\begin{align}
    \Xi \coloneqq \{(\taup, \Ap, \Bp) \in \mathbb{R}^3 \;|\; 0.2 \le \taup, \,
    0 \le \taup \sqrt{\Ap^2 + \Bp^2} \le 10.2 \}. 
\end{align}
The error function \( \epsp \) measures the mean normalized distance between model and empirical trajectories; its precise definition is given in the main text. 
The parameter estimates are given by \( \xip^* \coloneqq \argmin_{\xip \in \Xi} \epsp(\xip) \). 
Equivalently, we obtain \( \chip^* \) via the transformation defined in Eq.~\ref{eq:inverse_transformation}. 
Let \( \epsp^* = \min_{\xip \in \Xi} \epsp \) denote the attained minimum. 
We retain events in which the discrepancy between model and empirical trajectories is below a threshold for both players---that is, those satisfying 
\begin{align}
    \max(\epsa^*, \epsd^*) < 0.1. \label{eq:error_threshold}
\end{align}

The filtering process is summarized in Fig.~\ref{fig:pipeline}. 
The number of retained samples after each filtering is reported in Table~\ref{tab:filtering_summary}.

\subsection{\label{subsec:numerical_implementation} Numerical implementation}

Here we describe the numerical implementation of the methods presented in previous sections. 
We solve the equation of motion (Eq.~\ref{eq:oad}) using an adaptive Runge--Kutta 4(5) method (\texttt{RK45}) implemented in \texttt{SciPy}'s \texttt{solve\_ivp} solver to obtain model trajectories. 
Note that \texttt{LSODA} is used for obtaining phase diagrams to maintain robustness against potential stiffness. 
The opposing player's empirical trajectory (the smoothed data \( (\hat x, \hat y) \) obtained in Sec.~\ref{subsec:velocity_estimation}) is interpolated linearly. 
To avoid numerical instability, we use \( \mathbf{e}_{1, \varepsilon}(\ra, \rd) \) instead of \( \eo(\ra, \rd) \), defined by 
\begin{align}
    \mathbf{e}_{1, \varepsilon}(\ra, \rd) \coloneqq 
    \begin{dcases}
        \frac{\rr}{\|\rr\|} &\quad \| \rr \| \geq \varepsilon, \\
        \frac{\rr}{\varepsilon} &\quad \| \rr \| < \varepsilon,
    \end{dcases}
\end{align}
where \( \rr = \rd - \ra \), 
approximating the unit vector \( \eo \) when relative position, \( \rr \), falls within the \( \varepsilon \)-ball (\( \varepsilon = 10^{-9} \) in our simulations). 
Likewise, we replace \( \et(\ra, \rd) \) by \( \mathbf{e}_{2,\varepsilon}(\rr) \), obtained by rotating \( \mathbf{e}_{1,\varepsilon}(\rr) \) \( 90^\circ \) clockwise. 
Initial conditions are given by empirical data for both players, i.e., \( \rp(0) = \rtp(0) \) and \( \vp(0) = \vtp(0) \). 
We implement the optimization defined in Eq.~\ref{eq:optimization} using the \texttt{minimize} function in the \texttt{scipy.optimize} package. 
To minimize the multivariate objective function \( \epsp \), we choose the \texttt{COBYLA} method, which implements the Constrained Optimization BY Linear Approximation (COBYLA) algorithm \cite{gomez2013advances}.

To improve robustness against local minima, we sample 500 initial guesses \( \{ \xi_{\p,1}, \ldots, \xi_{\p,500} \} \) from a constrained parameter space 
\begin{align}
    \Xi_{\text{init}} \coloneqq \{\xip \in \mathbb{R}^3 \;|\; 0.2 \le \taup \le 300, \, 0 \le \taup \sqrt{\Ap^2 + \Bp^2} \le 10.2 \},
\end{align}
using Latin hypercube sampling (LHS); 
see Sec.~\ref{subsec:python_implementation} for more details on the sampling method and implementation in \textsc{Python}. 
The sampling space \( \Xi_{\text{init}} \) is illustrated in Figure~\ref{fig:sampling_initial_parameters}a. 
LHS is a stratified sampling method that provides more uniform coverage of a given parameter space with fewer samples than traditional methods such as random sampling. 
Starting from these initial guesses, \texttt{COBYLA} yields corresponding local minimizers \( \{ \xi_{\p,1}^*, \ldots, \xi_{\p,500}^* \} \). 
We then select the minimizer with the smallest objective value among them. 
That is, \( \xip^* = \xi_{{\p}, i^*}^* \), where \( i^* = \argmin_{i=1,\ldots,500} \epsp(\xi_{\p, i}^*) \). 
Optimization is terminated according to convergence and stopping criteria, including the prescribed final accuracy and the maximum number of iterations. 
Optimization runs for which \texttt{COBYLA} does not report successful convergence are discarded.

\begin{table}[t]
    \centering
    \begin{tabular*}{\columnwidth}{@{\extracolsep{\fill}}clcc}
        \toprule
        Filtering step & Events & Events retained \\
        \midrule 
        Action sequences listed in Table~\ref{tab:event_list} & Ball-possession events & \( \totballpossession \) \\
        One-on-one filters (Conditions~\ref{cond:duration}--\ref{cond:same_defender}) & One-on-one events & \( \totunfiltered \) \\
        Error thresholding (Eq.~\ref{eq:error_threshold}) & One-on-one events satisfying \( \max(\epsa^*,\epsd^*) < 0.1 \) & \( \totfiltered \) \\
        \bottomrule
    \end{tabular*}
    \caption{
    Number and fraction of remaining events after each filtering step. 
    We retained approximately \( \percentageremaining \% \) of the one-on-one events after applying the error threshold (\( \totfiltered / \totunfiltered \approx \fracremaining \)). 
    }
    \label{tab:filtering_summary}
\end{table}

\subsection{\label{subsec:python_implementation} \textsc{Python} Implementation}

The \texttt{COBYLA} method implemented in the \textsc{Python} package allows only a single optimization parameter, \texttt{rhobeg}, to control reasonable initial changes to the variables. 
Due to differences in scale across parameters and to minimize potential numerical issues associated with nonlinear constraints, we introduce 
\begin{align}
    \Up \coloneqq \Ap \taup, \quad 
    \Wp \coloneqq \Bp \taup, 
\end{align}
and rescale the parameters by 
\begin{align}
    \taup' \coloneqq \frac{\taup - \taumin}{\taumax - \taumin}, \quad
    \Up' \coloneqq \frac{\Up}{\alphamax}, \quad 
    \Wp' \coloneqq \frac{\Wp}{\alphamax}, 
\end{align}
so that the sampling space for LHS is given by  
\begin{align}
    0 \leq \taup' \leq 1, \quad 
    0 \le (\Up')^2 + (\Wp')^2 \le 1,
\end{align} 
which we denote by \( \Xi_{\text{init}}' \), and we set \( \taumin = 0.2 \), \( \taumax = 300 \), and \( \alphamax = 10.2 \). 
Furthermore, to search the vast range of \( \taup \in [0.2, 300] \) efficiently, we optimize \( \gammap' \coloneqq \ln \taup' \). 
To avoid divergence, we introduce \( \taumin^{\text{init}} = \taumin + \delta \), where \( \delta = 10^{-6} \), so that the sampling space for LHS becomes 
\begin{align}
    \frac{\taumin^{\text{init}} - \taumin }{\taumax - \taumin} \leq \taup' \leq 1, \quad 
    0 \le (\Up')^2 + (\Wp')^2 \le 1. 
\end{align}
We note that the constraints for optimization are given by 
\begin{align}
    \gammap' \leq 0, \quad 
    0 \le (\Up')^2 + (\Wp')^2 \le 1. 
\end{align}
To summarize, we reparameterize \( (\Ap, \Bp, \taup) \) as \( (\Up', \Wp', \taup') \) upon initialization to eliminate the nonlinear parameter constraints and optimize \( (\Up', \Wp', \gammap') \) for search efficiency.

\begin{figure}[t]
    \centering
    \includegraphics[width=\linewidth]{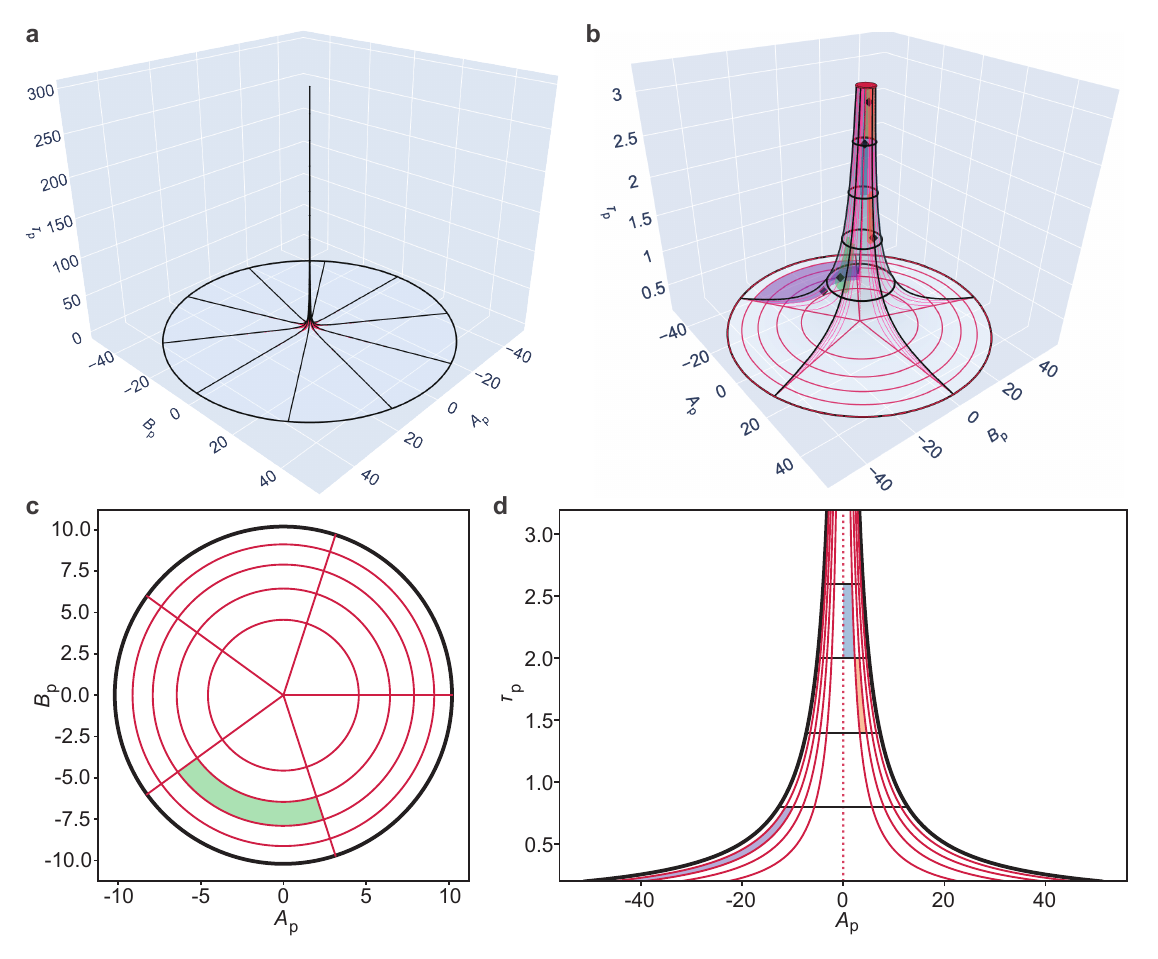}
    \caption{
    \pnl{a,} The parameter space \( \Xi_{\text{init}} \), in which 500 initial samples of \( (\Ap, \Bp, \taup) \) are generated (Sec.~\ref{subsec:python_implementation}).
    The spatial partitioning is much coarser than in the actual implementation for illustrative purposes (ten divisions for each parameter).
    \pnl{b--d,} Schematic illustration of the sampling method. 
    For illustration purposes, the upper bound of \( \taup \) is substantially lowered, and the radial and angular partitioning used for LHS is much coarser than in the actual implementation (\( \tau_{\min} = 0.2 \), \( \tau_{\max} = 3.198 \), and five samples). 
    The partitioning along the \( \taup \)-axis is shown using the same spacing rule as in the actual implementation.
    Across all panels, red and black lines denote the radial, angular, and \( \taup \)-axis partitions used in LHS. 
    The initial parameter sets are sampled from the partitioned strata such that every partition interval along each coordinate is sampled exactly once when projected onto that coordinate. 
    The sets of parameters sampled from colored subdivisions in (\pnl{b}) are indicated by black diamonds. 
    \pnl{c,} Cross-sectional view of the partitioning at \( \taup = 1 \). 
    \pnl{d,} Cross-sectional view of the partitioning at \( \Bp = 0 \). 
    }
    \label{fig:sampling_initial_parameters}
\end{figure}

Finally, we describe how the 500 initial guesses are sampled from the constrained parameter space \( \Xi_{\text{init}} \). 
We first generate 500 values of \( \taup \) in \( [\tau_{\min}, \tau_{\max}] \). 
To do so, we generate 500 samples of \( v_{\p} \in [0, 1] \) 
using LHS (implemented using \texttt{lhs} in \texttt{pyDOE}), randomly shuffle the samples, and transform them via \( \taup = (\tau_{\max} - \tau_{\min}) v_{\p} + \tau_{\min} \). 
Subsequently, we sample \( (\Ap, \Bp) \) from a \( \taup \)-dependent disk \( \{(\Ap, \Bp) \in \mathbb{R}^2 \,|\, \Ap^2 + \Bp^2 \leq (\alphamax/\taup)^2 \} \). 
To do so, we generate 500 samples each of \( u_{\p}, w_{\p} \in [0, 1] \) using LHS, randomly shuffle the two sample sets independently, and transform them via \( \fp = \alphamax\sqrt{u_{\p}} / \taup\) and \( \tp = 2\pi w_{\p} \), thereby ensuring uniform sampling over the disk. 
Finally, we transform \( (\fp, \tp) \) to \( (\Ap, \Bp) \) via Eq.~\ref{eq:transformation}. 
This stratified sampling method is summarized in Figs.~\ref{fig:sampling_initial_parameters}b--d.

\clearpage

\begin{table}[t]
  \centering
    \setlength{\tabcolsep}{3pt}
    \begin{tabular*}{\columnwidth}{@{\extracolsep{\fill}}c c cccccc c}
      \toprule
      \multicolumn{2}{c }{} & \multicolumn{6}{c }{End event} & \multirow{2}{*}{Total} \\
      \multicolumn{2}{c }{} 
      & home\_pass & away\_pass & cross & through & shoot & clear & \\
      \midrule
      \multirow{5}{*}{\rotatebox[origin=c]{90}{Start event}}
      & trap      
      & 78705 (44.97\%) & 76907 (43.94\%) & 4175 (2.39\%) & 5446 (3.11\%) & 2223 (1.27\%) & 511 (0.29\%) & 167967 (95.97\%) \\
      & dribble   
      & 168 (0.10\%) & 149 (0.09\%) & 2016 (1.15\%) & 247 (0.14\%) & 879 (0.50\%) & 1 (0.00\%) & 3460 (1.98\%) \\
      & tackle    
      & 908 (0.52\%) & 871 (0.50\%) & 44 (0.03\%) & 79 (0.05\%) & 31 (0.02\%) & 105 (0.06\%) & 2038 (1.16\%) \\
      & block     
      & 410 (0.23\%) & 384 (0.22\%) & 24 (0.01\%) & 35 (0.02\%) & 23 (0.01\%) & 112 (0.06\%) & 988 (0.56\%) \\
      & intercept 
      & 237 (0.14\%) & 251 (0.14\%) & 16 (0.01\%) & 43 (0.02\%) & 14 (0.01\%) & 4 (0.00\%) & 565 (0.32\%) \\
      \midrule
      \multicolumn{2}{c }{Total}
      & 80428 (45.95\%) & 78562 (44.89\%) & 6275 (3.59\%) & 5850 (3.34\%) & 3170 (1.81\%) & 733 (0.42\%) & 175018 (100\%) \\
      \bottomrule
    \end{tabular*}
    \caption{
    Breakdown of ball-possession events, identified based on the pairs of actions listed in Table~\ref{tab:event_list}. 
    Entries show the number of events, with percentages relative to the total number of ball-possession events (\( N = \totballpossession \)) in parentheses. 
    Percentages are rounded to two decimal places.
    }
    \label{tab:ball_possession_events}
\end{table}

\begin{table}[b]
  \centering
    \setlength{\tabcolsep}{3pt}
    \begin{tabular*}{\columnwidth}{@{\extracolsep{\fill}}c c cccccc c}
      \toprule
      \multicolumn{2}{c }{} & \multicolumn{6}{c }{End event} & \multirow{2}{*}{Total} \\
      \multicolumn{2}{c }{} 
      & home\_pass & away\_pass & cross & through & shoot & clear & \\
      \midrule
      \multirow{5}{*}{\rotatebox[origin=c]{90}{Start event}}
      & trap      
      & 8361 (43.48\%) & 7983 (41.51\%) & 670 (3.48\%) & 526 (2.74\%) & 140 (0.73\%) & 79 (0.41\%) & 17759 (92.35\%) \\
      & dribble   
      & 33 (0.17\%) & 27 (0.14\%) & 933 (4.85\%) & 54 (0.28\%) & 191 (0.99\%) & 0 (0\%) & 1238 (6.44\%) \\
      & tackle    
      & 65 (0.34\%) & 42 (0.22\%) & 6 (0.03\%) & 11 (0.06\%) & 2 (0.01\%) & 12 (0.06\%) & 138 (0.72\%) \\
      & block     
      & 36 (0.19\%) & 32 (0.17\%) & 4 (0.02\%) & 5 (0.03\%) & 4 (0.02\%) & 5 (0.03\%) & 86 (0.45\%) \\
      & intercept 
      & 3 (0.02\%) & 6 (0.03\%) & 0 (0\%) & 1 (0.01\%) & 0 (0\%) & 0 (0\%) & 10 (0.05\%) \\
      \midrule
      \multicolumn{2}{c }{Total}
      & 8498 (44.19\%) & 8090 (42.07\%) & 1613 (8.39\%) & 597 (3.10\%) & 337 (1.75\%) & 96 (0.50\%) & 19231 (100\%) \\
      \bottomrule
    \end{tabular*}
    \caption{
    Breakdown of one-on-one events, obtained by applying the five conditions introduced in Sec.~\ref{subsec:preprocessing} to the ball-possession events listed in Table~\ref{tab:ball_possession_events}. 
    Entries show the number of events, with percentages relative to the total number of one-on-one events (\( N = \totunfiltered \)) in parentheses. 
    Percentages are rounded to two decimal places. 
    The error threshold is not applied. 
    }
    \label{tab:1v1_breakdown}
\end{table}

\clearpage
\newpage

\section{Empirical Data Analysis}

\subsection{Estimated parameters and error distributions}

\begin{figure}[p]
    \centering
    \includegraphics[width=\linewidth]{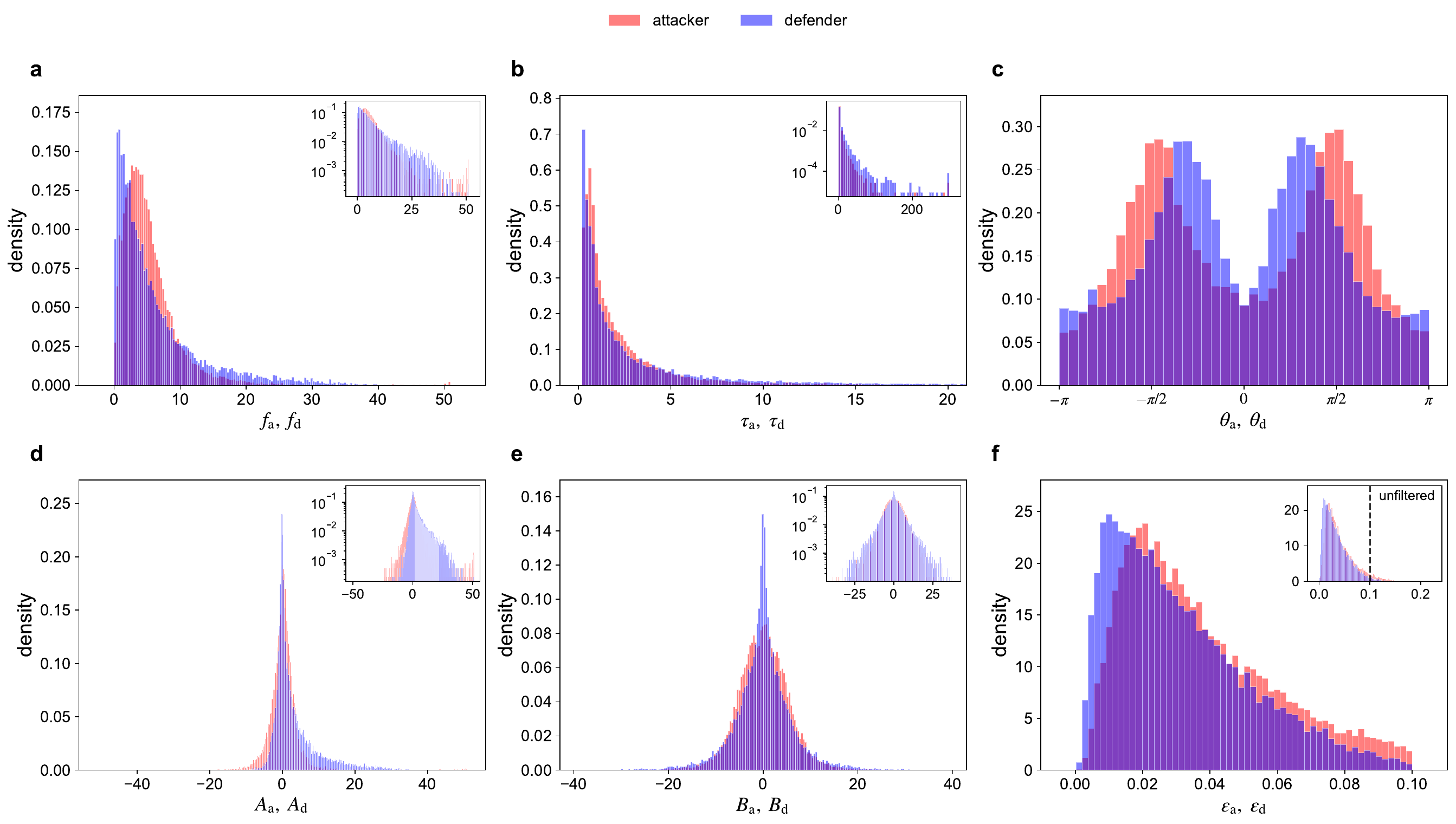}
    \caption{
    Probability densities of (\textbf{a--c}) estimated parameters, (\textbf{d,e}) fitted coefficients, and (\textbf{f}) resulting errors. 
    Insets in all panels but (\textbf{f}) are shown on a semi-logarithmic scale. 
    The inset in (\textbf{f}) shows the distribution of errors from all unfiltered one-on-one events (\totunfiltered events) before applying the error threshold; the vertical black dashed line indicates the threshold value (0.1), which applies jointly to \( \epsa \) and \( \epsd \). 
    The horizontal axis in the main panel of (\pnl{b}) is truncated at \( \taup = 20 \) due to its vast scale. 
    Non-periodic linear angular distributions are shown in (\textbf{c}); periodic circular representations appear in the main text. 
    Data from \totfiltered events (filtered one-on-ones). 
    }
    \label{fig:parameter_distributons}
\end{figure}

\begin{figure}[b]
    \centering
    \includegraphics[width=\linewidth]{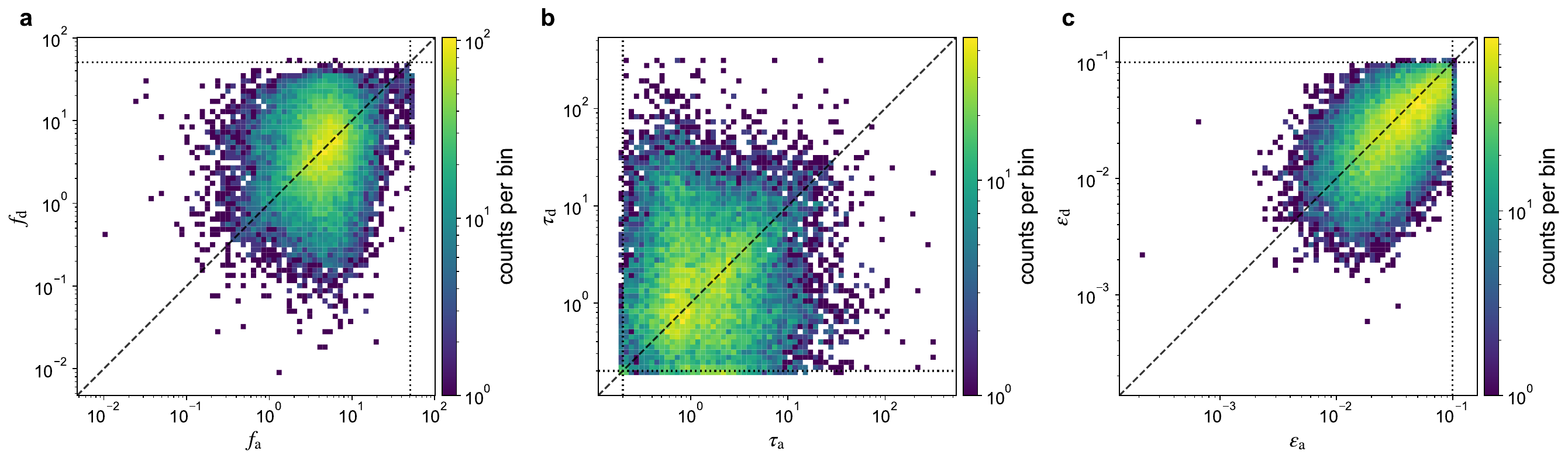}
    \caption{
    \textbf{a--c,} Two-dimensional histograms of the estimated parameters shown in Fig.~\ref{fig:parameter_distributons}, representing the joint distributions of attacker and defender parameter values for (\textbf{a}) driving-force magnitude \( \fp \), (\textbf{b}) relaxation time \( \taup \), and (\textbf{c}) trajectory error \( \epsp \). 
    Shared logarithmic bins are used on both axes; color indicates counts per bin on a logarithmic scale. 
    Dashed diagonal lines denote equality between attacker and defender parameter values. 
    Dotted lines indicate imposed or intrinsic parameter bounds: (\textbf{a}) \( \fp \le 51 \), (\textbf{b}) \( 0.2 \le \taup \), and (\textbf{c}) \( \epsp < 0.1 \). 
    Data from \totfiltered events. 
    }
    \label{fig:parameter_2d_distributions}
\end{figure}

Figure~\ref{fig:parameter_distributons} shows the distributions of estimated and fitted parameters for the filtered \totfiltered events (the error threshold applied), whereas Fig.~\ref{fig:parameter_2d_distributions} shows the joint distributions of attacker and defender parameter values. 
Figure~\ref{fig:parameter_distributons}f indicates that the attacker's trajectory error tends to exceed that of the defender. 
We speculate that this is because the attacker carries the ball; 
the model does not account for ball control, which could plausibly introduce additional variability in the attacker trajectory relative to model predictions.

\begin{figure}[p]
    \centering
    \includegraphics[width=\linewidth]{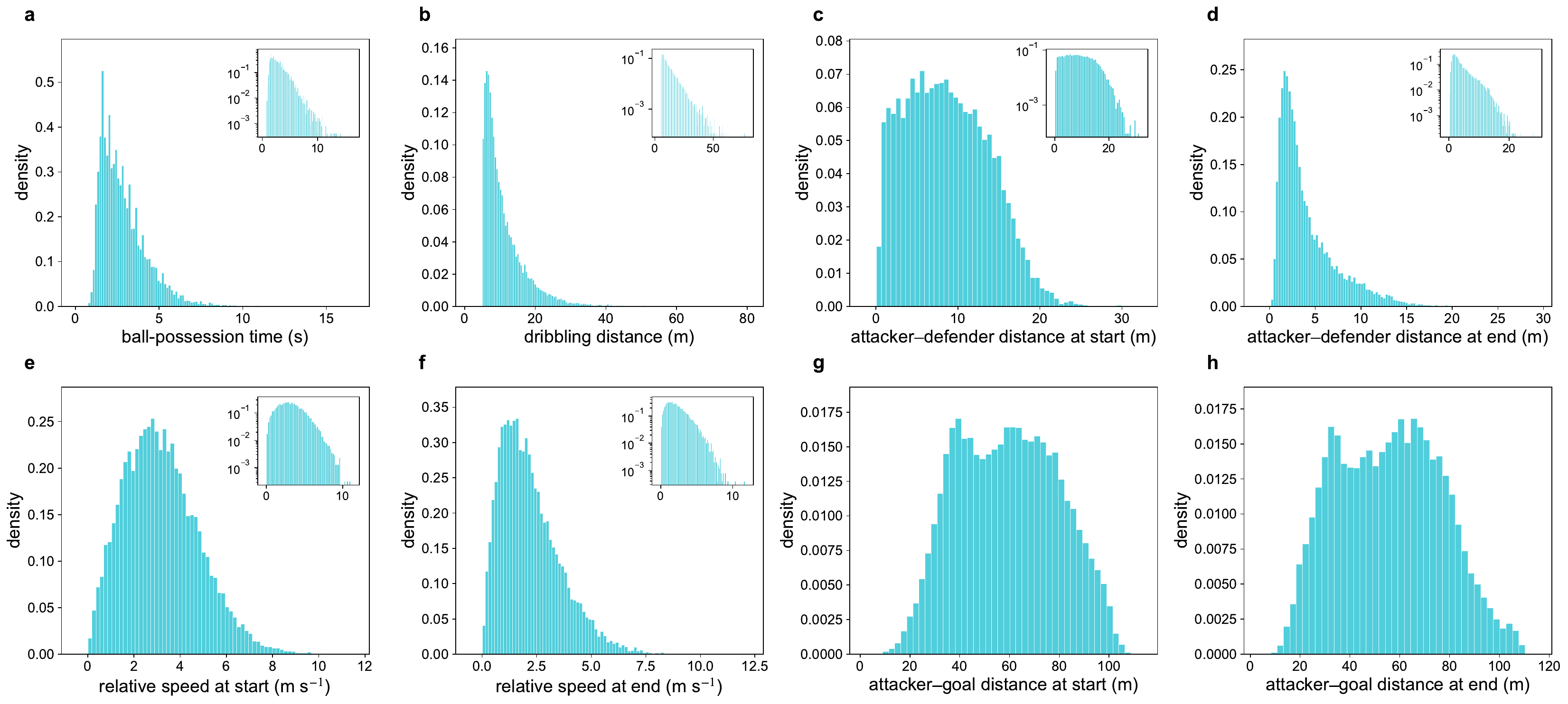}
    \caption{
    Probability densities of event-level statistics. 
    \textbf{a,} ball-possession time; 
    \textbf{b,} dribbling distance (distance traveled by the attacker over the course of an event; equivalent to the denominator of the error function \( \epsilon \)); 
    \textbf{c, d,} attacker--defender distance at the start and end of the event; 
    \textbf{e, f,} relative speed at the start and end of the event; 
    \textbf{g, h,} distance between attacker and the opponent's goal at the start and end of the event. 
    Insets are shown on a semi-logarithmic scale. 
    Data from \totfiltered events. 
    }
    \label{fig:event_distributions}
\end{figure}

Figure~\ref{fig:event_distributions} shows the distributions of event-related statistics, including ball-possession time, distance traveled by the attacker, and attacker--defender distance and relative speed at event onset and end. 
Table~\ref{tab:event_statistics} summarizes these data reporting their means, standard deviations, medians, and ranges. 
Both attacker--defender distance and relative speed are typically smaller at the end than at the start of an event (Table~\ref{tab:event_statistics}). 
This suggests that the defender tends to reduce distance to the attacker and relative speed over the course of a dribbling event. 
While relative-speed and distance are defined as objectives for minimization over an action set, the observed trend may reflect an intrinsic defensive tendency toward minimizing these quantities over both time and actions.

\subsection{Influence of the goal}

\begin{figure}[p]
    \centering
    \includegraphics[width=\linewidth]{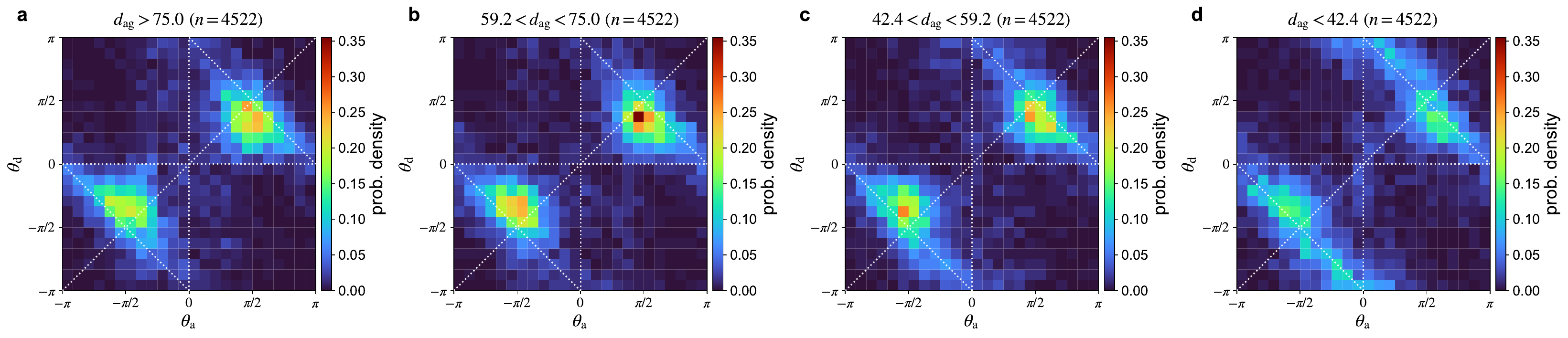}
    \caption{
    \textbf{a--d}, 
    Joint distributions of attacker and defender driving angles for different attacker--goal distance ranges. 
    Events were partitioned into four groups of approximately equal size according to the distance between the attacker and the opponent goal at the start of the event, denoted by \( d_{\mathrm{g}\at} \). 
    }
    \label{fig:ang_dist_attacker_goal}
\end{figure}

Our model does not explicitly account for the position of the goal. 
However, the angular constraint imposed during preprocessing (Condition~\ref{cond:angle}) implicitly constrains the geometric configuration of the attacker, defender, and the goal. 
Our analysis is therefore restricted to the cases where the defender is initially positioned roughly between their own goal and the attacker. 
In addition, the attacker--goal distance is approximately 7--14 times larger than the attacker--defender distance (Table~\ref{tab:event_statistics}), implying that the goal is not a dominant factor affecting player motion in the dataset. 
Figure~\ref{fig:ang_dist_attacker_goal} illustrates the joint distribution of attacker and defender angles as a function of attacker--goal distance at event outset. 
The empirical patterns are broadly consistent across panels, as well as with those reported in the main text. 
Near the goal [Fig.~\ref{fig:ang_dist_attacker_goal}d], however, the distribution becomes more diffuse and spreads along \( \mathcal{S}_2 \) (the anti-diagonal lines; see the main text for details). 
In other words, player motion is affected noticeably by the goal when they are close to it.
Case in point: defenders appear to favor the conservative strategy \( \mathcal{S}_2 \) more strongly than in the other panels, perhaps to play more cautiously. 
The slight increase in the joint density around \( (\ta, \td) \approx (0, \pm\pi) \) in Fig.~\ref{fig:ang_dist_attacker_goal}d also suggests that an attacker near the goal tends to accelerate toward the defender, as scoring probability is high and the risk of a counterattack is low; on the other hand, the defender has little choice but to retreat because of the increased defensive risk.

\clearpage

\begin{table}[p]
    \centering
    \caption{
    Summary of event-level statistics, reported as mean \( \pm \) standard deviation, median, and ranges, rounded to two decimal places. 
    Data from \totfiltered events.
    }
    \begin{tabular*}{\columnwidth}{@{\extracolsep{\fill}}lcccc}
        \toprule
        Quantity & Mean $\pm$ SD & Median & Min & Max \\
        \midrule
        Ball-possession time (s)
        & $2.85 \pm 1.40$
        & 2.52
        & 0.76
        & 15.96 \\
        
        Dribbling distance (distance traveled by the attacker) (m)
        & $11.09 \pm 6.14$
        & 9.11
        & 5.01
        & 77.02 \\
        
        Attacker--defender distance at start (m)
        & $8.76 \pm 4.99$
        & 8.40
        & 0.08
        & 31.09 \\
        
        Attacker--defender distance at end (m)
        & $4.11 \pm 3.20$
        & 3.01
        & 0.13
        & 27.98 \\
        
        Relative speed at start (\si{\meter\per\second})
        & $3.17 \pm 1.61$
        & 3.02
        & 0.01
        & 11.03 \\
        
        Relative speed at end (\si{\meter\per\second})
        & $2.21 \pm 1.39$
        & 1.95
        & 0.01
        & 11.66 \\
        
        Attacker--goal distance at start (m)
        & $59.30 \pm 20.22$
        & 59.26
        & 6.58
        & 108.80 \\
        
        Attacker--goal distance at end (m)
        & $55.88 \pm 21.03$
        & 56.00
        & 8.24
        & 110.34 \\
        \bottomrule
    \end{tabular*}
    \label{tab:event_statistics}
\end{table}

\clearpage

\clearpage

\section{Additional Supplementary Figures}

\subsection{Robustness of the geometric structure of modal angles}

\begin{figure}[H]
    \centering
    \includegraphics[width=0.8\linewidth]{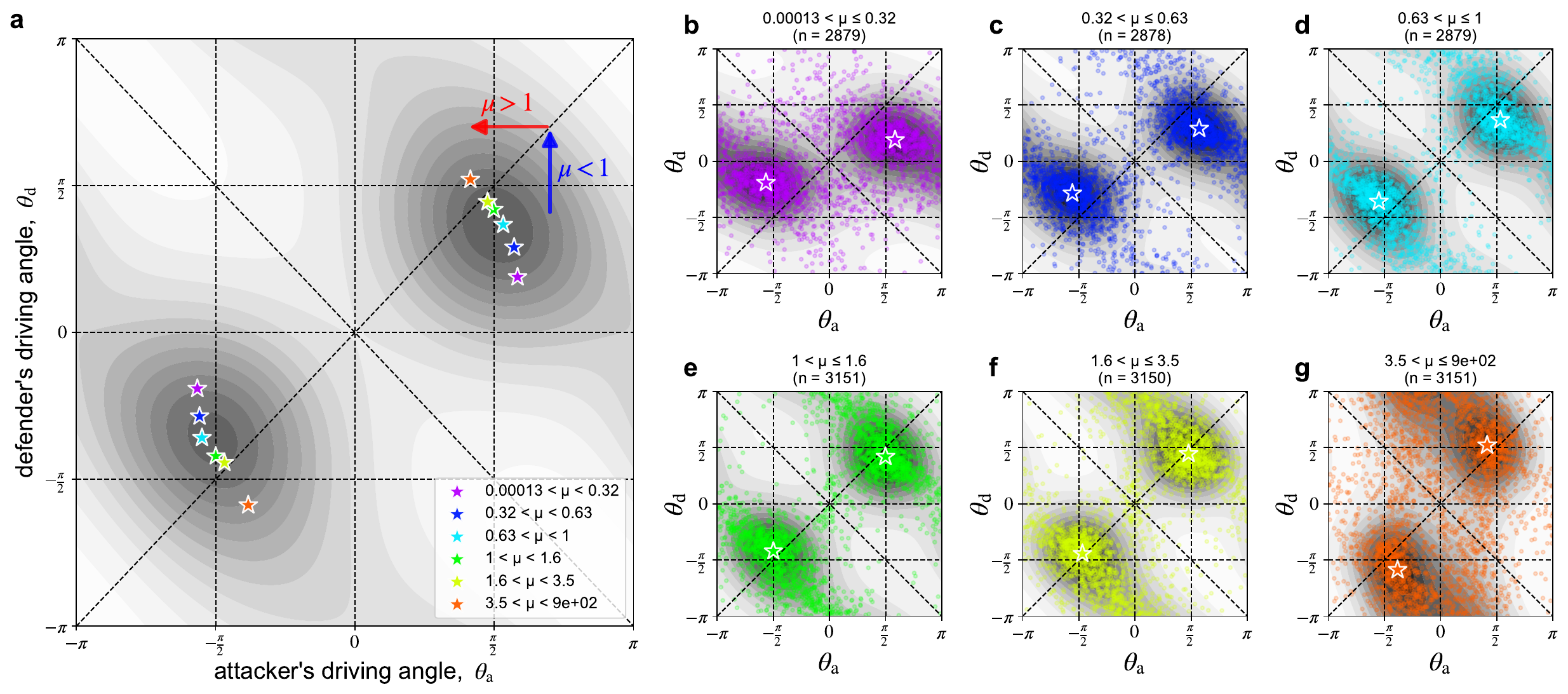}
    \caption{
    \( \kappa = 2 \).}
    \label{fig:kappa_2}
\end{figure}

\begin{figure}[H]
    \centering
    \includegraphics[width=0.8\linewidth]{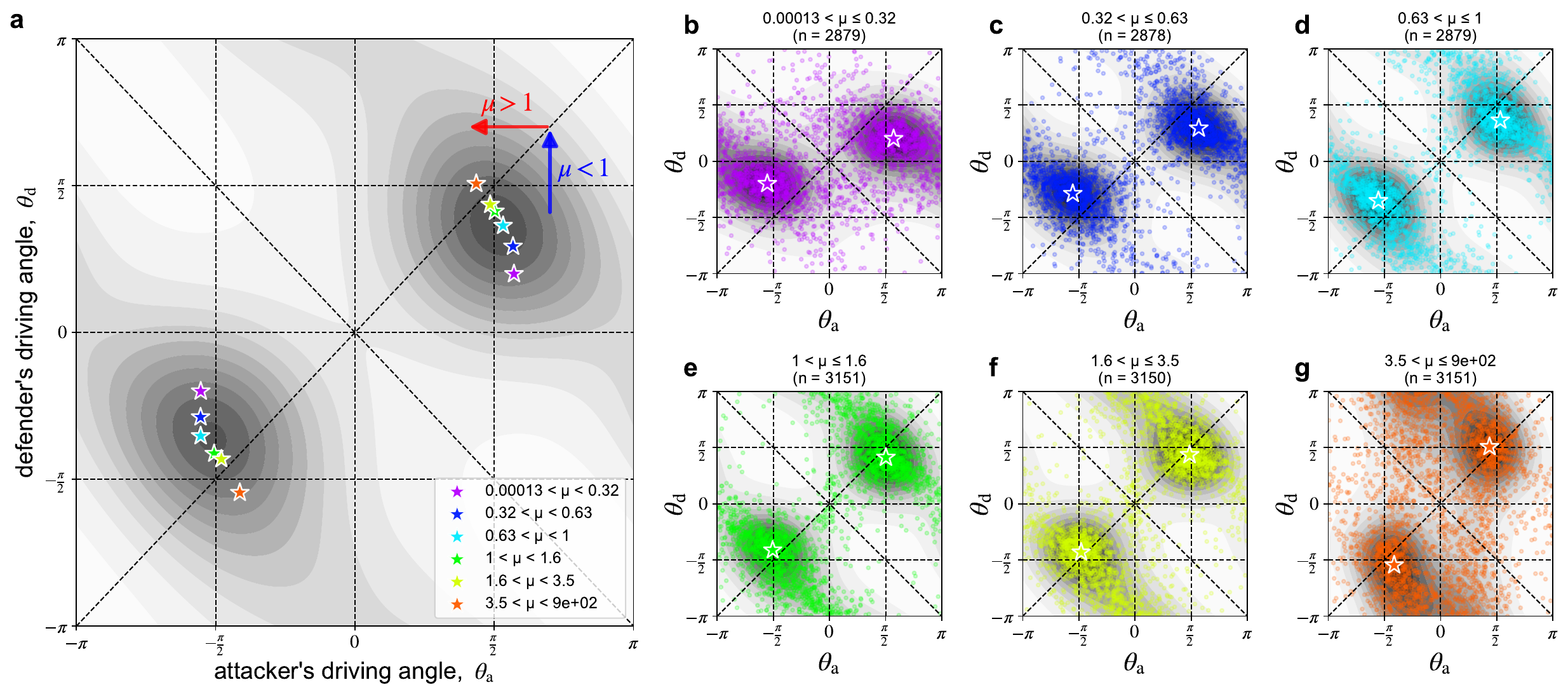}
    \caption{\( \kappa = 3 \).}
    \label{fig:kappa_3}
\end{figure}

\begin{figure}[H]
    \centering
    \includegraphics[width=0.8\linewidth]{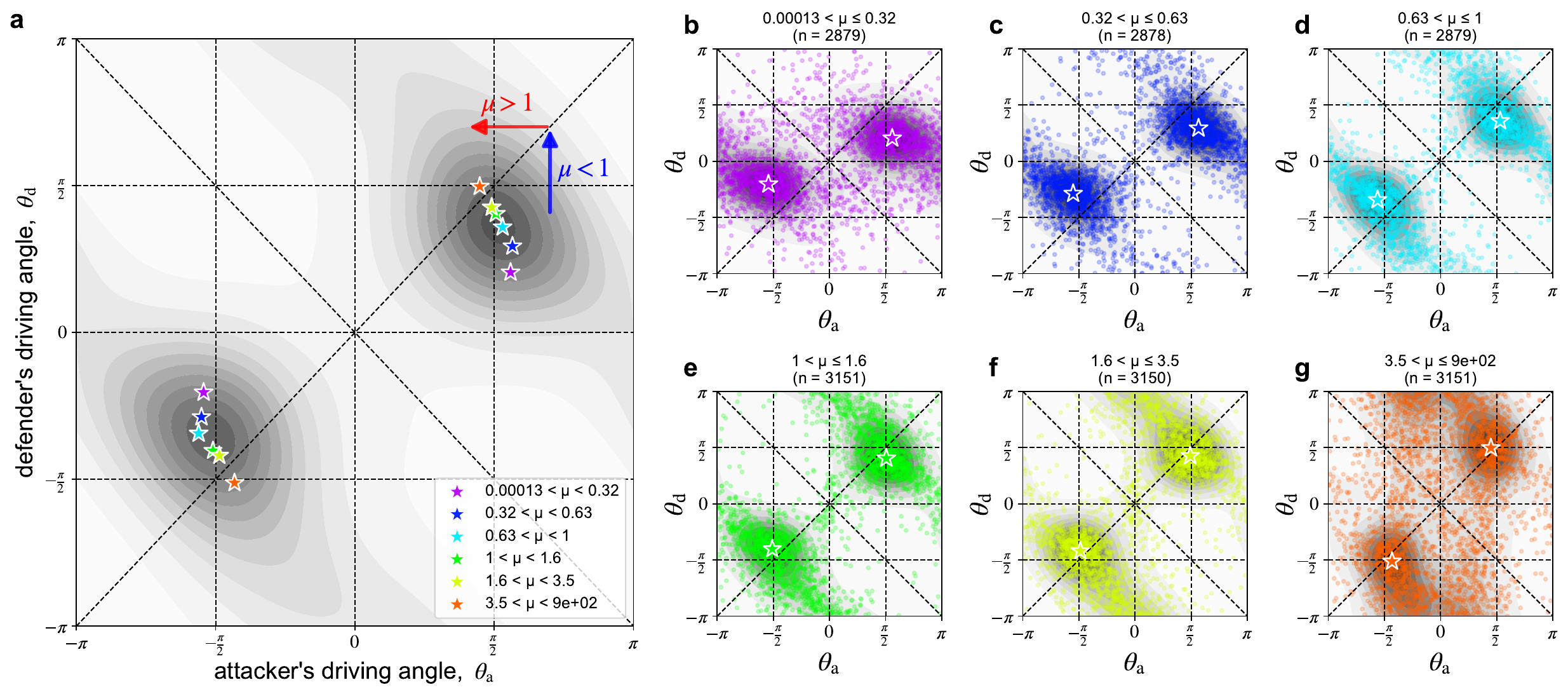}
    \caption{\( \kappa = 5 \).}
    \label{fig:kappa_5}
\end{figure}

\begin{figure}[H]
    \centering
    \includegraphics[width=0.8\linewidth]{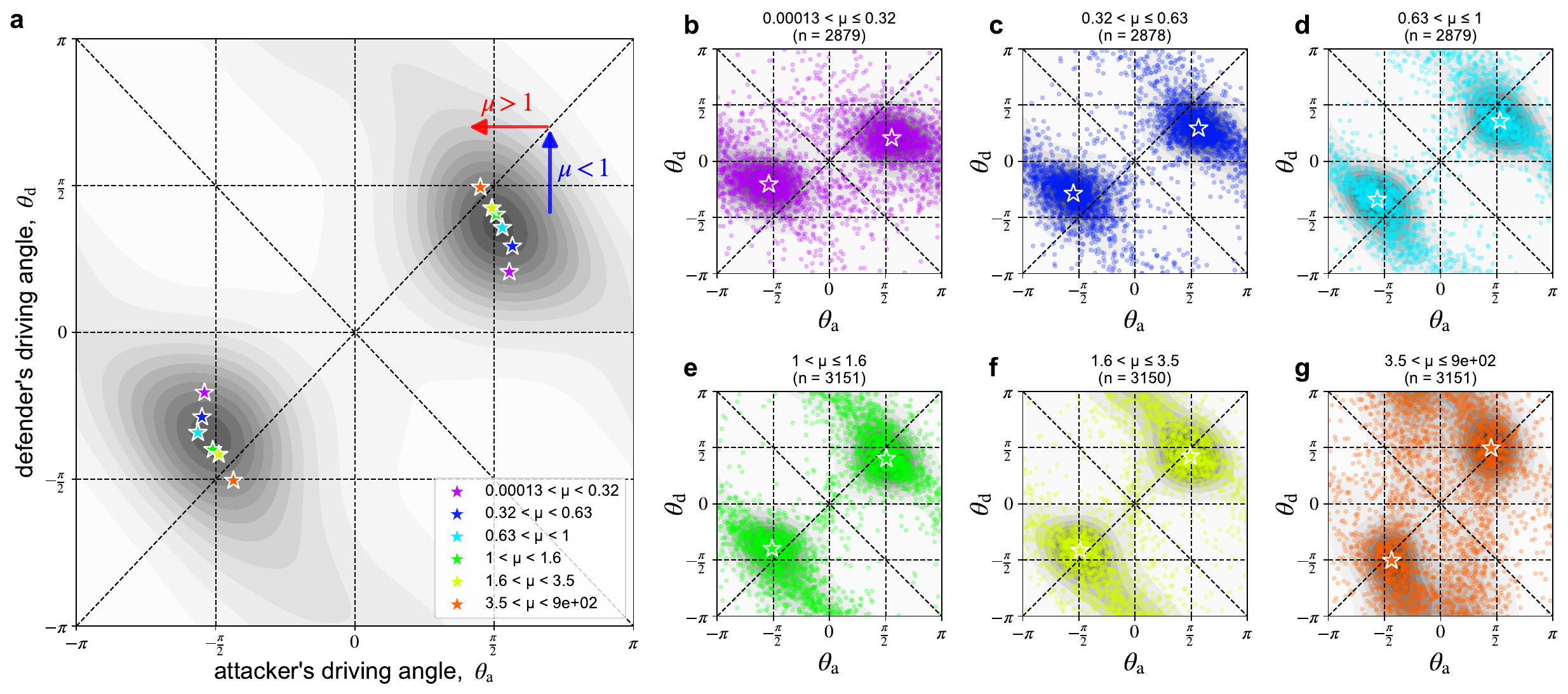}
    \caption{\( \kappa = 6 \).}
    \label{fig:kappa_6}
\end{figure}

\begin{figure}[H]
    \centering
    \includegraphics[width=0.8\linewidth]{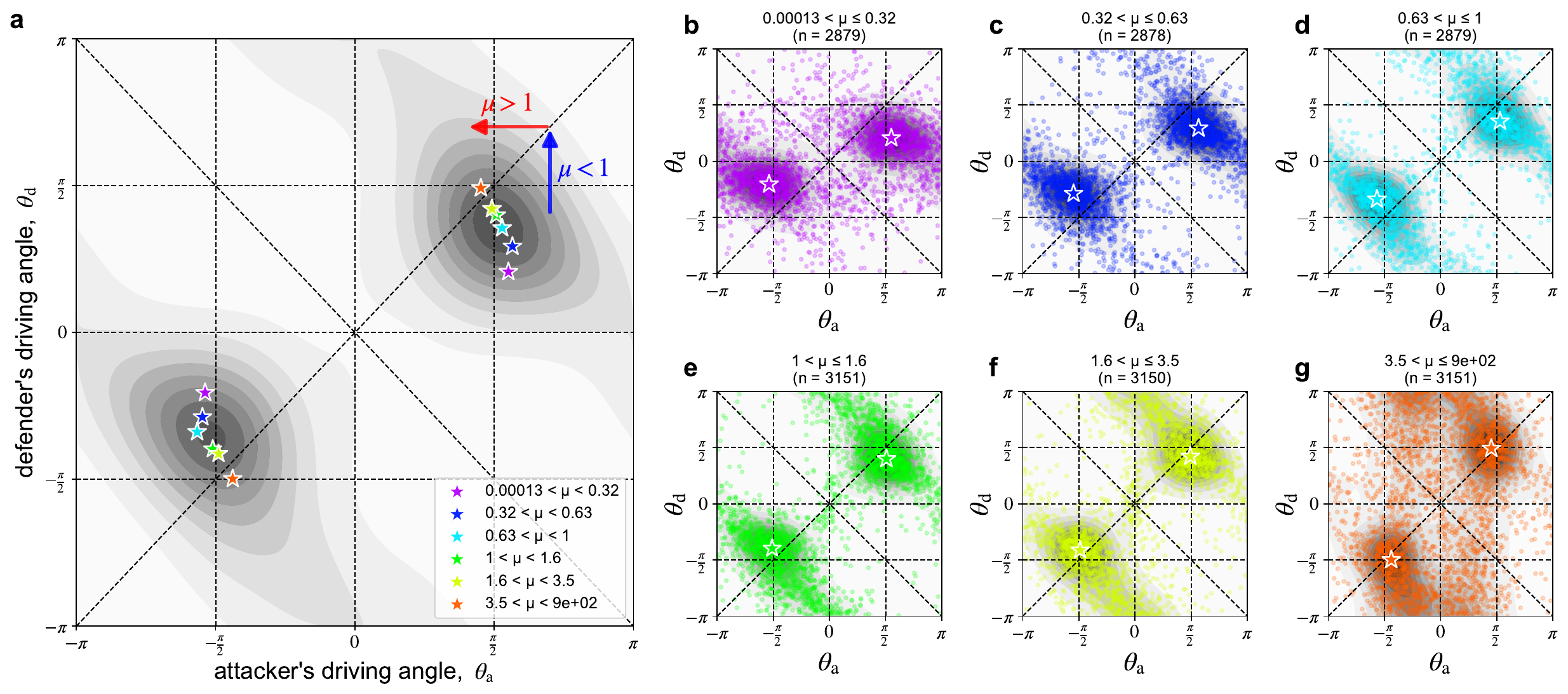}
    \caption{\( \kappa = 7 \).}
    \label{fig:kappa_7}
\end{figure}

\begin{figure}[H]
    \centering
    \includegraphics[width=0.8\linewidth]{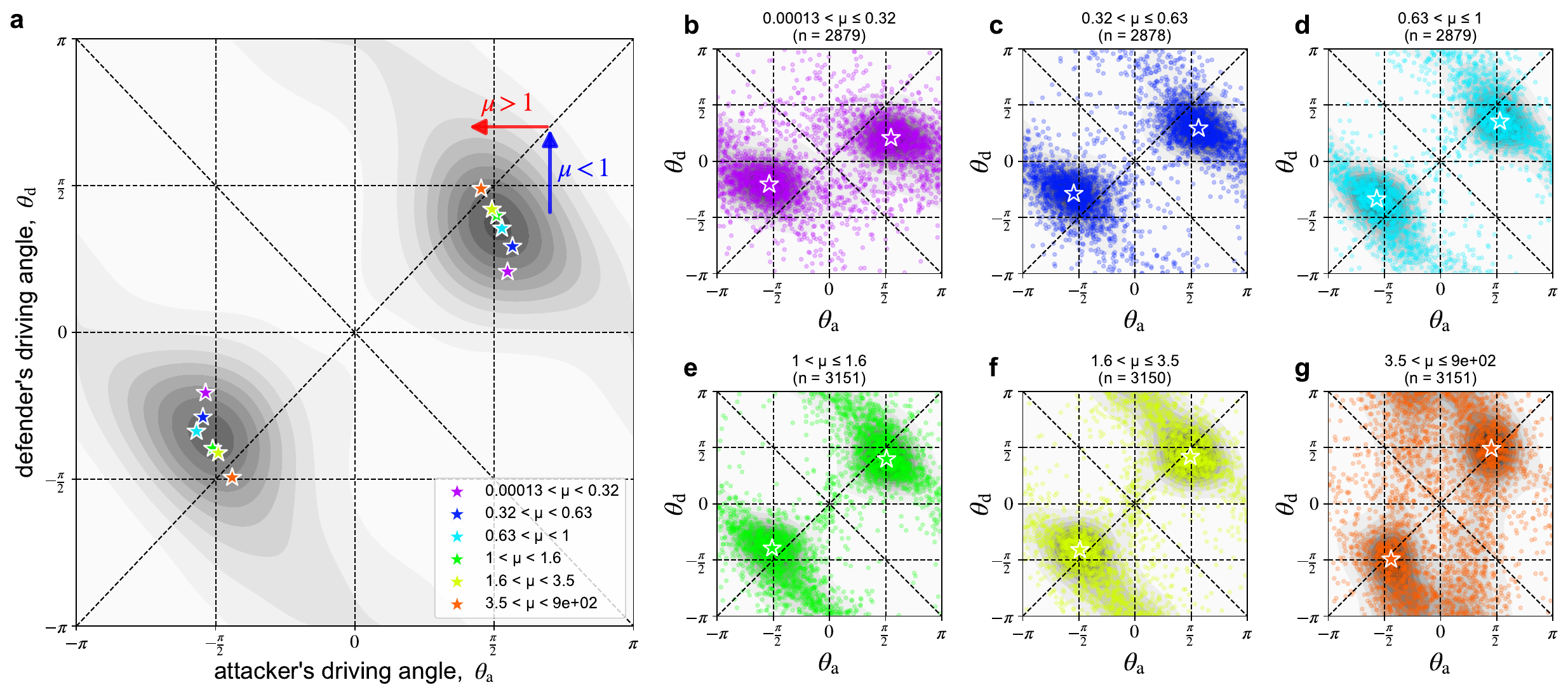}
    \caption{\( \kappa = 8 \).}
    \label{fig:kappa_8}
\end{figure}

\begin{figure}[H]
    \centering
    \includegraphics[width=0.8\linewidth]{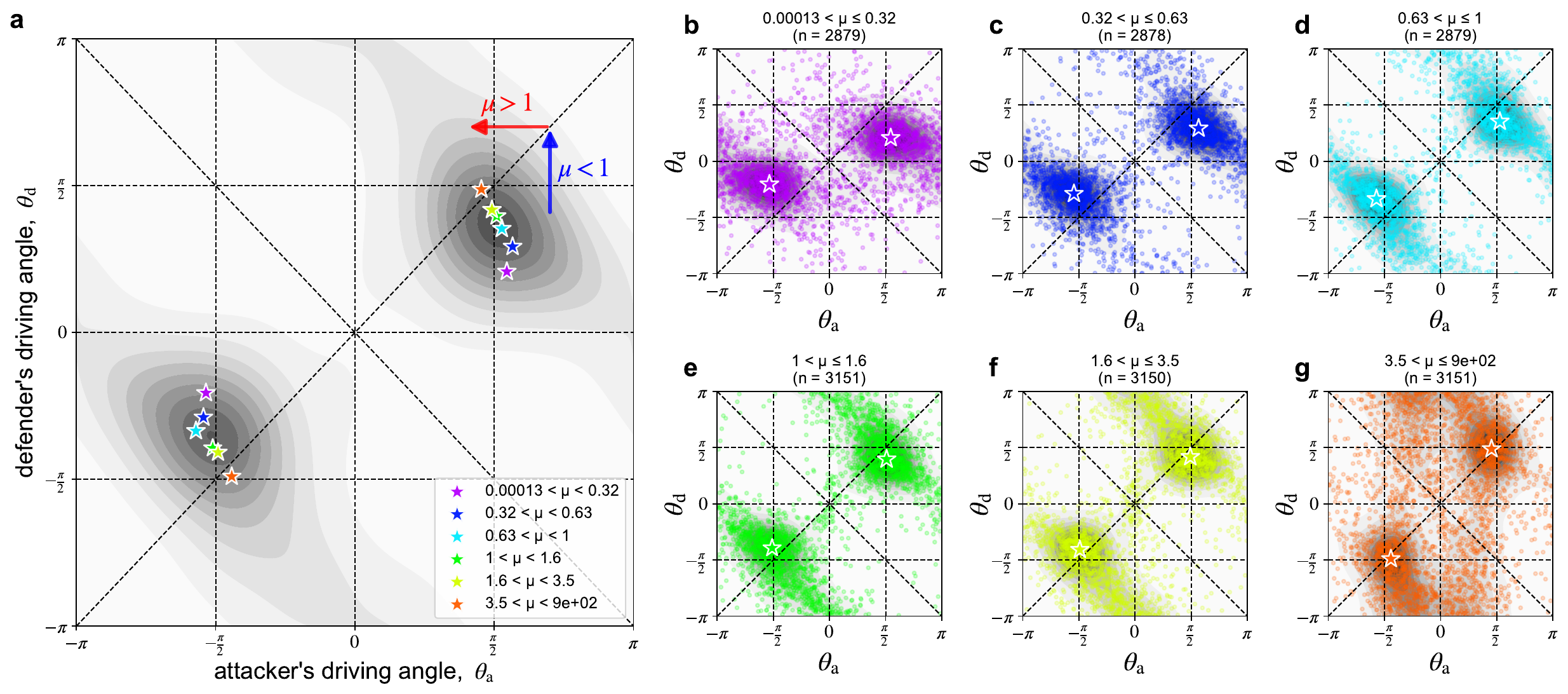}
    \caption{\( \kappa = 9 \).}
    \label{fig:kappa_9}
\end{figure}

\begin{figure}[H]
    \centering
    \includegraphics[width=0.8\linewidth]{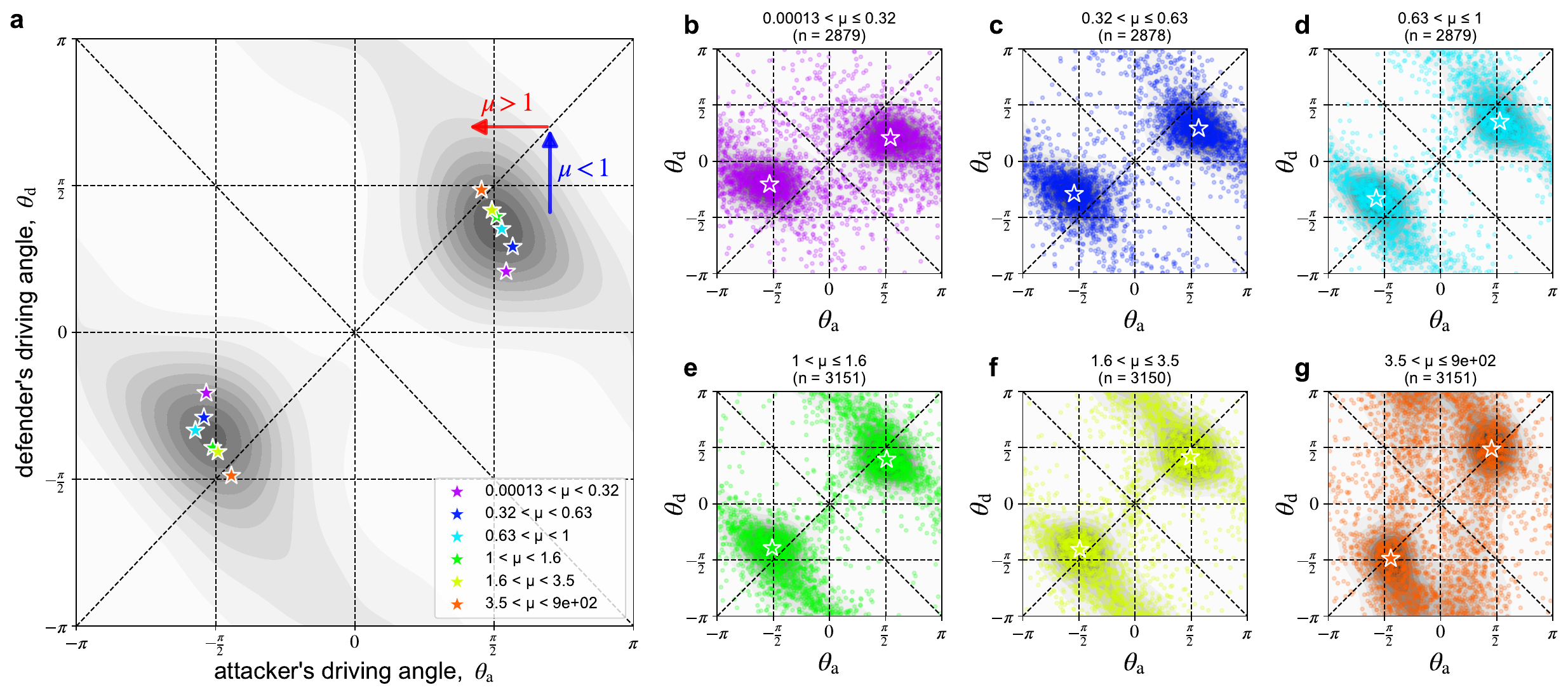}
    \caption{\( \kappa = 10 \).}
    \label{fig:kappa_10}
\end{figure}

Here we provide supplementary figures showing that the geometric structure of local angular peaks identified in the main text is robust across a wide range of smoothing factors \( \kappa \in [2,10] \) used for kernel density estimation (see Figs.~\ref{fig:kappa_2}--\ref{fig:kappa_10}). 
All plotting details are the same as in Fig.~\bl{4} of the main text.

\clearpage

\subsection{Finite-time objective landscapes}

\begin{figure}[H]
    \centering
    \includegraphics[width=0.9\linewidth]{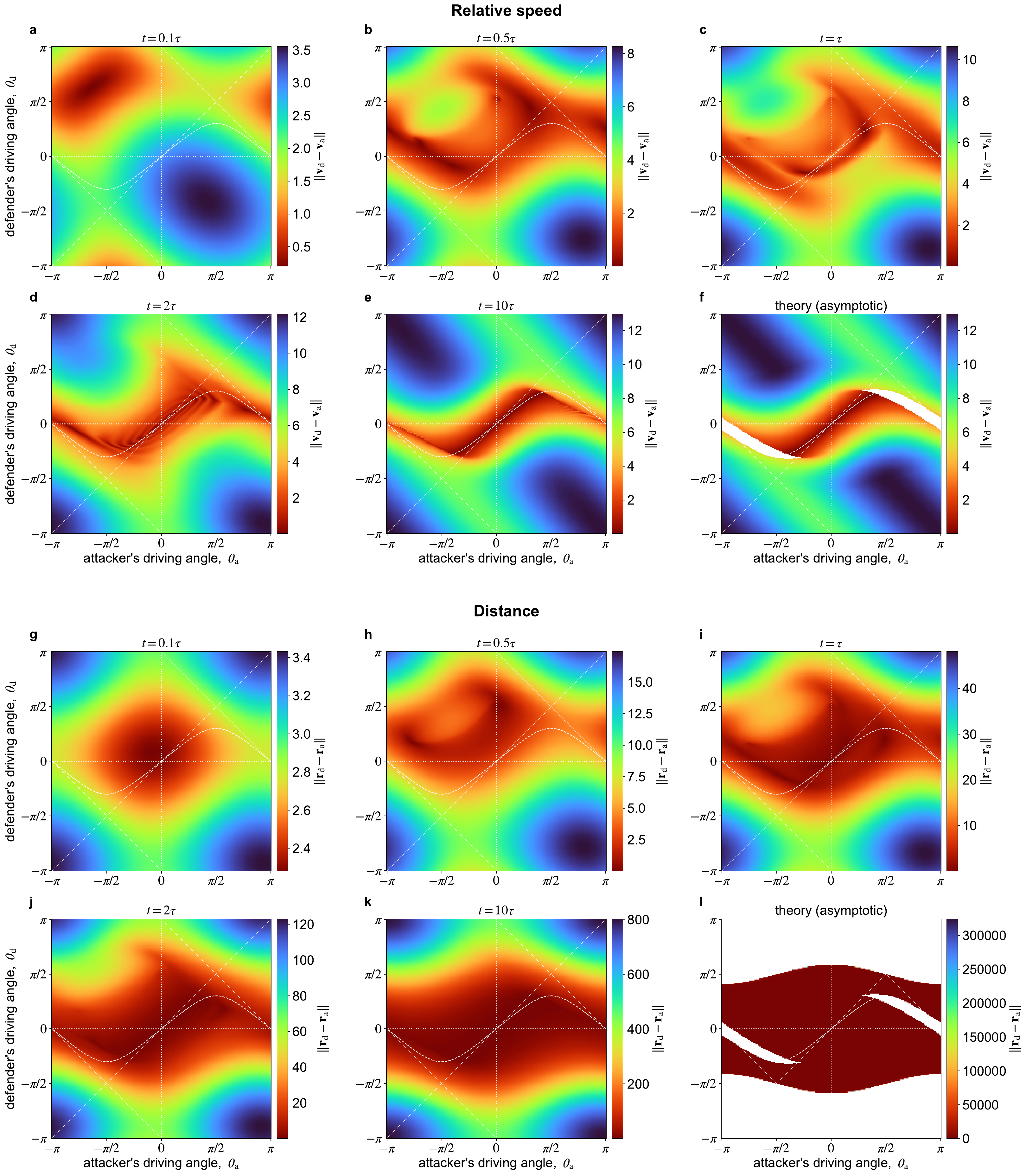}
    \caption{
    Time evolution of objective landscapes. 
    Plotting details are the same as in \ExFigs{4 and 5} of the main text. 
    Parameters and initial conditions: \( \fa = 1.25 \), \( \fd = 1.54 \), \( \ta = -2.81 \), \( \td = 1.13 \), \( \taua = 2.23 \), \( \taud = 6.64 \), \( \ra(0) = (74.82, 52.95) \), \( \rd(0) = (76.50, 50.15) \), \( \vat(0) = (3.66, 3.73) \), \( \vdf(0) = (4.07, 4.85) \). 
    These values were taken or estimated from a real one-on-one event in the dataset. 
    }
    \label{fig:landscape_1}
\end{figure}

\begin{figure}[H]
    \centering
    \includegraphics[width=0.9\linewidth]{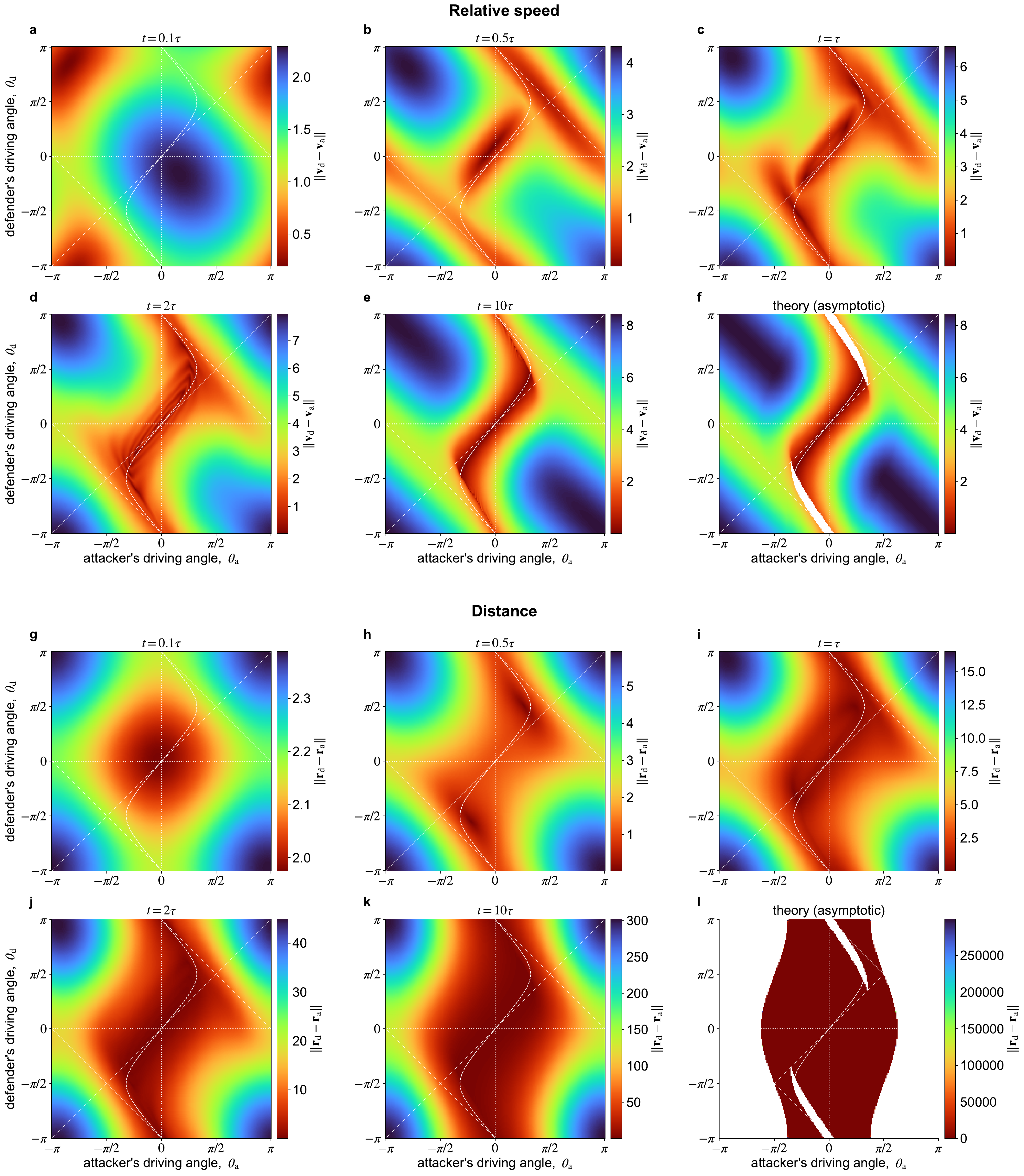}
    \caption{
    Time evolution of objective landscapes. 
    Plotting details are the same as in \ExFigs{4 and 5} of the main text. 
    Parameters and initial conditions: \( \fa = 1.59 \), \( \fd = 1.35 \), \( \ta = -2.25 \), \( \td = -0.37 \), \( \taua = 3.84 \), \( \taud = 1.71 \), \( \ra(0) = (37.14, 3.43) \), \( \rd(0) = (38.43, 5.83) \), \( \vat(0) = (-3.44, -1.27) \), \( \vdf(0) = (-5.13, -2.39) \). 
    These values were taken or estimated from a real one-on-one event in the dataset.  
    \label{fig:landscape_2}
    }
\end{figure}

We provide two additional examples of finite-time landscapes of relative speed and distance (Figs.~\ref{fig:landscape_1} and \ref{fig:landscape_2}).

\clearpage
\bibliography{ref}